\documentclass{IEEEtran}
\usepackage{multirow}
\usepackage{epsfig,graphicx,cite,amsmath}
\usepackage[algoruled, figure]{algorithm2e}
\begin{document}
%
\begin{titlepage}
\title{Distributed Rate Allocation Policies\\
for Multi-Homed Video Streaming\\
over Heterogeneous Access Networks}
\author{Xiaoqing Zhu$^{^{\ast}}$, Piyush Agrawal$^{\ast}$,
Jatinder Pal Singh$^{\dagger}$, Tansu Alpcan$^{\ddagger}$ and
Bernd Girod$^{^{\ast}}$ \\
       {$^{^{\ast}}  $Department of Electrical Engineering, Stanford University, Stanford, CA 94305, U.S.A.}\\
       {$^{\dagger}  $Deutsche Telekom R\&D Laboratories, Los Altos, CA 94022, U.S.A.}\\
       {$^{\ddagger} $Deutsche Telekom Laboratories, Ernst-Reuter Platz 7, Berlin 10587, Germany}%
       \thanks{This work appears in IEEE Transactions on Multimedia, 2009; applicable IEEE notice can be retrieved from IEEE website. A preliminary version of this work was presented in ACM 15th International Conference on Multimedia, September 2007, Augsburg, Germany~\cite{Zhu:MM07}. The authors can be contacted via email
       at $^{^{\ast}}$\{zhuxq,piyushag,bgirod\}@stanford.edu, $^{\dagger}$jatinder.singh@telekom.com,
       and $^{\ddagger}$tansu.alpcan@telekom.de.}}
\end{titlepage}
%
\maketitle \flushbottom \frenchspacing \sloppy

\begin{abstract}
We consider the problem of rate allocation among multiple
simultaneous video streams sharing multiple heterogeneous access
networks. We develop and evaluate an analytical framework for
optimal rate allocation based on observed available bit rate (ABR)
and round-trip time (RTT) over each access network and video
distortion-rate~(DR) characteristics. The rate allocation is
formulated as a convex optimization problem that minimizes the total
expected distortion of all video streams. We present a distributed
approximation of its solution and compare its performance against
\mbox{H$^\infty$-optimal} control and two heuristic schemes based on
TCP-style additive-increase-multiplicative-decrease (AIMD)
principles.
The various rate allocation schemes are evaluated in simulations of
multiple high-definition~(HD) video streams sharing multiple access
networks. Our results demonstrate that, in comparison with heuristic
AIMD-based schemes, both media-aware allocation and
H$^\infty$-optimal control benefit from proactive congestion
avoidance and reduce the average packet loss rate from 45\% to below
2\%. Improvement in average received video quality ranges between
1.5 to 10.7~dB in PSNR for various background traffic loads and
video playout deadlines. Media-aware allocation further exploits its
knowledge of the video DR characteristics to achieve a more balanced
video quality among all streams.
\end{abstract}
%
\section{Introduction}
With the proliferation of broadband access technologies such as
Ethernet, DSL, WiMax and IEEE 802.11a/b/g, portable devices tend to
possess multiple modes of connecting to the Internet. Most PDAs
provide both cellular and WLAN connectivity; laptops are typically
equipped with a built-in Ethernet port, an 802.11a/b/g card and a
phone jack for dial-up connections. Since a multitude of access
technologies will continue to co-exist, increasing efforts are
devoted to the standardization of architectures for network
convergence. Integration of heterogeneous access networks has been a
major consideration in the design of 4G
networks~\cite{Vidales:JSAC05}, IEEE 802.21~\cite{IEEE802.21}, and
the IP Multimedia Subsystem~(IMS) platform~\cite{IMS}. In addition,
multi-homed Internet access presents an attractive option from an
end-host's perspective. By pooling resources of multiple
simultaneously available access networks, it is possible to support
applications with higher aggregate throughput, lower latency, and
better error resiliency~\cite{Thompson:INFOCOM06}.\\
%
\indent In many applications, each end-host or device needs to
simultaneously support multiple application flows with heterogeneous
bit rate and latency requirements. One can easily imagine a
corporate user participating in a video conference call, while
uploading some relevant files to a remote server and browsing web
pages for reference. In the presence of many such users, each access
network can easily become congested with multiple competing
application flows from multiple devices. The problem of resource
allocation arises naturally, for determining the source rate of each
application flow, and for distributing the traffic among multiple
simultaneously available access networks. In this work, we focus on
video streaming applications as they impose the most demanding rate
and latency requirements. Flows from other applications, such as web
browsing and file transfer, are treated as background traffic. \\
%
\indent Challenges in the design of a rate allocation policy for
such a system are multi-fold. Firstly, access networks differ in
their attributes such as available bit rates (ABRs) and round trip
times (RTTs), which are time-varying in nature. Secondly, video
streaming applications differ in their latency requirements and
distortion-rate~(DR) characteristics. For instance, a
high-definition~(HD) video sequence containing dynamic scenes from
an action movie requires much higher data rate to achieve the same
quality as a static head-and-shoulder news clip for a mobile device.
Thirdly, unlike file transfer or web browsing, video streaming
applications require timely delivery of each packet to ensure
continuous media playout. Late packets are typically discarded at
the receiver, causing drastic quality degradation of the received
video due to error propagation at the decoder. In addition, the rate
allocation policy should also operate in a distributed manner to
avoid the traffic overhead and additional delay in collecting
global media and network information for centralized computation.\\
%
\indent This paper addresses the above considerations, and
investigates a suite of distributed rate allocation policies for
multi-homed video streaming over heterogeneous access
networks:\footnote{While our system model is general enough to
accommodate best-effort networks such as the Internet, data service
over cellular networks and 802.11 wireless home or corporate
networks, it may not apply to scenarios where the service provider
performs admission control or resource provisioning according to
traffic load, e.g., in carrier grade WLAN networks or properly
dimensioned UMTS networks for voice services. The extension of the
current work to accommodate more general network types and behaviors
is an interesting area of future research, and goes beyond the scope
of this paper.}\
%
\begin{itemize}
\item
\emph{Media-Aware Allocation}: When devices have information of both
video DR characteristics and network ABR/RTT attributes, we
formulate the rate allocation problem in a convex optimization
framework and minimize the sum of expected distortions of all
participating streams. A distributed approximation to the
optimization is presented, to enable autonomous rate allocation at
each device in a media- and network-aware fashion.\
\item
\emph{H$^\infty$-Optimal Control}: In the case where media-specific
information is not available to the devices, we propose a scheme
based on H$^\infty$-optimal control~\cite{Basar:95}. The scheme
achieves optimal bandwidth utilization on all access networks by
guaranteing a worst-case performance bound characterizing deviation
from full network utilization and excessive fluctuations in
allocated video rates.\
\item
\emph{AIMD-Based Heuristic}: For comparison, we present two
heuristic rate allocation schemes that react to congestion in the
network by adjusting the total rate of each stream following
TCP-style additive-increase-multiplicative-decrease (AIMD)
principle~\cite{Jacobson:SIGCOMM88}. They differ in their manners of
how rates are split among multiple access networks in accordance
with observed ABRs.\
\end{itemize}

%
\indent Performance of all four rate allocation policies are
evaluated in \texttt{ns-2}~\cite{ns2}, using ABR and RTT traces
collected from Ethernet, IEEE 802.11b and IEEE 802.11g networks in a
corporate environment. Simulation results are presented for the
scenario of simultaneous streaming of multiple high-definition~(HD)
video sequences over multiple access networks. We verify that the
proposed distributed media-aware allocation scheme approximates the
results from centralized computation closely. The allocation results
react quickly to abrupt changes in the network, such as arrival or
departure of other video streams. Both media-aware allocation and
H$^\infty$-optimal control schemes achieve significantly lower
packet delivery delays and loss ratios (less than 0.1\% for
media-aware allocation and below 2.0\% for H$^\infty$-optimal
control), whereas AIMD-based schemes incur up to 45\% losses, far
exceeding the tolerance level of video streaming applications. As a
result, media-aware allocation improves the average received video
quality by 1.5~-~10.7~dB in PSNR over the heuristic schemes in
various simulation settings. It further ensures equal utilization
across all access networks and more balanced video quality among
all streams. \\
%
\indent The rest of the paper is organized as follows.
Section~\ref{sec:RelatedWork} briefly reviews related work in
multi-flow, multi-network resource allocation. We present our system
model of the access networks and expected video distortion in
Section~\ref{sec:SystemModel}, followed by descriptions of the rate
allocation schemes in Section~\ref{sec:RateAllocation}. Performances
of the four schemes are evaluated in
Section~\ref{sec:PerformanceEvaluation} via simulations of three HD
video streaming sessions sharing three access networks under various
traffic conditions and latency requirements.
%
\section{Related Work}
\label{sec:RelatedWork}
%
Rate allocation among multiple flows that share a network is an
important and well-studied problem. Internet applications typically
use the TCP congestion control mechanism for regulating their
outgoing rate~\cite{Jacobson:SIGCOMM88}\cite{TCP_rfc}. For media
streaming applications over UDP, TCP-Friendly Rate Control (TFRC) is
a popular choice~\cite{Floyd:ITN99}\cite{TFRC_rfc}. Several
modifications have been proposed to improve its
media-friendliness~\cite{Wang:ACM04}. In \cite{Kelly:JORS98}, the
problem of rate allocation among flows with different utilities is
studied within a mathematical framework, where two classes of
pricing-based distributed rate allocation algorithms are analyzed.
In this work, the notion of utility of each flow corresponds to its
expected received video quality, measured in terms of
mean-squared-error~(MSE) distortion relative to the original
uncompressed video signals. We also extend the mathematical
framework in \cite{Kelly:JORS98} to consider rate allocation over
multiple networks.\\
\indent The problem of efficient utilization of multiple networks
via suitable allocation of traffic has been explored from different
perspectives. A game-theoretic framework to allocate bandwidth for
elastic services in networks with fixed capacities is described in
\cite{Yaiche:ITN00,Alpcan:ITN05,Alpcan:CDC03}. Our work, in
contrast, acknowledges the time-varying nature of the network
attributes and dynamically updates the allocation results according
to observed available bit rates and round-trip delays.
A solution for addressing the handoff, network selection, and
autonomic computation for integration of heterogeneous wireless
networks is presented in~\cite{Vidales:JSAC05}. The work, however,
does not address simultaneous use of heterogeneous networks and does
not consider wireline settings.
A cost-price mechanism is proposed for splitting traffic among
multiple IEEE 802.11 access points to achieve end-host
multi-homing~\cite{Shakkottai:INFOCOM06}\cite{Shakkottai:JSAC07}.
The work does not take into account existence of other types of
access networks or the characteristics of traffic, nor does it
specify an operational method to split the traffic.
In \cite{Thompson:INFOCOM06}, a flow scheduling framework is
presented for collaborative Internet access, based on modeling and
analysis of individual end-hosts' traffic behavior. The framework
mainly accounts for TCP flows and uses metrics useful for web
traffic including RTT and throughput for making scheduling
decisions. \\
\indent Rate adaptation of multimedia streams has been studied in
the context of heterogeneous networks in \cite{Szwabe:PV06}, where
the authors propose an architecture to allow online measurement of
network characteristics and video rate adaptation via transcoding.
Their rate control algorithm is based on TFRC and is oblivious of
the media content.
In \cite{Jurca:PV06}, media-aware rate allocation is achieved, by
taking into account the impact of both packet loss ratios and
available bandwidth over each link, on the end-to-end video quality
of a single stream, whereas in \cite{Zhu:PV06}, the rate allocation
problem has been formulated for multiple streams sharing one
wireless network.
Unlike our recent work where the multi-stream multi-network rate
allocation problem is addressed from the perspective of stochastic
control of Markov Decision Processes~\cite{Singh:WoWMoM07} and
robust H$^\infty$-optimal control of linear dynamic
systems~\cite{Alpcan:WiOpt07}\cite{Alpcan:ITMC}, in this paper we
stay within the convex optimization framework for media-aware
optimal rate allocation, and compare the performance of the scheme
with prior approaches. Preliminary results from this work have been
reported in \cite{Zhu:MM07} and \cite{Singh:MNCNA07}.\
%
\section{System Model}
\label{sec:SystemModel}
\begin{figure}
\centering
\includegraphics[width=0.7\columnwidth]{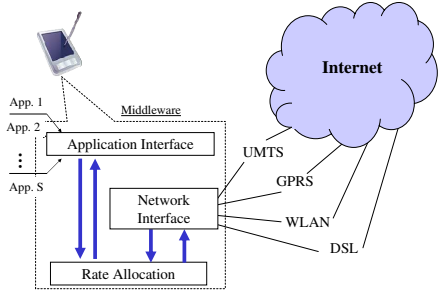}
\caption{\small Middleware functionality in a device. The rate
allocation module collects the observed media statistics and network
characteristics (e.g., ABR and RTT), and dictates the rate
allocation among application streams, over each network interface.}
\label{fig:SystemDiagram}
\end{figure}

%
In this section, we introduce the mathematical notations used for
modeling the access networks, and for estimating expected received
video distortion of each stream. We envision a middleware
functionality as depicted in Fig.~\ref{fig:SystemDiagram}, which
collects characteristic parameters of both the access networks and
video streams, and performs the optimal rate allocation according to
one of the schemes described in Section~\ref{sec:RateAllocation}. A
more detailed discussion of the middleware functionality can be
found in~\cite{Singh:MNCNA07}.\
%
%
\subsection{Network Model}
\label{subsec:NetworkModel}
%
Consider a set of access networks ${\cal N}=\{1, 2, \ldots, N\}$,
simultaneously available to multiple devices. Each access network
$n$ is characterized by its available bit rate $c_n$ and round trip
time $\tau_n$, which are measured and updated periodically. For each
device, the set of video streams is denoted as ${\cal S}=\{1, 2,
\ldots, S\}$. Traffic allocation can be expressed in matrix form:
${\mathbf{r}}$ = $\{r^s_n\}_{S \times N}$, where each element
$r^s_n$ corresponds to the allocated rate of Stream $s$ over Network
$n$. Consequently, the total allocated rate over Network $n$ is $r_n
= \sum_{s}{r^s_n}$, and the total allocated rate for Stream $s$ is
$r^s = \sum_{n}r^s_n$. We denote the \emph{residual bandwidth} over
Network $n$ as:\
%
\begin{equation}
\label{eqn:Bn}
e_n=c_n-\sum_{s \in {\cal S}}r^s_n = c_n - r_n.
\end{equation}
%
\noindent From the perspective of Stream $s$, the observed available
bandwidth is:\
%
\begin{equation}
\label{eqn:ABRPerStream}
c^s_n=c_n-\sum_{s' \neq s}r^{s'}_n.
\end{equation}
\noindent Note that $e_n = c_n - r_n = c^s_n - r^s_n$. \\
\indent As the allocated rate on each network approaches the maximum
achievable rate, average packet delay typically increases due to
network congestion. We use a simple rational function to approximate
the non-linear increase of packet delay with traffic rate over each
network:\
%
\begin{equation}
\label{eqn:Delay}
t_n = \frac{{\alpha}_n}{e_n}=\frac{{\alpha}_n}{c_n - r_n}
=\frac{{\alpha}_n}{c^s_n - r^s_n},
\end{equation}
%
\noindent The value of ${\alpha}_n$ is estimated from past
observations of $\tau_n$ and $e_n$, assuming equal delay on both
directions:\footnote{For multi-homed end hosts, acknowledgement
packets for traffic sent over each network interface are returned
over the same network. Therefore RTT is a good indication of network
congestion, occurring either on the forward or backward path.}\
%
\begin{equation}
\label{eqn:alpha}
{\alpha}_n = \frac{e_n \tau_n}{2}.
\end{equation}
\noindent We note that despite oversimplification in this delay
model, it is still effective in driving a rate allocation scheme
with proactive congestion avoidance, as can be verified later by
simulation results in Section~\ref{sec:PerformanceEvaluation}.\
%
\subsection{Video Distortion Model}
\label{subsec:VideoDistortionModel}
%
Expected video distortion at the decoder comprises of two terms:\
%
\begin{equation}
\label{eqn:DistortionDec}
d_{dec} = d_{enc} + d_{loss},
\end{equation}
%
\noindent where $d_{enc}$ denotes the distortion introduced by lossy
compression performed by the encoder, and $d_{loss}$ represents the
additional distortion caused by packet loss~\cite{Stuhlmuller:00}.\\
\indent The distortion-rate~(DR) characteristic of the encoded video
stream can be fit with a parametric model~\cite{Stuhlmuller:00}:\
%
\begin{equation}
\label{eqn:DistortionEnc}
d^s (r^s) = d^s_0 + \frac{{\theta}^s}{(r^s - r^s_0)},
\end{equation}
%
\noindent where the parameters $d^s_0$, ${\theta}^s$ and $r^s_0$
depend on the coding scheme and the content of the video. They can
be estimated from three or more trial encodings using non-linear
regression techniques. To allow fast adaptation of the rate
allocation to abrupt changes in the video content, these parameters
are updated for each group of pictures (GOP) in the encoded
video sequence, typically once every 0.5~second.\\
\indent The distortion introduced by packet loss due to transmission
errors and network congestion, on the other hand, can be derived
from \cite{Zhu:EURASIP05} as:\
%
\begin{equation}
\label{eqn:DistortionLoss}
d^s_{loss} = {\kappa}^s p^s_{loss},
\end{equation}
%
\noindent where the sensitivity factor ${\kappa}^s$ reflects the
impact of packet losses $p^s_{loss}$, and depends on both the video
content and its encoding structure. In general, packet losses are
caused by both random transmission errors and overdue delivery due
to network congestion. Since looses over the former type cannot be
remedied by means of mindful rate allocation, we choose to omit its
contribution in modeling decoded video distortion. For simplicity,
$p^s_{loss}$ comprises solely of late losses due to network
congestion in the rest of this paper. \
%
\section{Distributed Rate Allocation}
\label{sec:RateAllocation}
%
In this section, we address the problem of rate allocation among
multiple streams over multiple access networks from several
alternative perspectives. We first present a convex optimization
formulation of the problem in
Section~\ref{subsec:DistortionMinimized}, and explain how to
approximate the media- and network-aware optimal solution with
decentralized calculations. In the case that video DR
characteristics are unavailable, we resort to a formulation of
H$^\infty$-optimal control in Section~\ref{subsec:Hinfinity}, which
dynamically adjusts the allocated rate of each stream according to
fluctuations in observed network available bandwidth. For
comparison, we include in Section~\ref{subsec:AIMD} two heuristic
allocation schemes following TCP-style
additive-increase-multiplicative-decrease~(AIMD) principle. All four
schemes are distributed in nature, in that the rate allocation
procedures performed by each stream does not need coordination or
synchronization with other streams. Rather, interactions between the
streams are \emph{implicit}, as the ABRs and RTTs observed by one
stream are affected by the allocated rates of other competing
streams sharing the same interface networks. \
%
\subsection{Media-Aware Allocation}
\label{subsec:DistortionMinimized}
%
We seek to minimize the total expected distortion of all video
streams sharing multiple access networks:\
%
\begin{eqnarray}
\label{eqn:Objective}
\min_{{\mathbf{r}}} &  \sum_s {d^s_{dec}(r^s, p^s_{loss})} \\
s. t.        & r^s = \sum_n {r^s_n}, \:\:\:  \forall s \in {\cal S}  \\
             & r_n = \sum_s {r^s_n} < c_n, \:\:\:\forall n \in {\cal N} \\
             & r^s_n = {\rho}_n r^s, \:\:\: \forall n \in {\cal N}
             \label{eqn:RateAllocRatio}.
\end{eqnarray}
%
\noindent In \eqref{eqn:Objective}, the expected distortion
$d^s_{dec}$ is a function of the allocated rate $d^s$ and average
packet loss $p^s_{loss}$ according to \eqref{eqn:DistortionDec}. The
constraint \eqref{eqn:RateAllocRatio} is introduced to impose
uniqueness of the optimal solution. We choose $\rho_n =
c_n/\sum_{n'}c_{n'}$ to ensure balanced utilization over each
interface:\
%
\begin{equation}
\label{eqn:EqualUtilization}
\frac{r_n}{c_n} = \frac{\sum_s r^s_n}{c_n}
= \frac{{\rho}_n \sum_s r^s}{c_n}
= \frac{\sum_{n'} r_{n'}}{ \sum_{n'} c_{n'}},
\forall n \in {\cal N}.
\end{equation}
%
\noindent It can also be shown that ${\rho}_n =
c^s_n/\sum_n{c^s_n}$, $\forall{s}\in{\cal{S}}$. Each stream can
therefore calculate the value of ${\rho}_n$ independently, based on
its own ABR observation $c^s_n$ for Network~$n$.\\
\indent The average packet loss $p^s_{loss}$ for each stream is the
weighted sum of packet losses over all networks:\
%
\begin{equation}
\label{eqn:Loss}
p^s_{loss} = \sum_n {\rho}_n e^{-t^s_0/t_n}.
\end{equation}
%
\noindent Following the derivations in \cite{Zhu:EURASIP05}, the
percentage of late packets is estimated as $e^{-t^s_0/t_n}$,
assuming exponential delay distributions with average $t_n$ for
Network~$n$ and playout deadline $t^s_0$ for Stream~$s$. Given
\eqref{eqn:Delay}, $p^s_{loss}$ is expressed as:\
%
\begin{equation}
\label{eqn:P}
p^s_{loss} = \sum_{n} {{\rho}_n e^{-t^s_0(c^s_n-r^s_n)/{\alpha}_n}}.
\end{equation}
%
\indent Combining \eqref{eqn:DistortionDec}-\eqref{eqn:P}, it can be
easily confirmed that the optimization objective is a convex
function of the variable matrix ${\mathbf{r}}$. If all the
observations and parameters were available in one place, the
solution could be found
by a suitable convex optimization method~\cite{Boyd:04}.\\
\indent We desire to minimize the objective~\eqref{eqn:Objective} in
a distributed manner, with as little exchange of information among
the devices as possible. One approach is to consider the impact of
network congestion on one stream at a time, and alternate between
the streams until convergence. From the perspective of Stream~$s$,
its contribution to \eqref{eqn:Objective} can be rewritten as:\
%
\begin{eqnarray}
\label{eqn:AlternateStream}
\min_{r^s} & d^s(r^s)  + {\kappa}^s\sum_n {\rho}_n
e^{-t^s_0(c^s_n-r^s_n)/{\alpha}_n} \nonumber \\
& + \sum_{n}\sum_{s' \neq s} {\rho}_n{\kappa}^{s'} e^{-t^{s'}_0(c^s_n - r^s_n)/{\alpha}_n} \\
s. t.  &  r^s_n = {\rho}_n r^s, \forall n \in {\cal N}  \nonumber \\
\:     &  r^s_n < c^s_n ,     \forall n \in {\cal N}. \nonumber
\end{eqnarray}
\noindent In \eqref{eqn:AlternateStream}, optimization of rate
allocation for Stream~$s$ requires knowledge of not only its own
distortion-rate function~$d^s(r^s)$ and packet loss sensitivity
${\kappa}^s$, but also its impact on the late loss of other streams
via the parameters ${\kappa}^{s'}$ and $t^{s'}_0$. While each stream
can obtain information regarding its own packet loss sensitivity and
playout deadline, exchange of such information among different
streams is undesirable for a distributed scheme.\\
\indent We therefore further simplify the optimization to:\
%
\begin{eqnarray}
\label{eqn:AlternateStreamSimplified}
\min_{r^s} & d^s(r^s)  +
\sum_{n}{\kappa}' {\rho}_n e^{-t^s_0 (c^s_n - r^s_n)/{\alpha}_n} \\
s.t.    &   r^s_n = {\rho}_n r^s,  \forall n \in {\cal N}  \nonumber \\
\:      &   r^s_n < c^s_n,       \forall n \in {\cal N}, \nonumber
\end{eqnarray}
\noindent where ${\kappa}'$ is empirically tuned to control the
scheme's aggressiveness. Even though
\eqref{eqn:AlternateStreamSimplified} does not necessarily lead to
an optimal solution for \eqref{eqn:Objective}, it nevertheless
incorporates considerations of both network congestion and encoder
video distortion in choosing the optimal rates. The impact on other
streams is captured implicitly by the second term in
\eqref{eqn:AlternateStreamSimplified},reflecting congestion
experienced by all streams traversing that network. Effectiveness of
this distributed approximation will be verified later in
Section~\ref{subsec:Cent}. \\
\indent In essence, optimization of
\eqref{eqn:AlternateStreamSimplified} involves a one-dimensional
search of $r^s$, thus can be solved efficiently using numerical
methods. Computational complexity of the scheme increases linearly
with the number of competing streams $S$ and the number of available
access networks $N$, on the order of ${\cal O}(NS)$.
In practice, each stream needs to track its observations of
$c^s_n$'s and $\tau_n$'s over all available access networks, and to
observe its video DR parameters $\theta^s$ and $r^s_0$. At each time
instance, the scheme would update its estimate of ${\alpha}_n$
according to \eqref{eqn:alpha}. It then determines the allocated
rate $r^s$ by minimizing \eqref{eqn:AlternateStreamSimplified}, and
divides up the rate in proportion to ${\rho}_n$ over respective
networks. Figure~\ref{fig:MediaAwareAlgorithm} summarizes these
procedures.\
%
\begin{algorithm}
\KwIn{ABR and RTT measurements of available access networks $c^s_n$, $\tau_n$;\\
$\:\:\:\:\:\:\:\:\:\:\:$  video DR characteristic $\theta^s$,
$r^s_0$ for the current GOP;}
\KwSty{Parameters: $\:$}{level of aggressiveness $\kappa'$;}\\
\KwOut{Allocated rate $r^s_n$ for each access network $n$;}
\BlankLine
\ForEach{Network $n$ available to Stream $s$}{
  Update estimate of $\alpha_n$ according to \eqref{eqn:alpha}\;
}
Update $\rho_n$ as $c^s_n/\sum_{n}c^s_n$\;
Update $r^s$ to minimize \eqref{eqn:AlternateStreamSimplified}\;
\ForEach{Network $n$ available to Stream $s$}{
 Update $r^s_n$ as $\rho_n r^s$\;
 }
\caption{\small Procedures of the media-aware allocation scheme run
by Stream~$s$. \label{fig:MediaAwareAlgorithm}}
\end{algorithm}
\subsection{H$^\infty$-Optimal Control}
\label{subsec:Hinfinity}
%
In the case when media-specific knowledge is unavailable to the
wireless devices, the rate allocation problem can be addressed using
H$^\infty$-optimal control~\cite{Alpcan:WiOpt07}. In this approach,
we track current and past observations of available bit rate (ABR)
of each network, and model variations in ABR as unknown disturbances
to a continuous-time linear system. The design goal is to achieve
full network utilization while preventing excessive fluctuations in
allocated video rates. An optimal rate controller is derived based
on H$^\infty$-optimal analysis~\cite{Basar:95} to bound the
\emph{worst-case} system performance. The scheme is distributed by
nature, in that it treats the dynamics of each stream as unknown
disturbance for others, thereby decoupling interactions between
different streams. \\
\indent Each stream estimates via various online measurement
tools~\cite{AbURL} the \emph{measured residual bandwidth} as:\
\begin{equation}
\label{e:w}
w_n=\begin{cases}
e_n, & \text{if } e_n \geq 0 \\
\mu(t_f-t_i),  & \text{if } e_n < 0
        \end{cases},
\end{equation}
\noindent in which $e_n = c_n - r_n$ is defined by \eqref{eqn:Bn};
$t_i$ and $t_f$ denote the initial and final time instance
when $e_n$ is negative and $\mu$ is a negative scaling constant.\\
\indent We next define a continuous-time linear system from the
perspective of a single stream keeping track of a single network.
For notational simplicity we subsequently drop the subscript $n$ and
omit the time index $t$. The extension to multiple access networks
is discussed in the Appendix. Since each stream is independent of
others in the H$^\infty$-optimal control formulation, the scheme
also generalizes immediately to the case with multiple
streams~\cite{Alpcan:WiOpt07}\cite{Alpcan:ITMC}.\\
%
\indent From the perspective of Stream $s$, its rate update system
can be expressed as:\
\begin{eqnarray}
%
\dot x^s & = & a x^s + b u^s + w,  \label{eqn:SystemStateX}\\
\dot r^s & = & -\phi r^s + u^s.  \label{eqn:RateUpdateR}
\end{eqnarray}
\noindent where the system state variable $x^s$ reflects roughly
residual network bandwidth for Stream~$s$ and $u^s$ represents the
rate control action. In \eqref{eqn:SystemStateX}, the parameters $a
< 0$ and $b < 0$ adjust the memory horizon and the expected
effectiveness of control actions, respectively, on the system state
$x^s$. A smaller value of $a$ corresponds to a longer horizon, i.e.,
smoother values of $x^s$ over time. A higher value of $b$ means a
more responsive system, where the rate control action of an
individual stream has greater impact on total network utilization.
In \eqref{eqn:RateUpdateR}, the rate update is approximately in
proportion to the control action, with $\phi > 0$ sufficiently small
to guarantee stability~\cite{Alpcan:WiOpt07}\cite{Alpcan:ITMC}.
Recall that $w$ is function of residual bandwidth $e$, which, in
turn, is function of aggregate rates from all video streams.
Therefore the evolutions \eqref{eqn:SystemStateX} and
\eqref{eqn:RateUpdateR} are connected via a feedback loop. \\
%
\indent Ideally, if the network is fully utilized at equilibrium,
$w$ is zero while $u^s$ and $x^s$ approach zero for $\phi$
sufficiently small. To prevent excessive fluctuations in the
allocated rate of each video stream, however, fluctuations in the
measured available bandwidth cannot be tracked perfectly. Design of
the rate controller $u^s$ therefore needs to balance the incentive
for full network utilization against the risk of excessive
fluctuation in allocated video rates. Such design objective can be
expressed in mathematical terms, in the form of a cost function\
%
\begin{equation}
\label{eqn:ControllerCostL}
L^s(x^s,u^s,w) = \frac{\|z^s\|}{\|w\|},
\end{equation}
\noindent where $z^s := [ h x^s \; g u^s ]^T$ denotes system output
with user-specified weights $h>0$ and $g>0$ on relative importance
of full network utilization and video rate smoothness. In
\eqref{eqn:ControllerCostL}, $\|z^s\|^{2} := \int_{0}^{\infty}
|z^s(t)|^2 dt$ and $\|w\|^2 := \int_{0}^{\infty} |w(t)|^2 dt$. The
cost function captures the proportional change of the system output
$z^s$ with respect to system input $w$. Intuitively, when variations
in the observed residual bandwidth $w$ is large, larger
variations are allowed in the allocated video rates.\\
%
\indent From H$^\infty$-optimal control theory~\cite{Basar:95}, one
can choose the optimal rate controller as:\
\begin{equation}
\label{eqn:ControllerU}
u^s_{\gamma}(x)= -\left(\dfrac{b}{g^2}{\sigma}_{\gamma}\right) x^s,
\end{equation}
\noindent with
$\sigma_\gamma=(-a\pm\sqrt{a^2-\lambda{h}^2})/\lambda$ and
$\lambda=1/\gamma^2-b^2/g^2$ to ensure a \emph{worst-case}
performance factor $\gamma:=\sup_{w} L^s(u^s,w)$. The lowest
possible performance factor is calculated as:
$\gamma^*=[\sqrt{{a^2}/{h^2}+{b^2}/{g^2}}]^{-1}$. In other words,
for any given value of $\gamma>\gamma^*$, one can find an optimal
rate controller according to \eqref{eqn:ControllerU} to ensure that
in the worst case, the cost function
\eqref{eqn:ControllerCostL} will not exceed $\gamma$.\\
\indent Although analysis and controller design are conducted around
the equilibrium point, the streams do not have to compute the actual
equilibrium values. In practice, the H$^\infty$-optimal rate control
scheme is implemented through the procedures summarized in
Fig.~\ref{fig:HinfAlgorithm}. Similar as for media-aware allocation,
computational complexity of the H$^\infty$-optimal control scheme
scales linearly with number of competing streams and number of
available access networks, on the order of ${\cal{O}}(NS)$.\
%
\begin{algorithm}
\KwIn{ABR measurements of available access networks;}
\KwSty{Parameters: $\:$}{Stream-specific weighting parameters $(a, b)$ and $(h,g)$;}\\
\KwOut{Feedback control $u^s$ and allocated rate $r^s$;}
\BlankLine
\ForEach{Access network available to Stream~$s$}{
  Measure current ABR ($w$) and delay\;
  Update $x^s$ according to \eqref{eqn:SystemStateX}\;
  Compute $u^s$ according to \eqref{eqn:ControllerU}\;
  Update rate $r^s$ according to \eqref{eqn:RateUpdateR}\;
}
\caption{\small Procedures of the H$^\infty$-optimal rate control
scheme run by Stream~$s$. \label{fig:HinfAlgorithm}}
\end{algorithm}
%
\subsection{AIMD-Based Heuristics}
\label{subsec:AIMD}
\begin{figure}
\centering
\includegraphics[width=0.6\columnwidth]{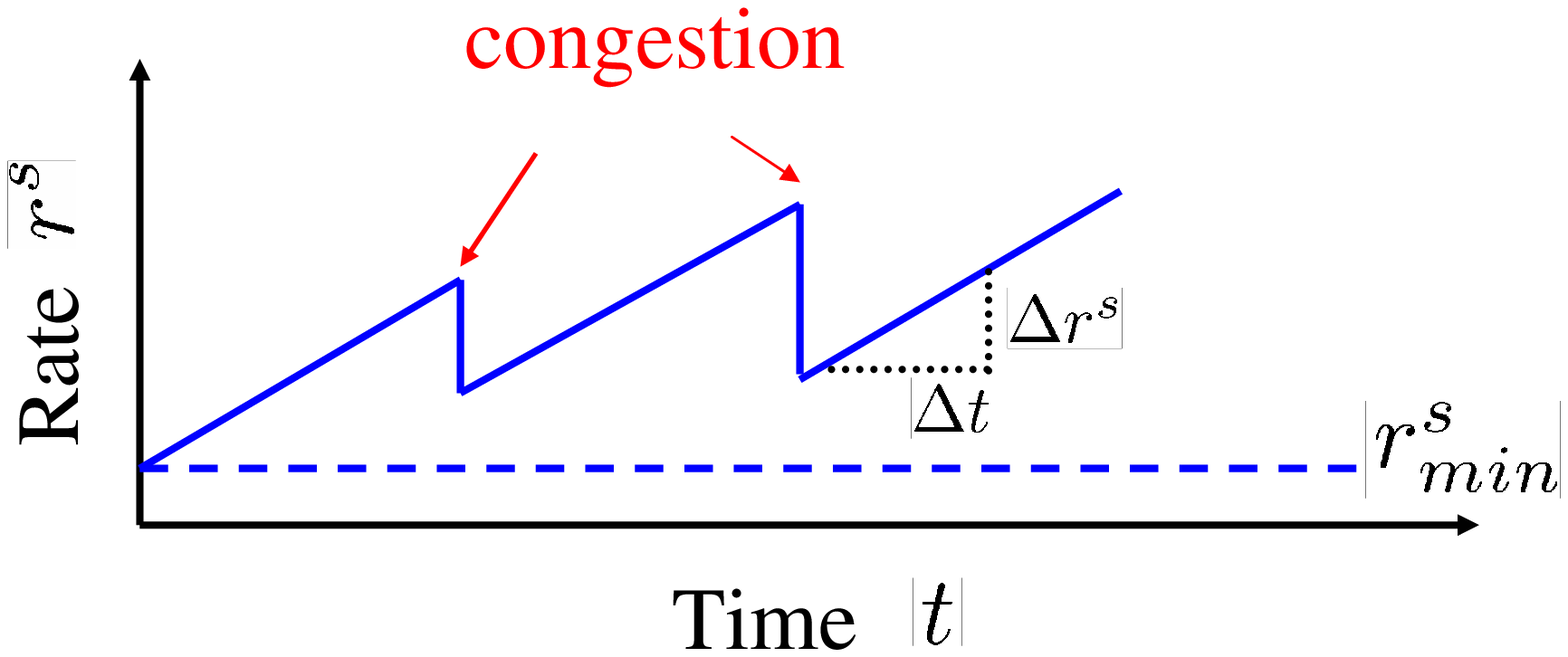}
\caption{\small Illustration of AIMD-based heuristic schemes.
Allocated rate for each stream $r^s$ keeps increasing at a rate of
$\Delta r^s/\Delta t$ until congestion is detected from packet
losses or excessive round trip delay. In that case, the rate $r^s_n$
is cut down by $(r^s_n-r^s_{min})/2$ over the congestion network
$n$. } \label{fig:AIMDHeuristics}
\end{figure}

%
For comparison, we introduce in this section two heuristic rate
allocation schemes based on the
additive-increase-multiplicative-decrease~(AIMD) principle used by
TCP congestion control~\cite{Jacobson:SIGCOMM88}. Instead of
performing proactive rate allocation by optimizing a chosen
objective according to observed network attributes and video
characteristics, the AIMD-based schemes are reactive in nature, in
that they probe the network for available bandwidth and
reduce the allocated rates only \emph{after} congestion is detected.\\
\indent As illustrated in Fig.~\ref{fig:AIMDHeuristics}, each stream
initiates at a specified rate $r^s_{min}$ corresponding to the
minimum acceptable video quality, and increases its allocation by
${\Delta}r^s$ every ${\Delta}t$ seconds unless network congestion is
perceived, in which case the allocated rate is
dropped by $(r^s_n - r^s_{min})/2$ over the congested network $n$.\\
\indent We consider two variations of the AIMD-based schemes. They
differ in how the total allocated stream rate $r^s$ is distributed
across multiple access networks during the additive-increase phase:\
%
\begin{itemize}
\item \emph{Greedy AIMD}:  The increase in rate allocation
${\Delta}r^s$ is allocated to the network interface offering the
maximum instantaneous available bit rate: $r^s_n = r^s$, if $c^s_n
\geq c^s_{n'}, \forall n' \neq n \in {\cal N}$.\
\item \emph{Rate Proportional AIMD}:  The increase in rate
allocation ${\Delta}r^s$ is allocated to all available networks in
proportion to their instantaneous available bit rates $r^s_n =
\frac{c^s_n}{\sum_n c^s_n} r^s$.\
\end{itemize}
%
\indent In both schemes, congestion over Network $n$ is indicated
upon detection of a lost packet or when the observed RTT exceeds a
prescribed threshold $\tau^s_{th}$. The value of $\tau^s_{th}$, in
turn, is adjusted according to the video playout deadline.\
%
%
\section{Performance Evaluation}
\label{sec:PerformanceEvaluation}
\subsection{Simulation Methodology}
\label{subsec:SimulationMethodology}
\begin{figure}
\centering
\includegraphics[width=0.6\columnwidth]{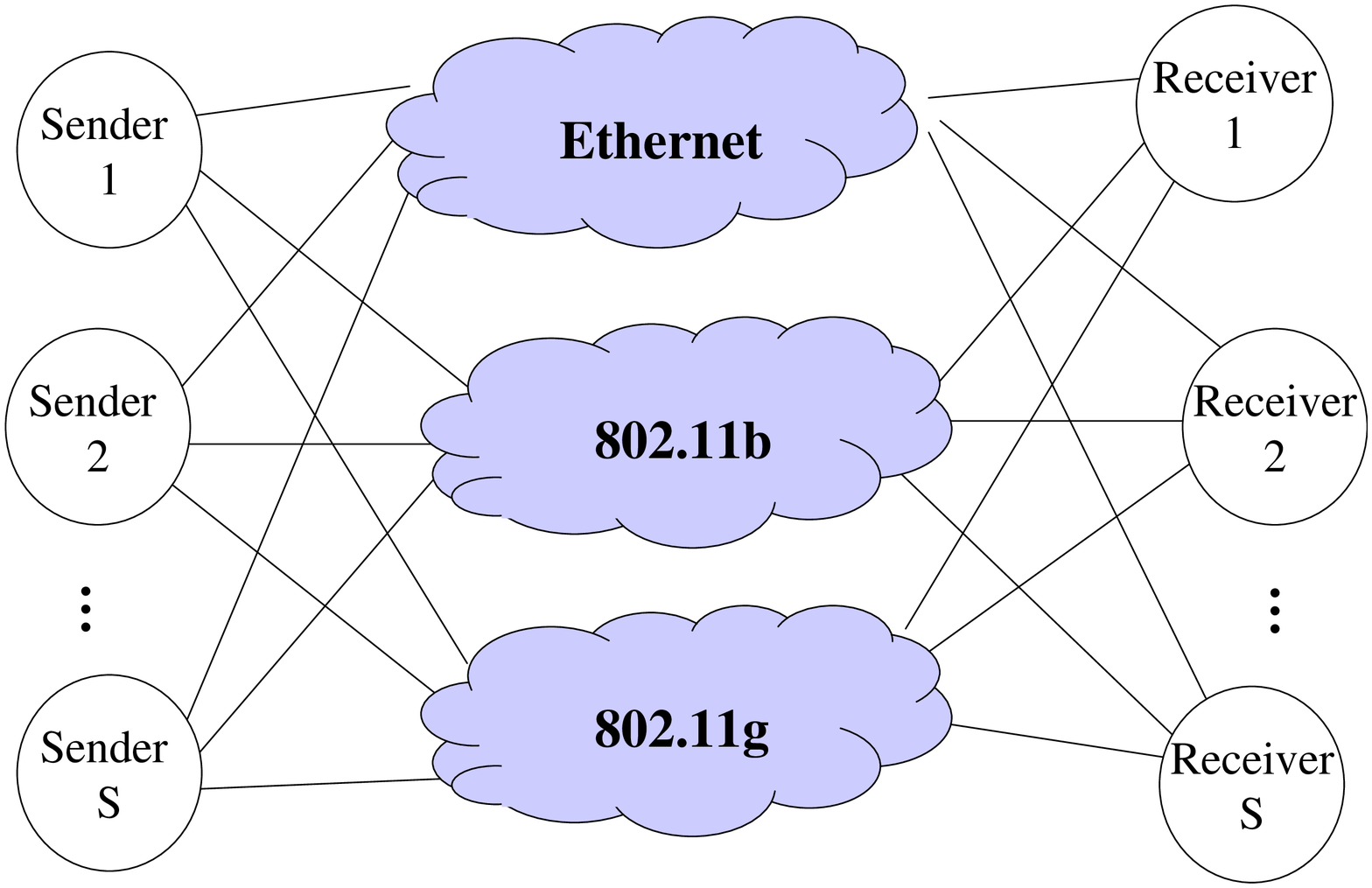}
\caption{\small Topology for network simulations}
\label{fig:NetworkTopology}
\end{figure}
\begin{figure}
   \begin{minipage}{0.495\columnwidth}
     \begin{center}
       \leavevmode
       \includegraphics[width=1.0\columnwidth]{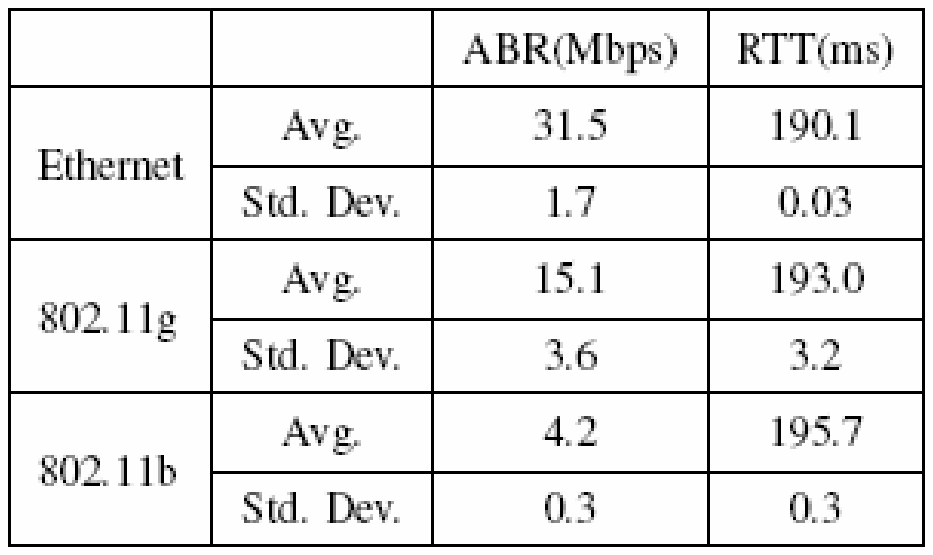} \\(a)
     \end{center}
   \end{minipage}
   \begin{minipage}{0.495\columnwidth}
     \begin{center}
       \leavevmode
       \includegraphics[width=1.0\columnwidth]{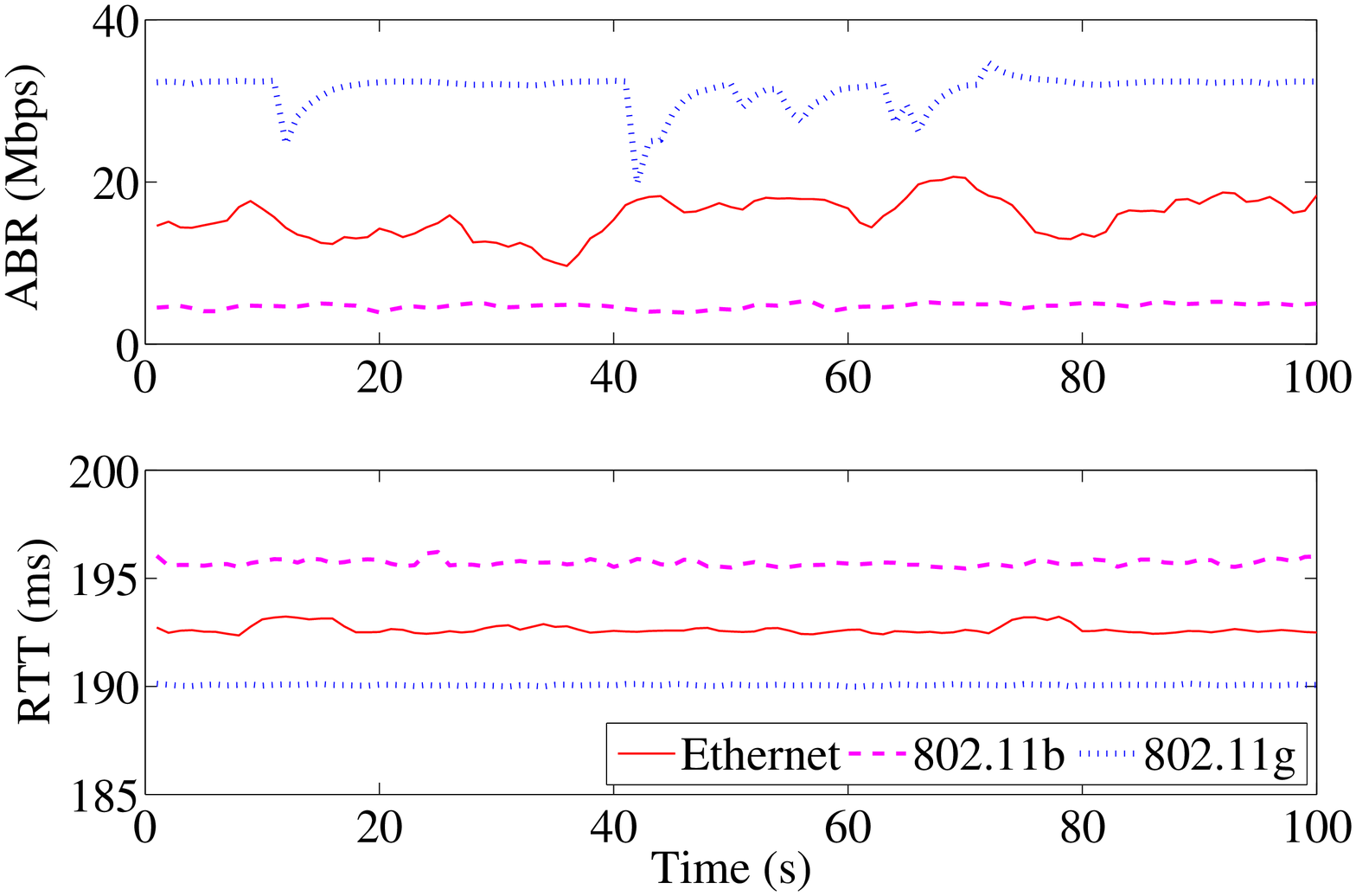} \\(b)
     \end{center}
   \end{minipage}
\caption{\small Statistics (a) and sample trace segments (b) of
measured Available Bit Rate~(ABR) and round-trip-time~(RTT) from
Deutsche Telekom Laboratories to Stanford University. The traces are
collected over a two-hour duration in a week day afternoon. } 
\label{fig:Traces}
\end{figure}
\begin{figure}
   \begin{minipage}{0.3\columnwidth}
     \begin{center}
       \leavevmode
\includegraphics[width=1.0\columnwidth]{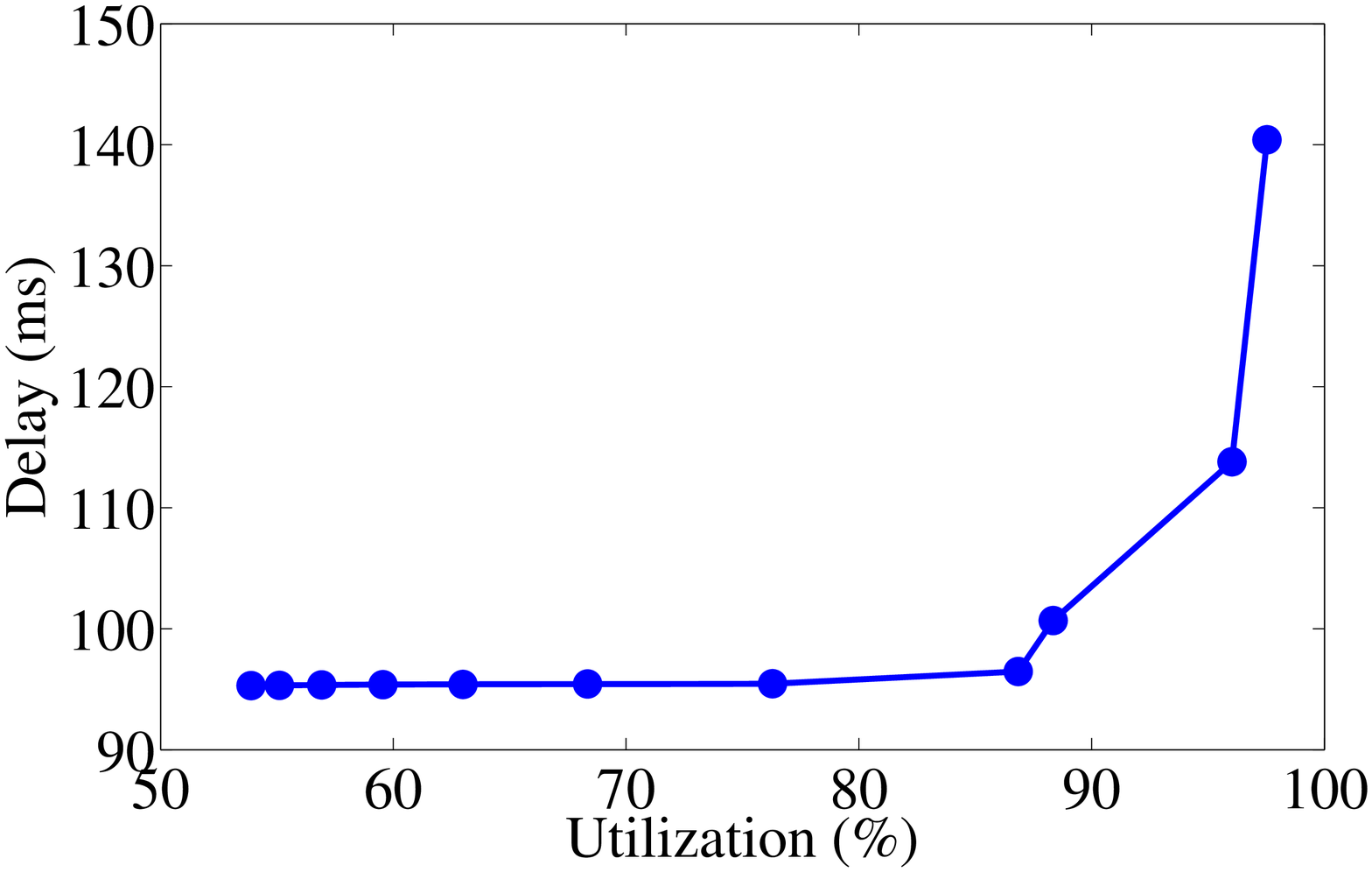}
     \end{center}
   \end{minipage}
   \hfill
   \begin{minipage}{0.3\columnwidth}
     \begin{center}
       \leavevmode
\includegraphics[width=1.0\columnwidth]{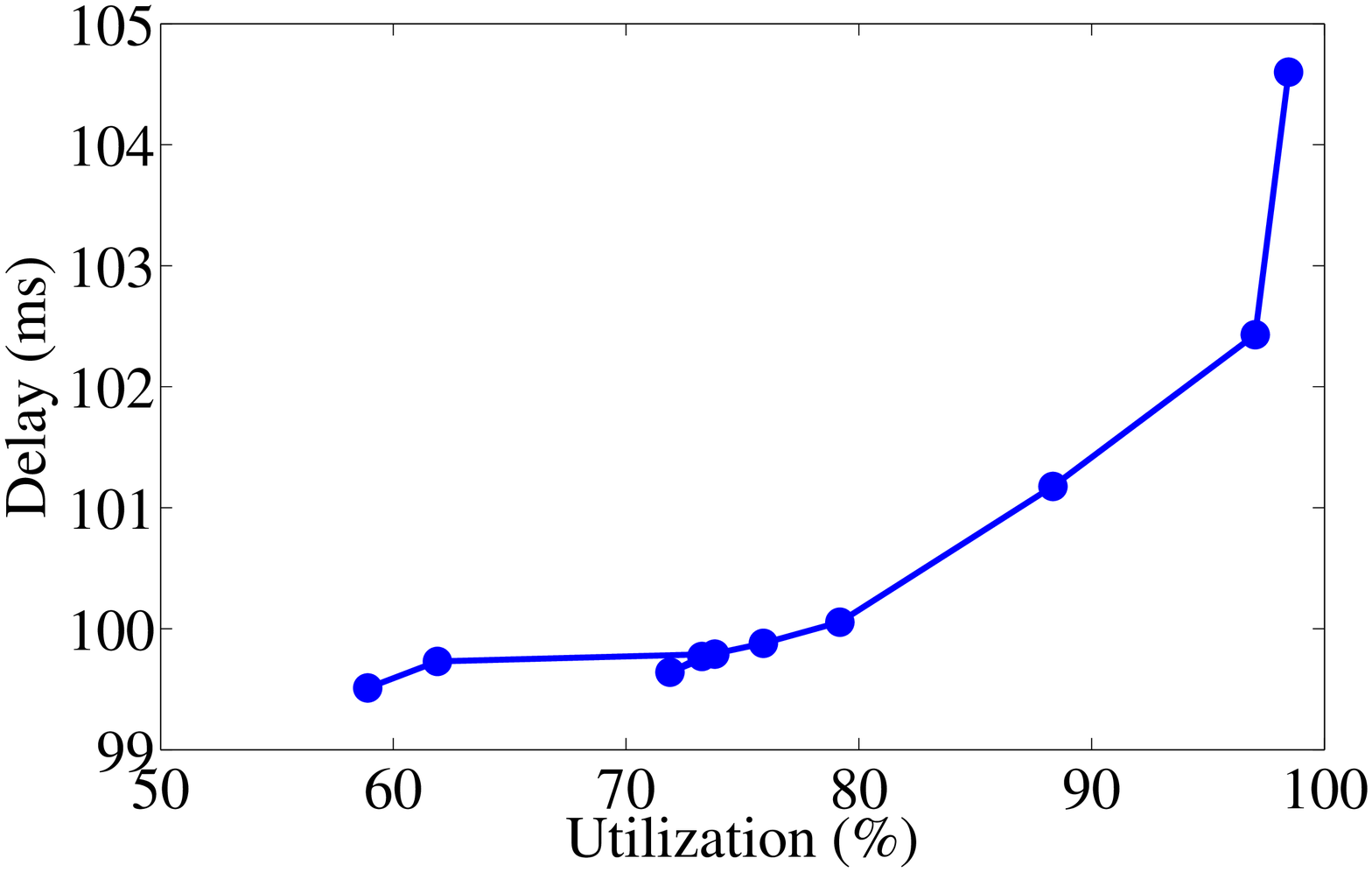}
     \end{center}
   \end{minipage}
   \hfill
   \begin{minipage}{0.3\columnwidth}
     \begin{center}
       \leavevmode
\includegraphics[width=1.0\columnwidth]{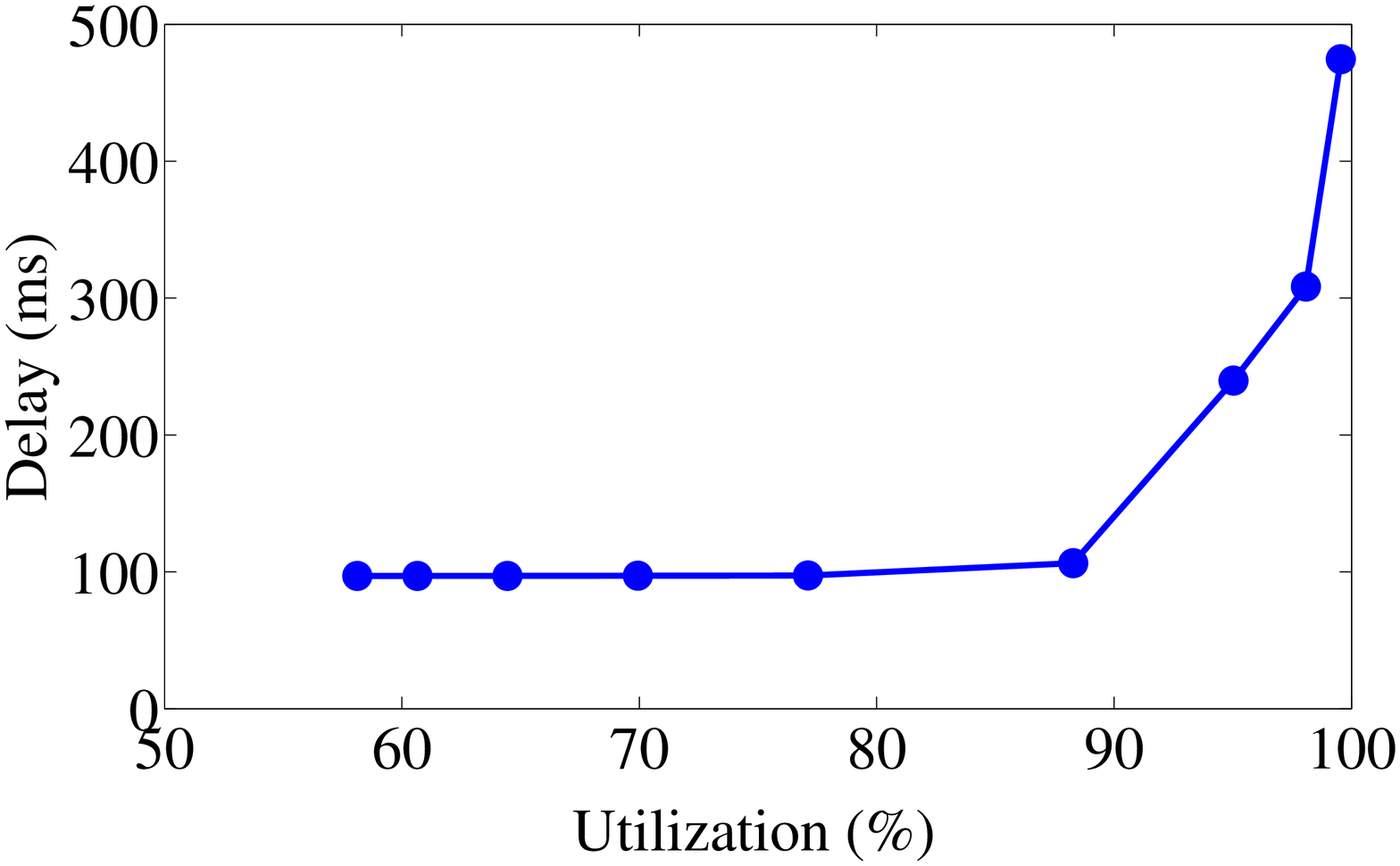}
     \end{center}
   \end{minipage}
   \vfill
   \begin{minipage}{0.3\columnwidth}
     \begin{center}
       \leavevmode
\includegraphics[width=1.0\columnwidth]{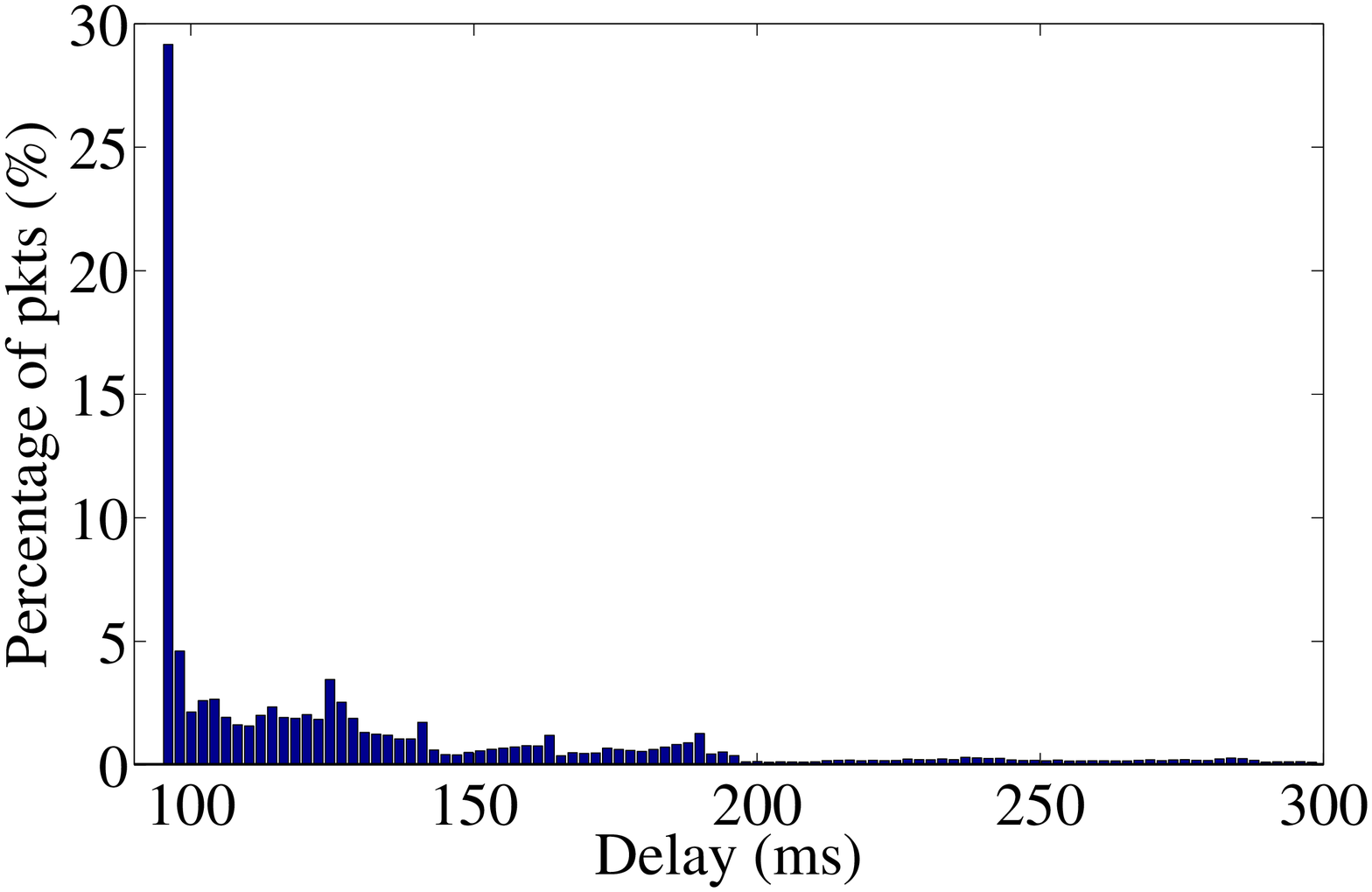}
       \\(a) Ethernet
     \end{center}
   \end{minipage}
   \hfill
   \begin{minipage}{0.3\columnwidth}
     \begin{center}
       \leavevmode
\includegraphics[width=1.0\columnwidth]{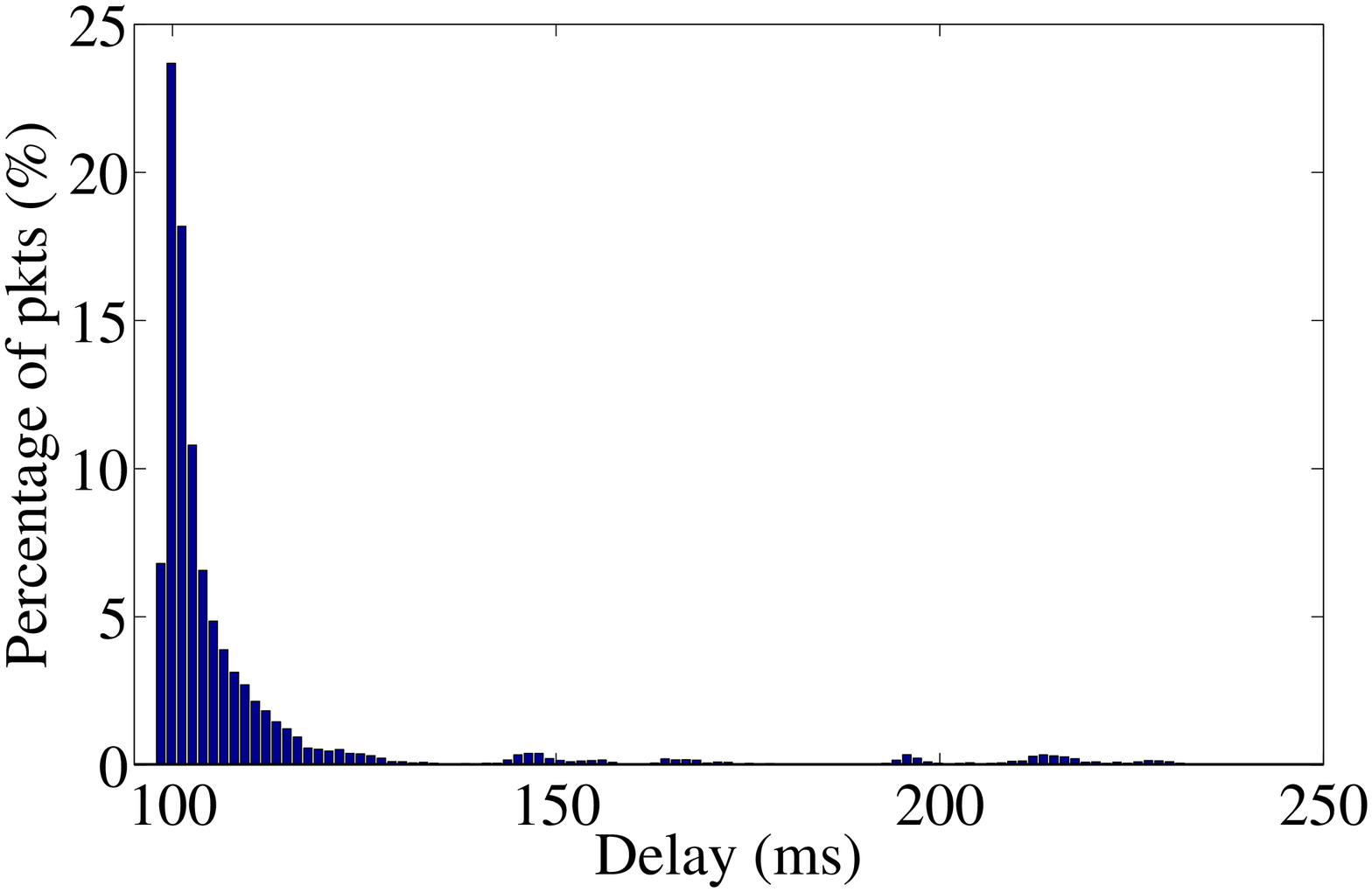}
       \\(b) 802.11b
     \end{center}
   \end{minipage}
   \hfill
   \begin{minipage}{0.3\columnwidth}
     \begin{center}
       \leavevmode
\includegraphics[width=1.0\columnwidth]{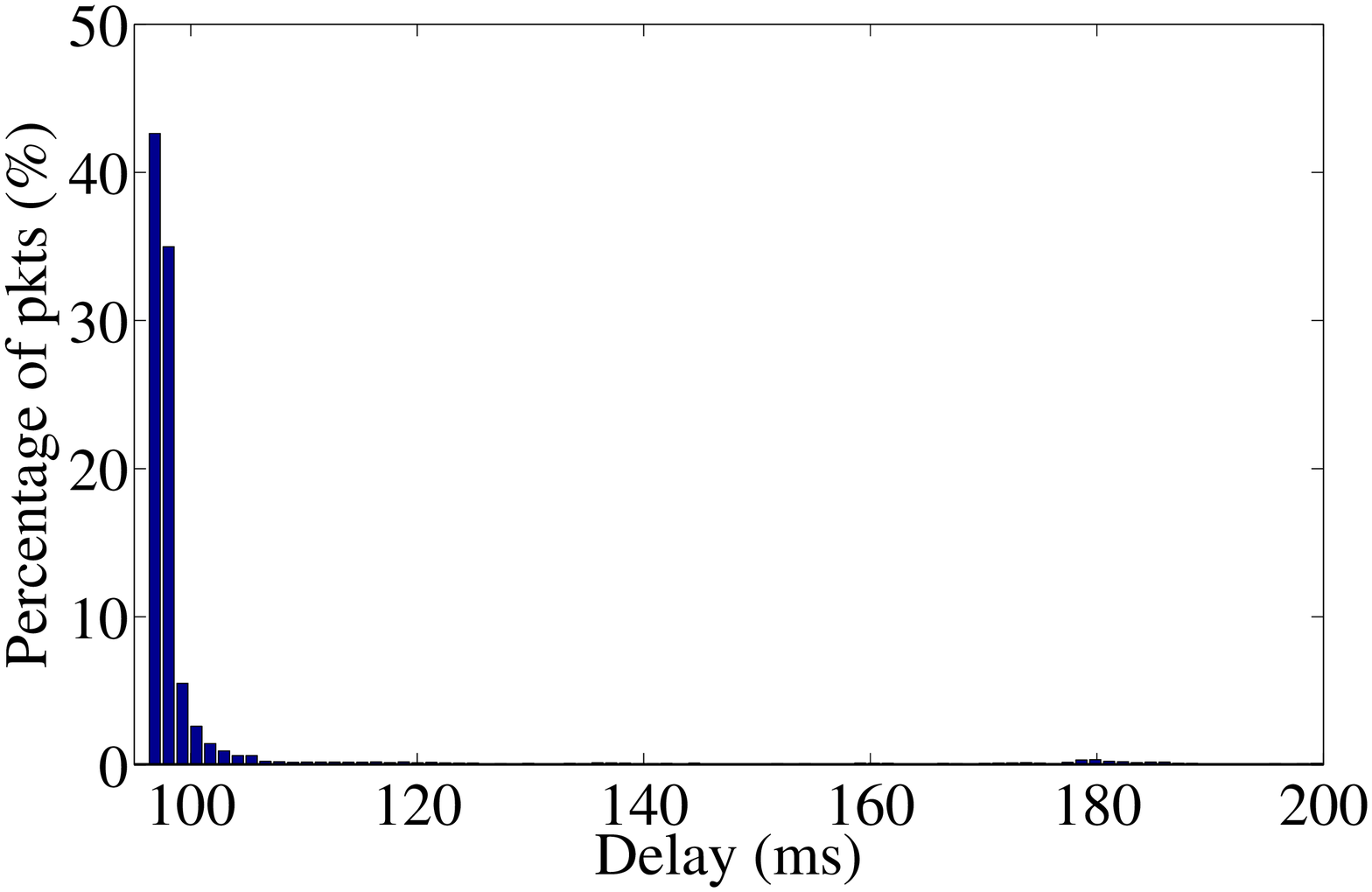}
       \\(c) 802.11g
     \end{center}
   \end{minipage}
\caption{\small Average packet delivery delay as function of network
utilization (top), as well as example packet delay distributions at
a given utilization level (bottom), from the three interfaces used
in \texttt{ns-2} simulations. The packet delay distributions are
plotted with a utilization level of 87\%, 88\%, and 88\% over the
three interfaces, respectively. \label{fig:Delay}}
\end{figure}
\begin{figure}
   \begin{minipage}{0.495\columnwidth}
   \begin{center}
       \leavevmode
\includegraphics[width=1.0\columnwidth]{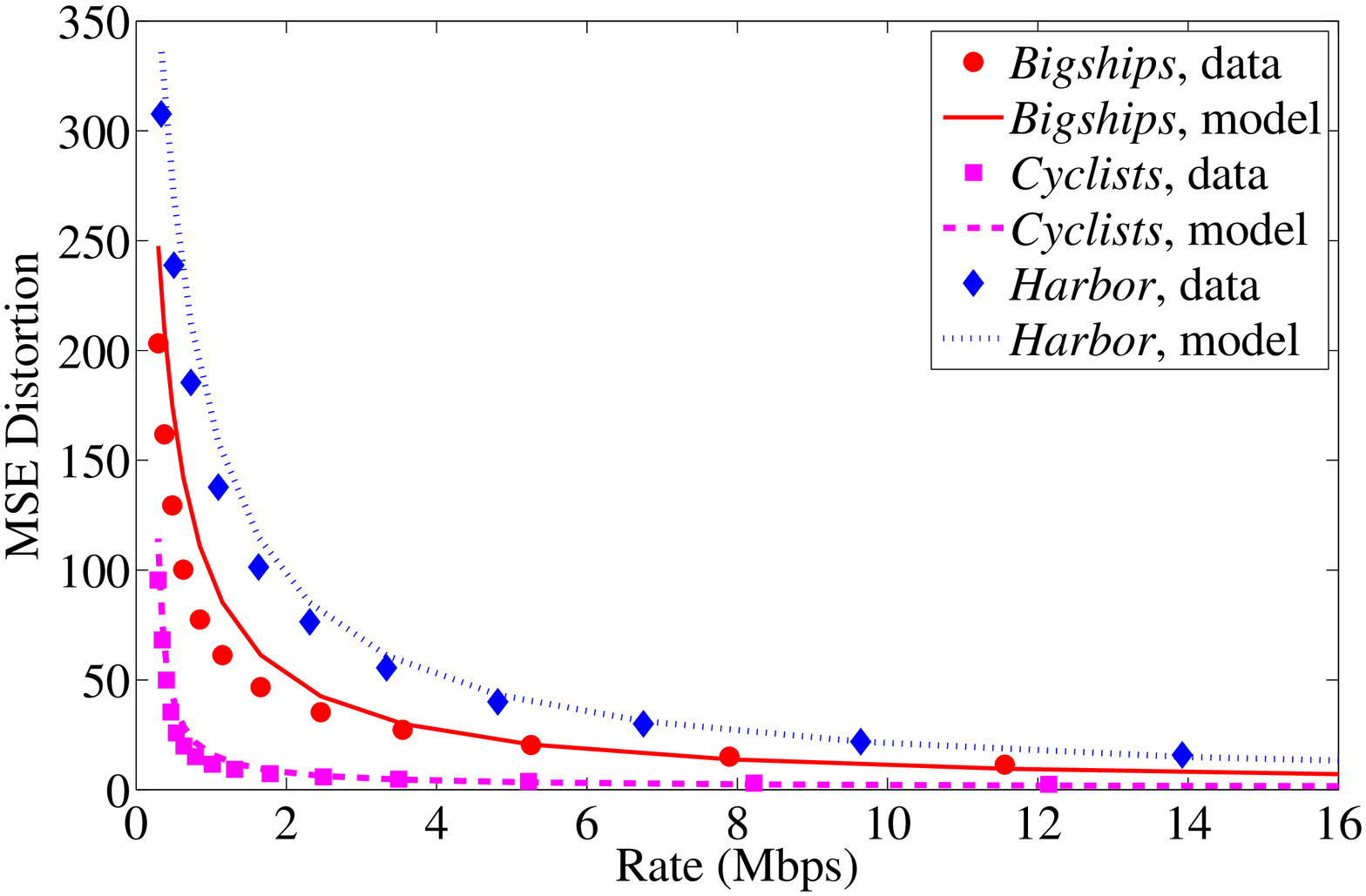}\\ (a)
   \end{center}
   \end{minipage}
   \begin{minipage}{0.495\columnwidth}
     \begin{center}
       \leavevmode
\includegraphics[width=1.0\columnwidth]{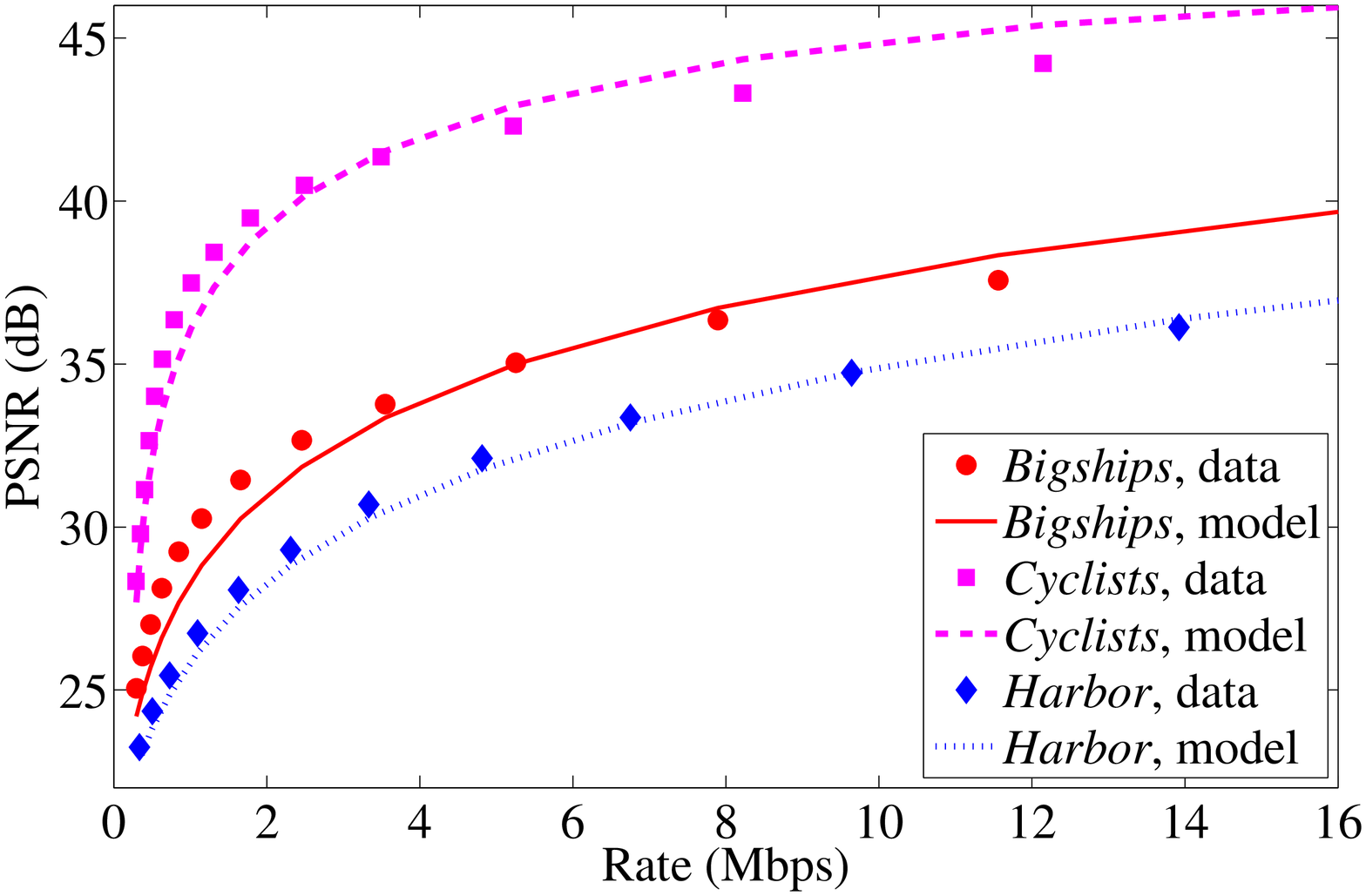}\\ (b)
     \end{center}
   \end{minipage}
\caption{\small Rate-distortion (a) and rate-PSNR (b) curves of 3 HD
video sequences used in the experiments: \emph{Bigships},
\emph{Cyclists} and \emph{Harbor}, all encoded using the H.264/AVC
codec at 60 frames per second, GOP length of 30. The measured data
points obtained from encoding are plotted against model curves
fitted according to \eqref{eqn:DistortionEnc}. } 
\label{fig:Stream}
\end{figure}

%
Performance of all four rate allocation policies are evaluated in
\texttt{ns-2}~\cite{ns2}, for an example network topology shown in
Fig.~\ref{fig:NetworkTopology}. Each sender streams one HD video
sequence via all three access networks to its receiver. Rate
allocation over each network is determined by the middleware
functionality depicted in Fig.~\ref{fig:SystemDiagram}. We collect
available bit rate (ABR) and round-trip-time (RTT) measurement from
three real-world access networks (Ethernet, 802.11b and 802.11g) in
a corporate environment using
\texttt{Abing}~\cite{AbURL}\cite{Navratil:PAM03}.\footnote{In both
802.11b and 802.11g networks, the transmission rate over each
interface is automatically adjusted according to wireless channel
conditions. The effect of link rate adaptation is reflected in
fluctuations in the ABR traces observed by \texttt{Abing}. Note that
the rate allocation schemes under discussion only passively
\emph{react} to, instead of interact with, such fluctuations.} The
ABR and RTT values are measured once every 2 seconds. The traces are
then used to drive the capacity and delay over each simulated access
network in \texttt{ns-2}.\footnote{Both forward and backward trip
delays are simulated as half of measured RTTs.} Statistics of the
network measurement, together with a sample segment of the measured
traces are presented in Fig.~\ref{fig:Traces}.
Figure~\ref{fig:Delay} shows how average packet delivery delay
varies with utilization percentage over each access network, as well
as sample packet delay distributions at a given utilization level.
In all three interface networks, the average packet delay increases
drastically as the utilization level approached 100\%, as described
in \eqref{eqn:Delay}. In accordance with our assumptions, the
example packet delay distributions also exhibit exponential shapes.
We refer to~\cite{Singh:WoWMoM07} for further details of the trace
collection procedures and bandwidth and delay measurements using \texttt{Abing}.\\
\indent Three high-definition~(HD) video sequences: \emph{Bigships},
\emph{Cyclists}, \emph{Harbor} are streamed by three senders,
respectively. The sequences have spatial resolution of $1280 \times
720$ pixels, and temporal resolution of 60 frames per second (fps).
Each stream is encoded using a fast implementation of the H.264/AVC
codec~\cite{H.264}\cite{X264} at various quantization step sizes,
with GOP length of 30 and IBBP... structure similar to that often
used in MPEG-2 bitstreams. Figure~\ref{fig:Stream} shows the
tradeoff of encoded video quality measured in MSE distortion and
PSNR versus average bit rate over the entire sequence durations. The
measured data points are plotted against fitted model curves
according to \eqref{eqn:DistortionEnc}. Encoded video frames are
segmented into packets with maximum size of 1500 bytes. Transmission
intervals of each packet in the entire GOP are spread out evenly to
avoid unnecessary queuing delay due to the large sizes of intra
coded frames. \\
\indent In addition to the video streaming sessions, additional
background traffic is introduced over each network interface by the
exponential traffic generator in \texttt{ns-2}. The background
traffic rate varies between 10\% and 50\% of the total ABR of each
access network. We also employ an implementation of the
\texttt{Abing} agent in \texttt{ns-2} to perform online ABR and RTT
measurement over each access network for each stream. This allows
the simulation system to capture the interaction among the three
competing HD streams as they share the three access networks
simultaneously. For consistency, measurement frequency of the
\texttt{Abing} agents in \texttt{ns-2} is also once every 2 seconds.
Update of video rate allocation is in sync with the time instances
when new network measurements are obtained for each stream. Note
that no coordination or synchronization is required across rate
updates in different streams, due to the distributed
nature of the rate allocation schemes.\\
\indent In the following, we first focus on the media-aware
allocation scheme. Its allocation results are compared against
optimal solutions for \eqref{eqn:AlternateStream} in
Section~\ref{subsec:Cent} and its convergence behavior is compared
against H$^\infty$-optimal control in
Section~\ref{subsec:Convergence}. Performance of all four allocation
schemes are evaluated with 20\% of background traffic load over each
network and a playout deadline of 300~ms in
Section~\ref{subsec:AllocationComparison}.
Section~\ref{subsec:RandomPacketLoss} compares allocation results
from networks with or without random packet losses. The impact of
background traffic load on the allocation results obtained from
different schemes is studied in
Section~\ref{subsec:BackgroundTraffic}. The effect of different
video streaming playout deadlines is investigated in
Section~\ref{subsec:PlayoutDeadline}.\
%
\subsection{Comparison with Optimal Allocation}
\label{subsec:Cent}
\begin{figure}
   \begin{minipage}{\columnwidth}
     \begin{center}
       \leavevmode
\includegraphics[width=0.7\columnwidth]{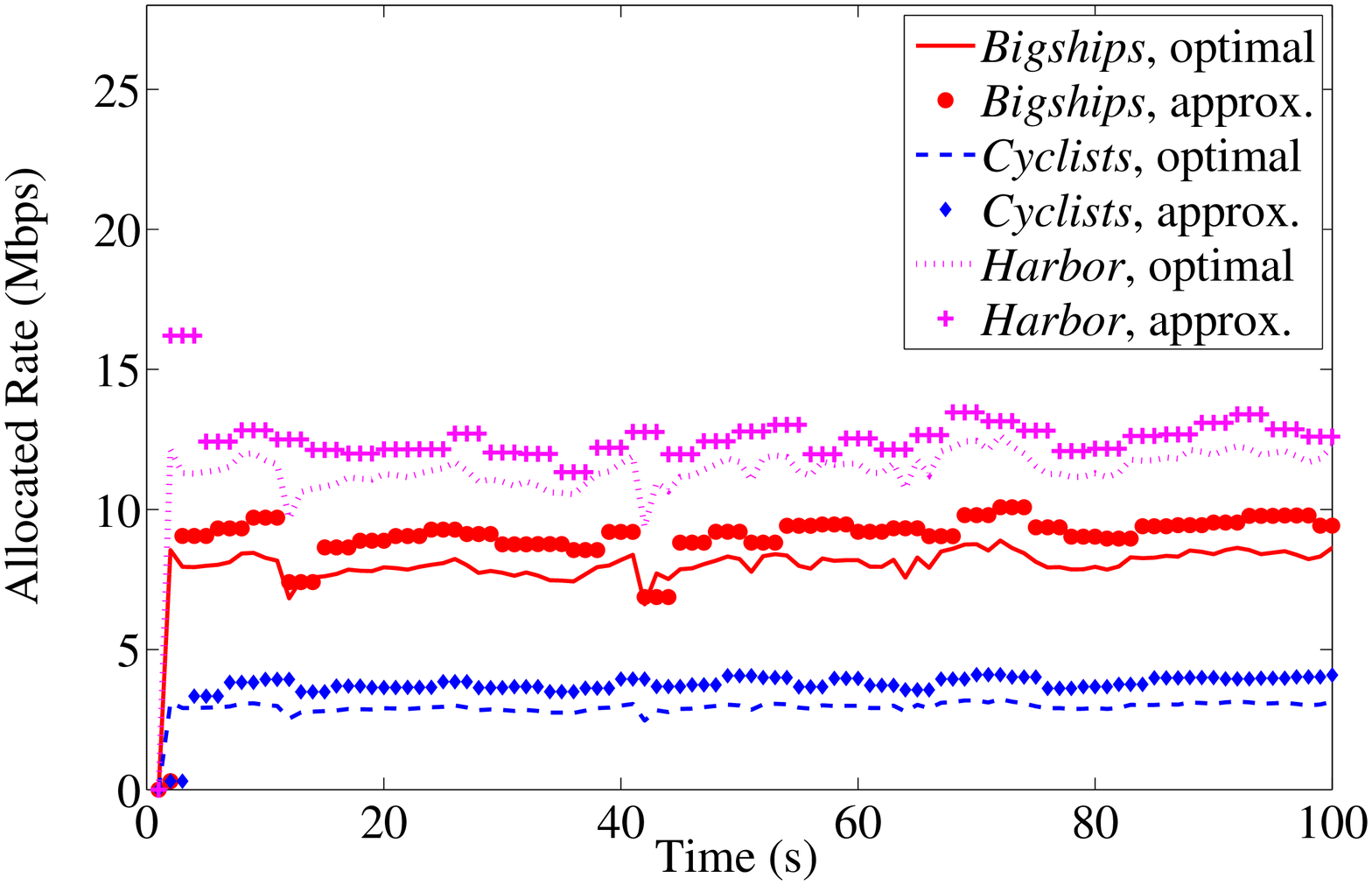}\\
     \end{center}
   \end{minipage}
 \caption{\small Comparison of allocated rate to each video stream, from
 the optimal solution for \eqref{eqn:AlternateStream} and its distributed
 approximation \eqref{eqn:AlternateStreamSimplified}.
 Background traffic load is 20\% and the playout deadline is 300~ms.}%
\label{fig:CompareCent}
\end{figure}

%
We first verify how well the distributed solution from
\eqref{eqn:AlternateStreamSimplified} can approximate optimal
solution for \eqref{eqn:AlternateStream}.
Figure~\ref{fig:CompareCent} compares the traces of allocated rate
to each video stream calculated from both solutions. The value of
$\kappa'$ used in the distributed approximation corresponds to the
sum of $\kappa^s$ for all three streams: $\kappa' =
\sum_{s}\kappa^s$. It can be observed that allocation from the
distributed approximation tracks the optimal solution closely. Since
the congestion term in \eqref{eqn:AlternateStreamSimplified} ignores
the impact of a stream on the expected distortion of other streams,
the distributed approximation achieves slightly higher rates.\
%
\subsection{Comparison of Convergence Behavior}
\label{subsec:Convergence}
\begin{figure}
   \begin{minipage}{\columnwidth}
     \begin{center}
       \leavevmode
        \includegraphics[width=0.7\columnwidth]{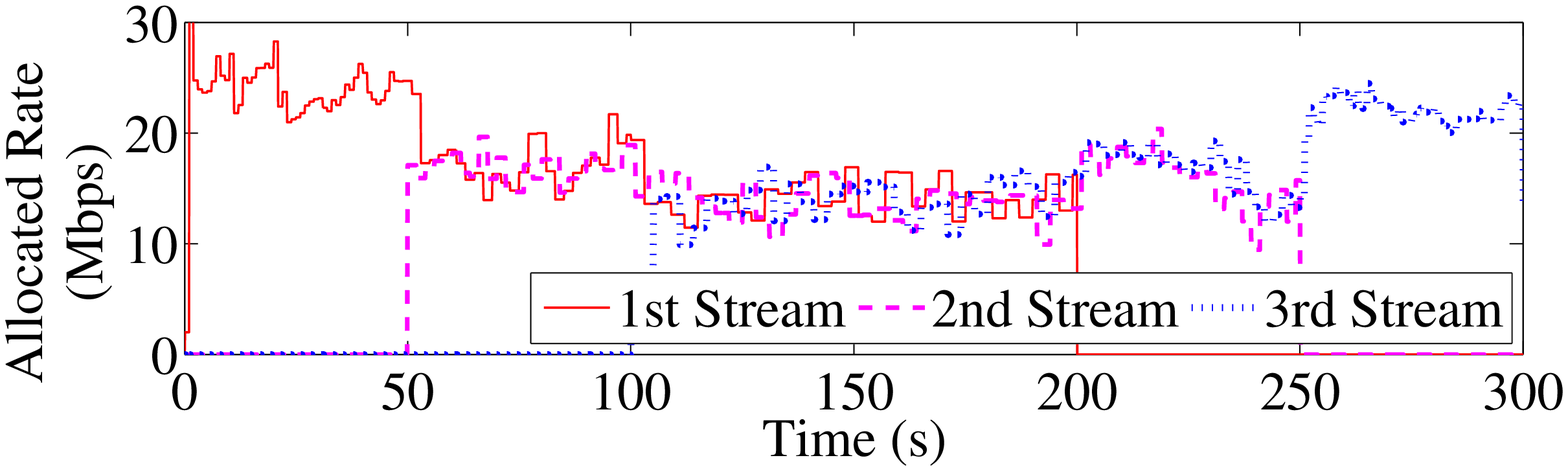} \\
       (a) Media-Aware Allocation
     \end{center}
   \end{minipage}
   \begin{minipage}{\columnwidth}
     \begin{center}
       \leavevmode
        \includegraphics[width=0.7\columnwidth]{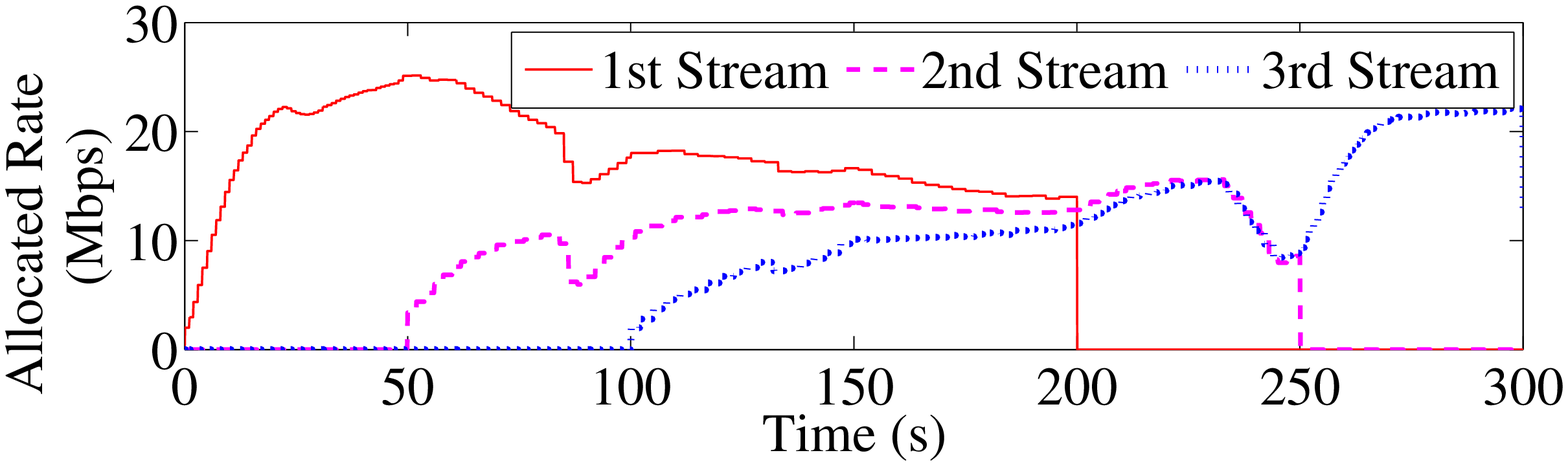}\\
       (b) H$^\infty$-Optimal Control
     \end{center}
   \end{minipage}
 \caption{\small Trace of total allocated rate to each stream, as the number of competing
video streams, all \emph{Harbor}, increases from 1 to 3. Background
traffic load is 20\% and the playout deadline is 300~ms.}%
\label{fig:VaryingFlow}
\end{figure}

\begin{figure}
   \begin{minipage}{\columnwidth}
     \begin{center}
       \leavevmode
    \includegraphics[width=0.5\columnwidth]{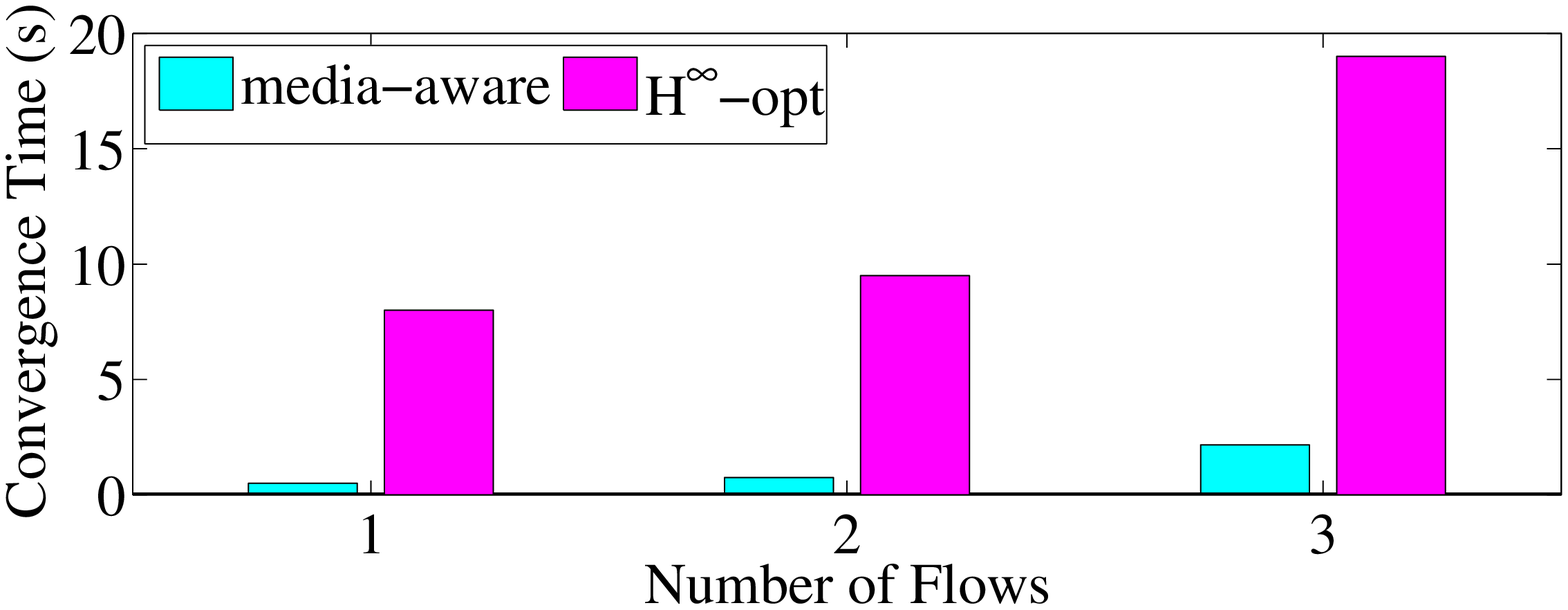}
            \end{center}
   \end{minipage}
   \begin{minipage}{\columnwidth}
     \begin{center}
       \leavevmode
    \includegraphics[width=0.5\columnwidth]{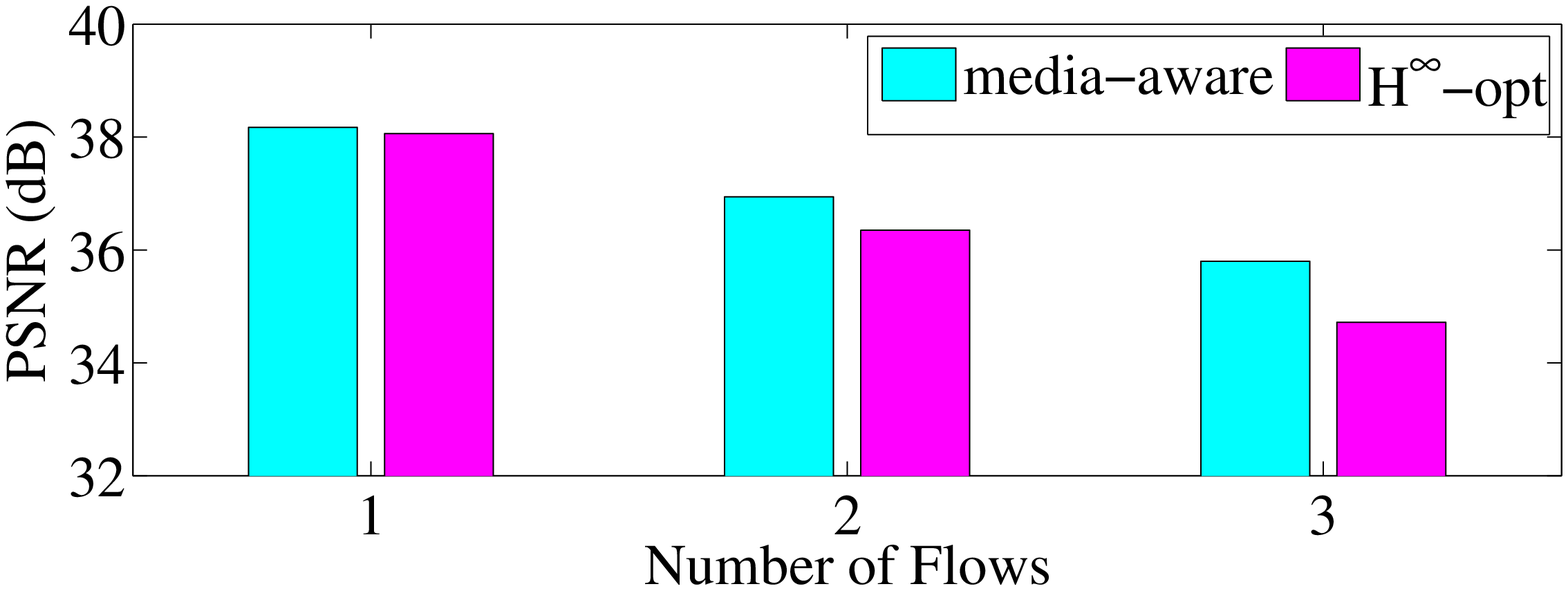}
     \end{center}
   \end{minipage}
 \caption{\small Convergence time and corresponding average video quality,
 achieved by media-aware and \mbox{H$^\infty$-optimal} control schemes, with
 1, 2, or 3 competing streams of the HD video sequence \emph{Harbor}. Background
 traffic load is 20\%; the playout deadline is 300~ms.}
 \label{fig:VaryingFlowConvergence}
\end{figure}

%
Figure~\ref{fig:VaryingFlow} shows traces of allocated rate, when
the number of competing streams over the three access networks
increases from 1 to 3. In this experiment, all three streams are the
\emph{Harbor} HD video sequence, hence the allocated rate to each
stream is expected to be the same after convergence. The second and
third streams start at 50 and 100 seconds, and complete at 200 and
250 seconds respectively. Correspondingly, abrupt drops and rises in
allocated rate can be observed in Fig.~\ref{fig:VaryingFlow}~(a) for
media-aware allocation. It is also interesting to note the
fluctuations in the allocated rates after convergence, reflecting
slight variations in the video contents and network attributes. The
H$^\infty$-optimal control scheme, on the other hand, requires
longer time for the allocation to converge, as shown in
Fig.~\ref{fig:VaryingFlow}~(b). \\
\indent Next, we measure the allocation convergence times when 1, 2
or 3 competing streams join the network simultaneously. Convergence
time is defined as the duration between the start of the streams and
the time at which allocated video rates settle between adjacent
quality levels. Figure~\ref{fig:VaryingFlowConvergence} compares
results from media-aware allocation against H$^\infty$-optimal
control. While both schemes yield similar allocated rates and video
qualities, convergence time from media-aware allocation is shorter
than H$^\infty$-optimal control.\
%
\subsection{Comparison of Allocation Traces}
\label{subsec:AllocationComparison}
\begin{figure}
   \begin{minipage}{0.495\columnwidth}
     \begin{center}
       \leavevmode
    \includegraphics[width=1.0\columnwidth]{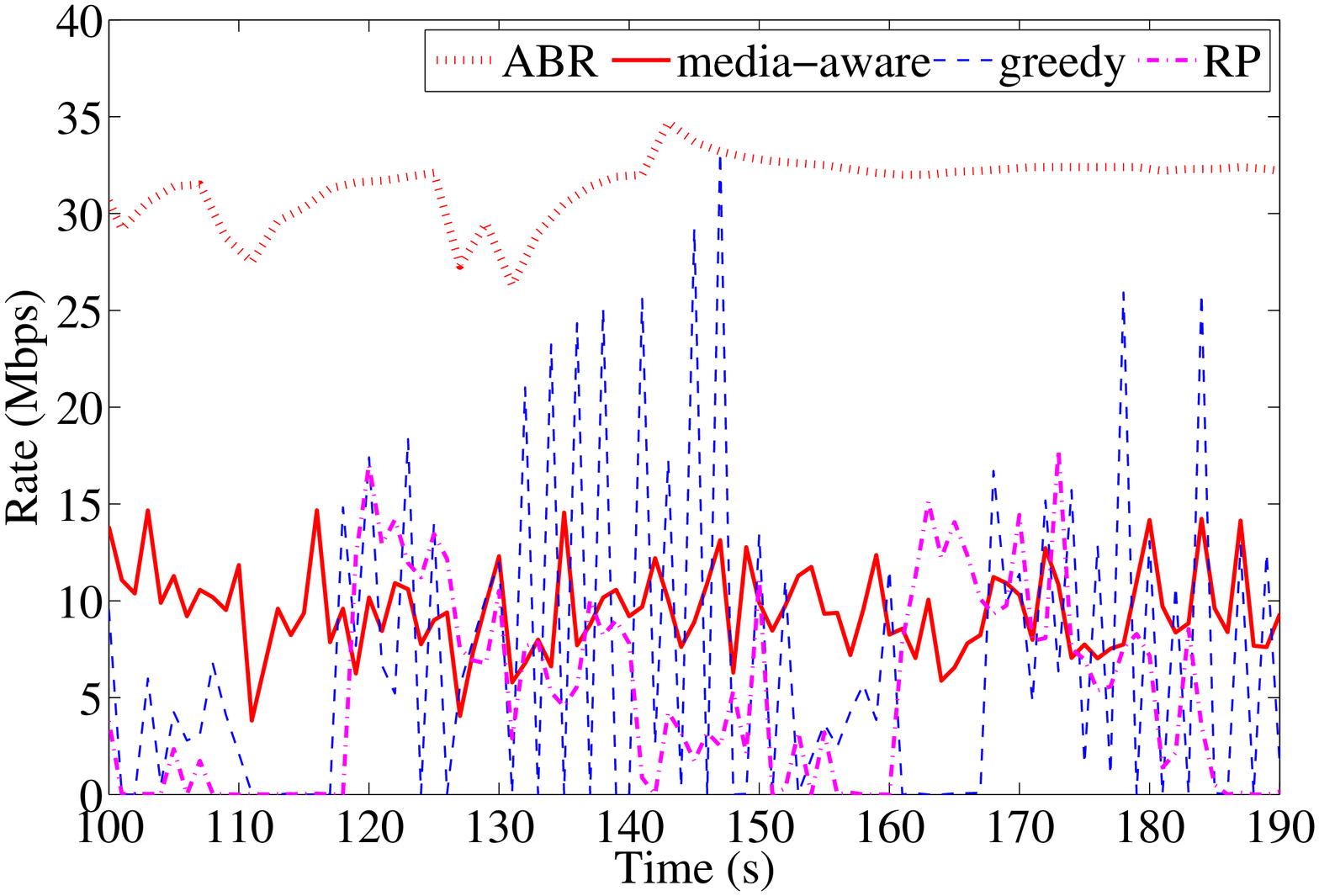}    \\(a)
     \end{center}
   \end{minipage}
   \begin{minipage}{0.495\columnwidth}
     \begin{center}
       \leavevmode
    \includegraphics[width=1.0\columnwidth]{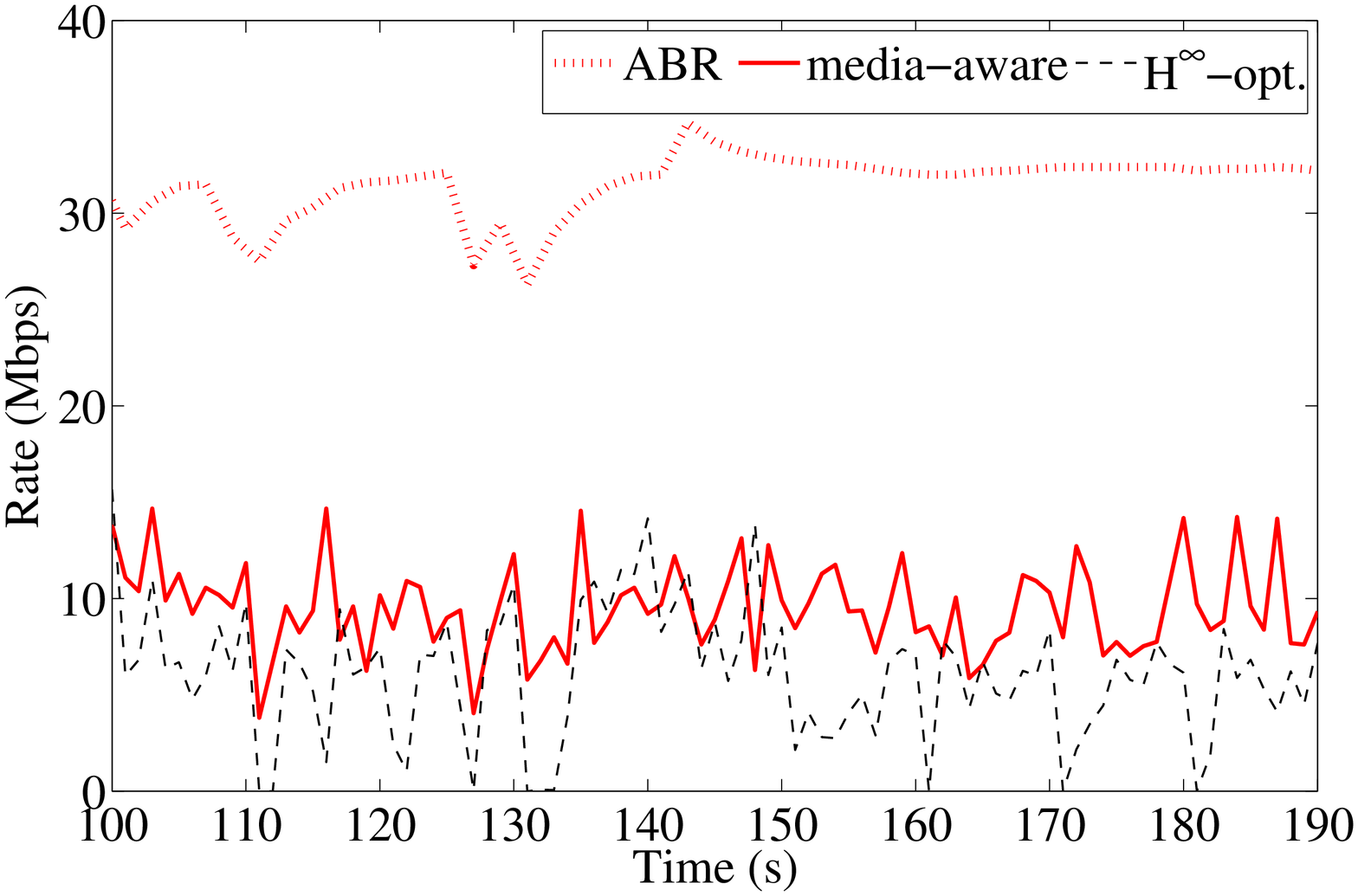}    \\(b)
     \end{center}
   \end{minipage}
 \caption{\small Trace of aggregated rate over the Ethernet interface.
(a) Media-aware allocation versus AIMD-based heuristics; (b)
Media-aware allocation versus H$^\infty$-optimal control. In this
experiment, background traffic load is 20\% and the playout deadline
is 300~ms. The network available bit rate is also plotted as a
reference.} \label{fig:UtilTrace}
\end{figure}

\begin{figure}
\begin{minipage}{\columnwidth}
     \begin{center}
    \includegraphics[width=0.7\columnwidth]{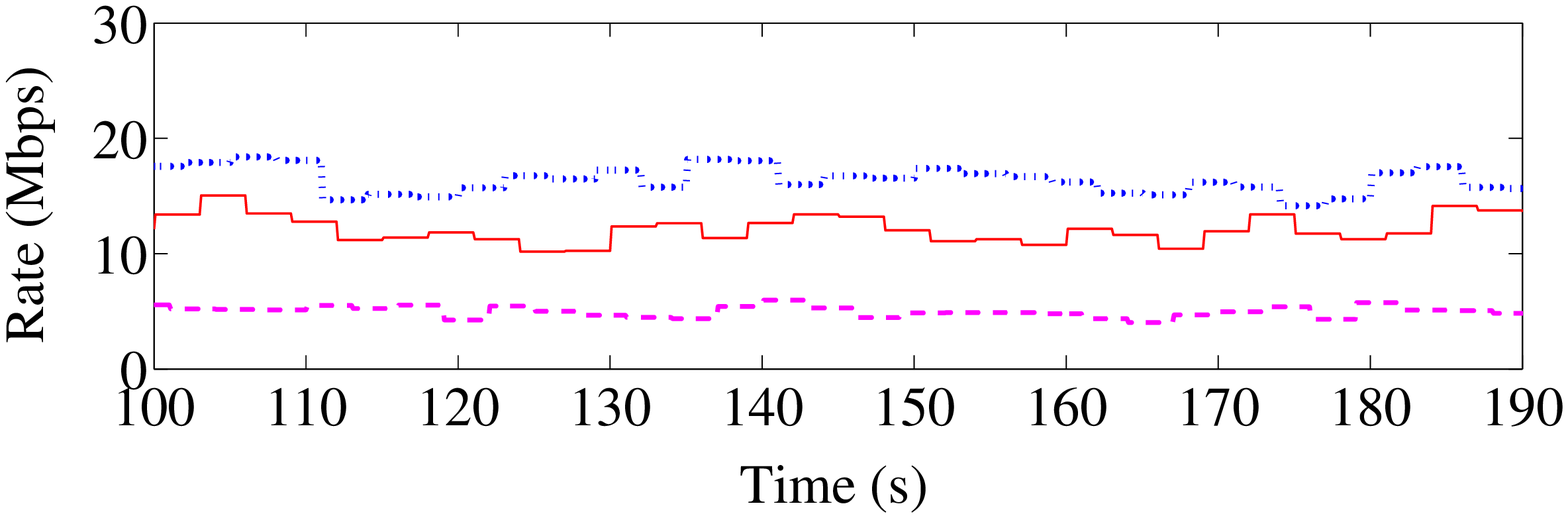}\\
    (a) Media-Aware
\end{center}
\end{minipage}
\begin{minipage}{\columnwidth}
\begin{center}
    \includegraphics[width=0.7\columnwidth]{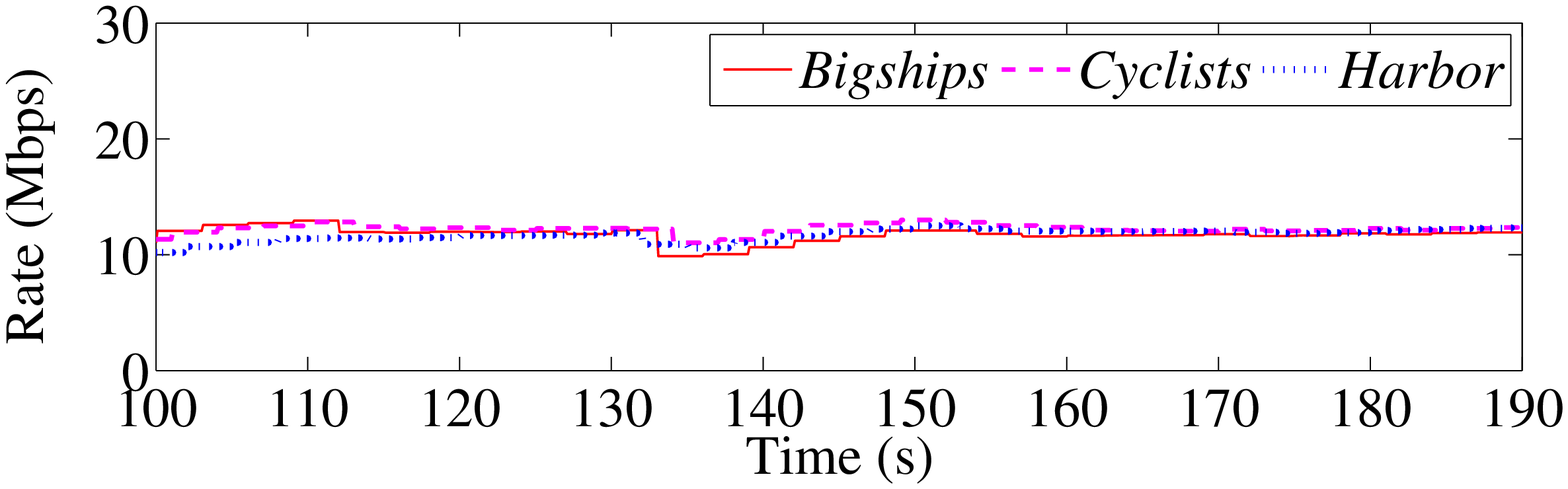}\\
(b) H$^\infty$-Optimal
\end{center}
\end{minipage}
\begin{minipage}{\columnwidth}
\begin{center}
    \includegraphics[width=0.7\columnwidth]{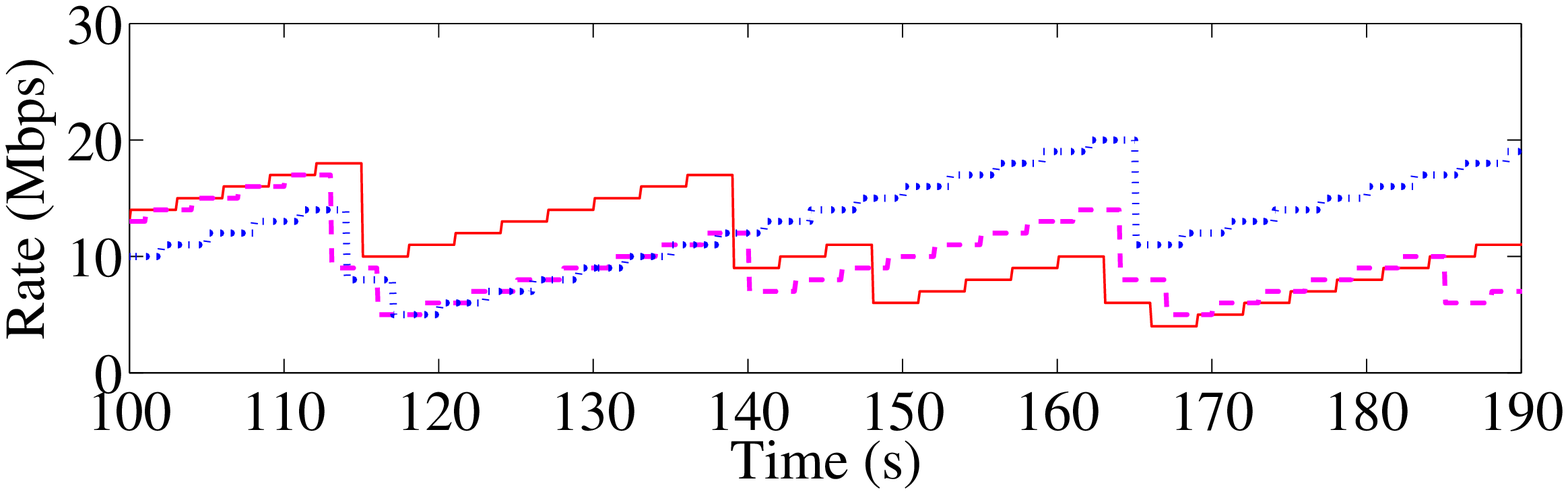}\\
(c) Greedy AIMD
\end{center}
\end{minipage}
\begin{minipage}{\columnwidth}
\begin{center}
    \includegraphics[width=0.7\columnwidth]{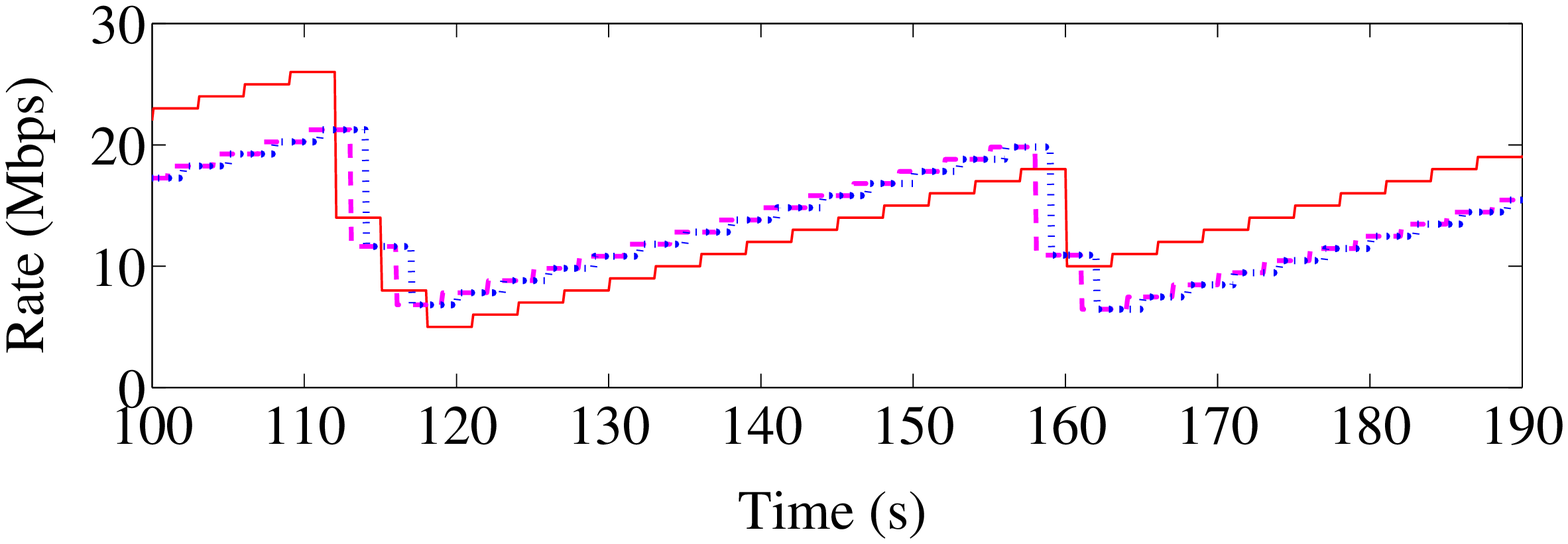}\\
(d) Rate Proportional AIMD
\end{center}
\end{minipage}
\caption{\small Trace of allocated rate to each video stream,
aggregated over three interfaces. Background traffic load is 20\%
and the playout deadline is 300~ms.} %
\label{fig:FlowTrace}
\end{figure}

%
Figure~\ref{fig:UtilTrace} plots the traces of aggregate rate
allocated over the Ethernet interface for all four allocation
schemes, together with the available bit rate over that network. It
can be observed in Fig.~\ref{fig:UtilTrace}~(a) that media-aware
allocation avoids much of the fluctuations in the two AIMD-based
heuristics. Figure~\ref{fig:UtilTrace}~(b) shows that it achieves
higher network utilization than H$^\infty$-optimal control, as the
latter is designed to optimize for the worst-case scenario. Similar
observations also hold for traces of aggregate allocated
rate over the other two interfaces.\\
\indent In Fig.~\ref{fig:FlowTrace}, we compare the traces of total
allocated rate for each video stream, resulting from the various
allocation schemes. In greedy AIMD allocation, the total rate of
each stream increases until multiplicative decrease is triggered by
either packet losses or increase in the observed RTTs from one of
the interfaces. Therefore traces of the allocated rates bear a
saw-tooth pattern. Behavior of the rate proportional AIMD scheme is
similar, except that rate drops tend to occur at around the same
time. The \mbox{H$^\infty$-optimal} control scheme yields less
fluctuations in the allocated rates. In both the rate proportional
AIMD allocation and the \mbox{H$^\infty$-optimal} control schemes,
allocated rates are almost identical to each video stream, since all
flows are treated with equal importance. The media-aware convex
optimization scheme, in contrast, consistently allocates higher rate
for the more demanding \emph{Harbor} stream, with reduced allocation
for \emph{Cyclists} with less complex contents.\
%
\subsection{Impact of Random Packet Loss}
\label{subsec:RandomPacketLoss}
\begin{figure}
   \begin{minipage}{\columnwidth}
     \begin{center}
       \leavevmode
    \includegraphics[width=0.6\columnwidth]{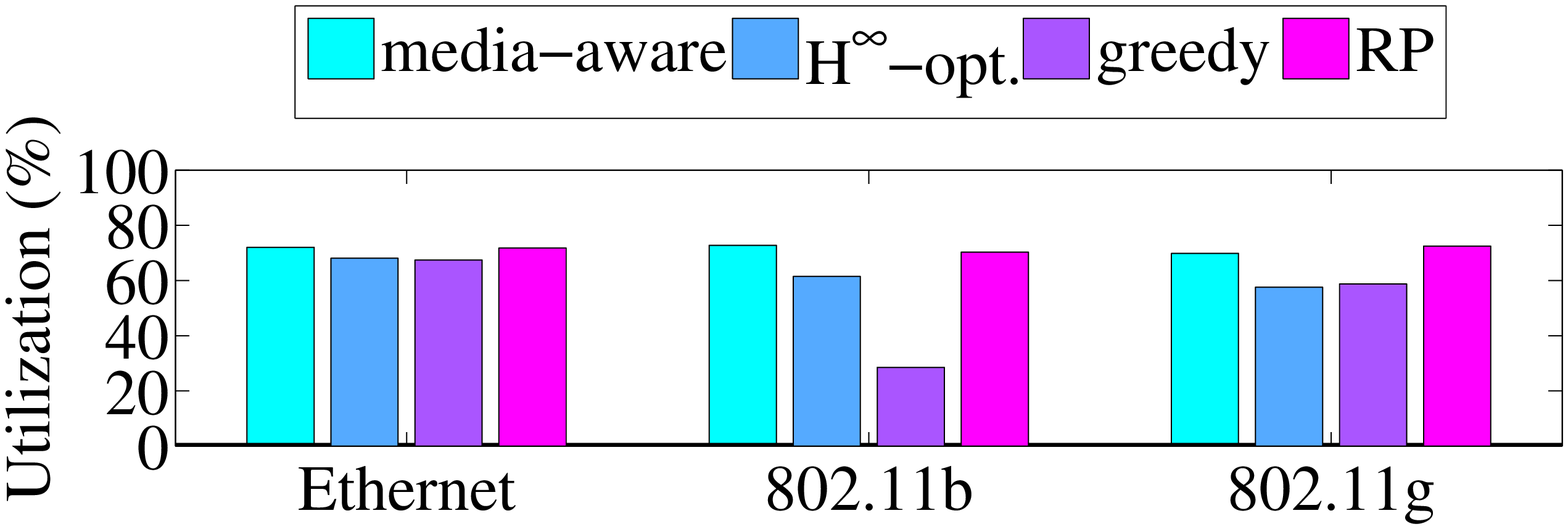}
     \end{center}
   \end{minipage}
   \begin{minipage}{\columnwidth}
     \begin{center}
       \leavevmode
        \includegraphics[width=0.6\columnwidth]{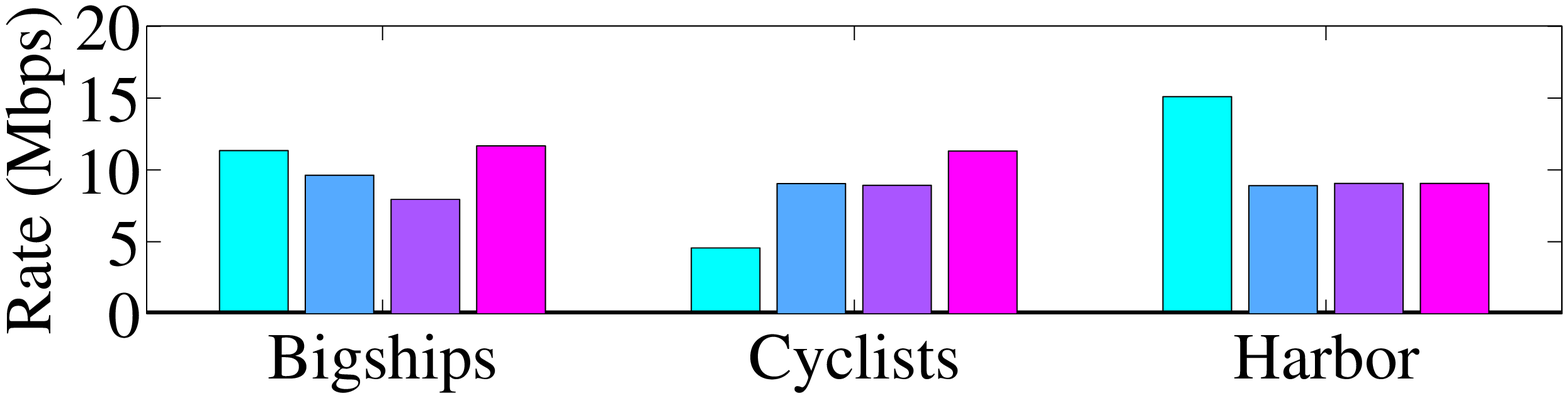}
     \end{center}
   \end{minipage}
   \begin{minipage}{\columnwidth}
     \begin{center}
       \leavevmode
        \includegraphics[width=0.6\columnwidth]{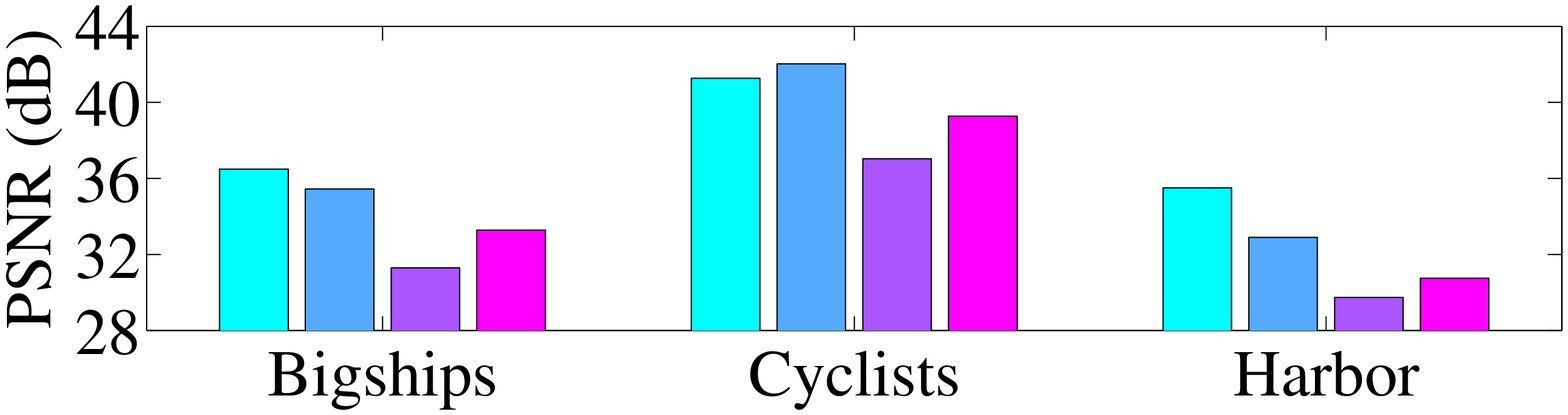}
     \end{center}
   \end{minipage}
 \caption{\small Comparison of allocation results from different
 schemes with background traffic load of 30\% and playout deadline
 of 300~ms. Aggregated network utilization over each interface (top); allocated
video rate for each stream (middle); received video quality in PSNR
(bottom).}
\label{fig:NoPacketLoss}
\end{figure}

\begin{figure}
   \begin{minipage}{\columnwidth}
     \begin{center}
       \leavevmode
     \includegraphics[width=0.6\columnwidth]{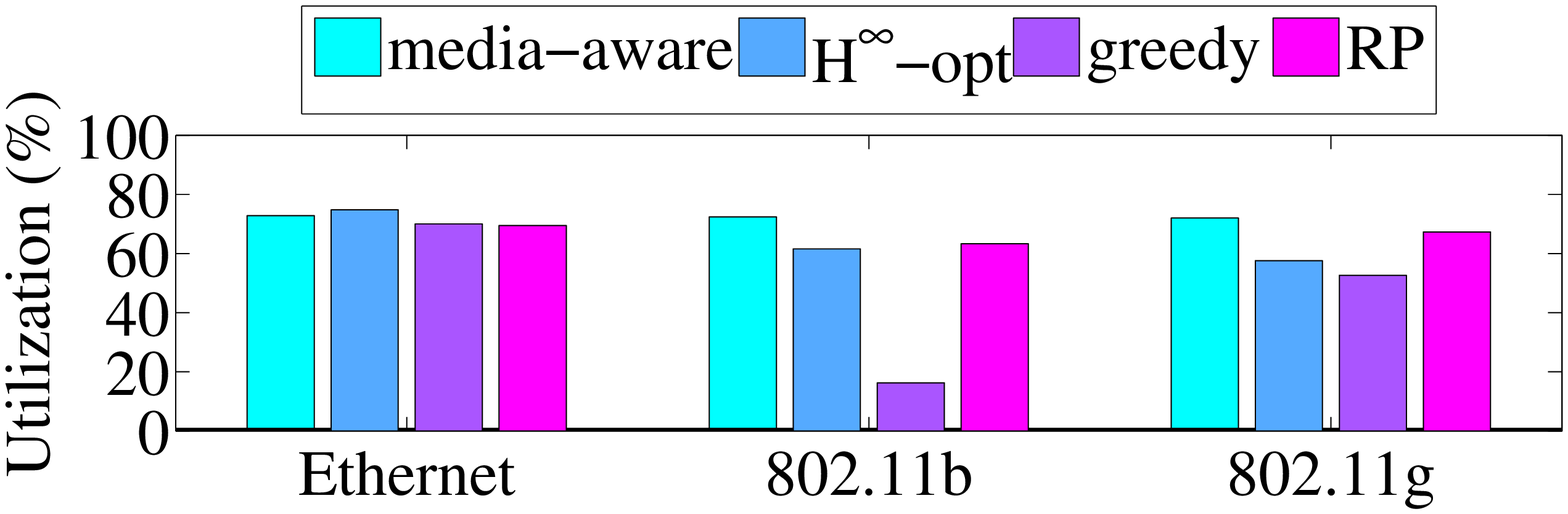}
     \end{center}
   \end{minipage}
   \begin{minipage}{\columnwidth}
     \begin{center}
       \leavevmode
 \includegraphics[width=0.6\columnwidth]{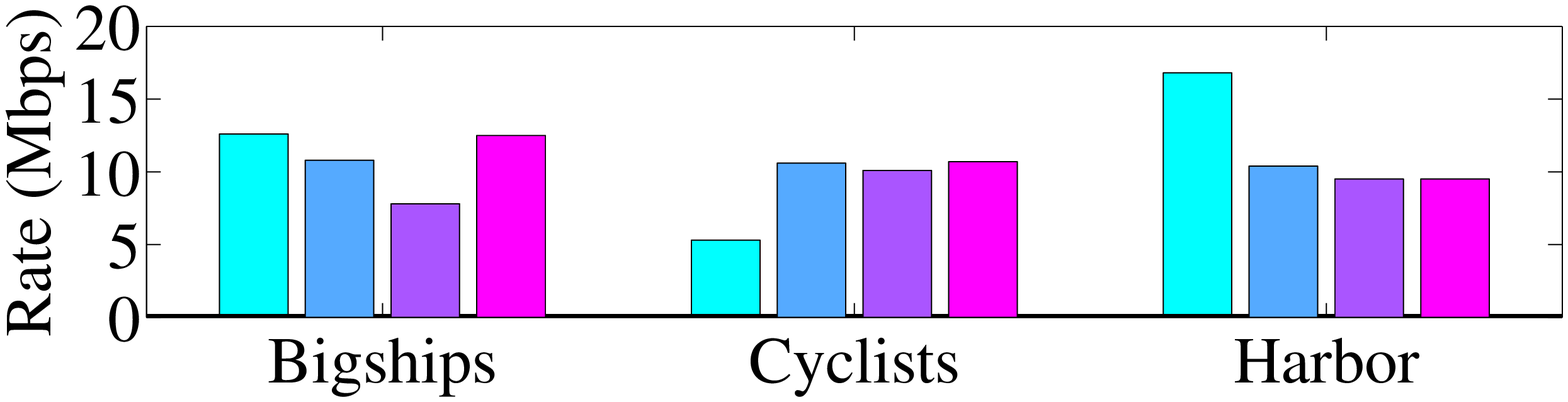}
     \end{center}
   \end{minipage}
   \begin{minipage}{\columnwidth}
     \begin{center}
       \leavevmode
 \includegraphics[width=0.6\columnwidth]{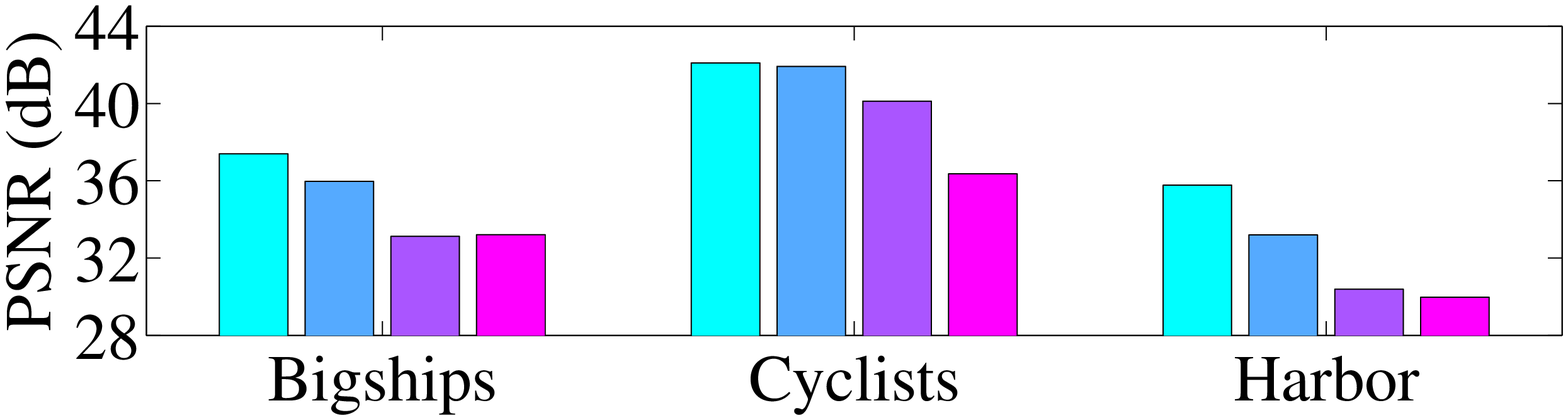}
     \end{center}
   \end{minipage}
 \caption{\small Comparison of allocation results from different
 schemes with background traffic load of 30\% and playout deadline
 of 300~ms. There is 1\% of random packet loss over each
 interface.
Aggregated network utilization over each interface (top); allocated
video rate for each stream (middle); received video quality in PSNR
(bottom). }
\label{fig:RandomPacketLoss}
\end{figure}

%
Figure~\ref{fig:NoPacketLoss} compares the average utilization over
each interface, allocated rate to each stream, and corresponding
received video quality achieved by the four allocation schemes, for
background traffic load of 30\%. The media-aware scheme allocates
lower rate for \emph{Cyclists} and higher rate for \emph{Harbor},
compared to the other schemes. This improves the video quality of
\emph{Harbor}, the stream with the lowest PSNR amongst the three, at
the expense of reducing the quality of the less demanding
\emph{Cyclists}. Consequently, the video quality is more balanced among
all three streams.\\
\indent A similar graph is shown in Fig.~\ref{fig:RandomPacketLoss},
for the same simulation with 1\% random packet loss over each
network interface. While the presence of random packet losses tend
to reduce received video quality, its impact cannot be mitigated by
means of careful rate allocation. Consequently, relative performance
of the four rate allocation schemes remain the same in both
scenarios. This justifies the absence of a term representing random
packet losses when formulating the media-aware rate allocation
problem. For the rest of the simulations, we therefore focus on
comparisons without random packet losses.\
%
\subsection{Varying Background Traffic Load}
\label{subsec:BackgroundTraffic}
\begin{figure*}
   \begin{minipage}{0.5\columnwidth}
     \begin{center}
        \leavevmode
    \includegraphics[width=1.0\columnwidth]{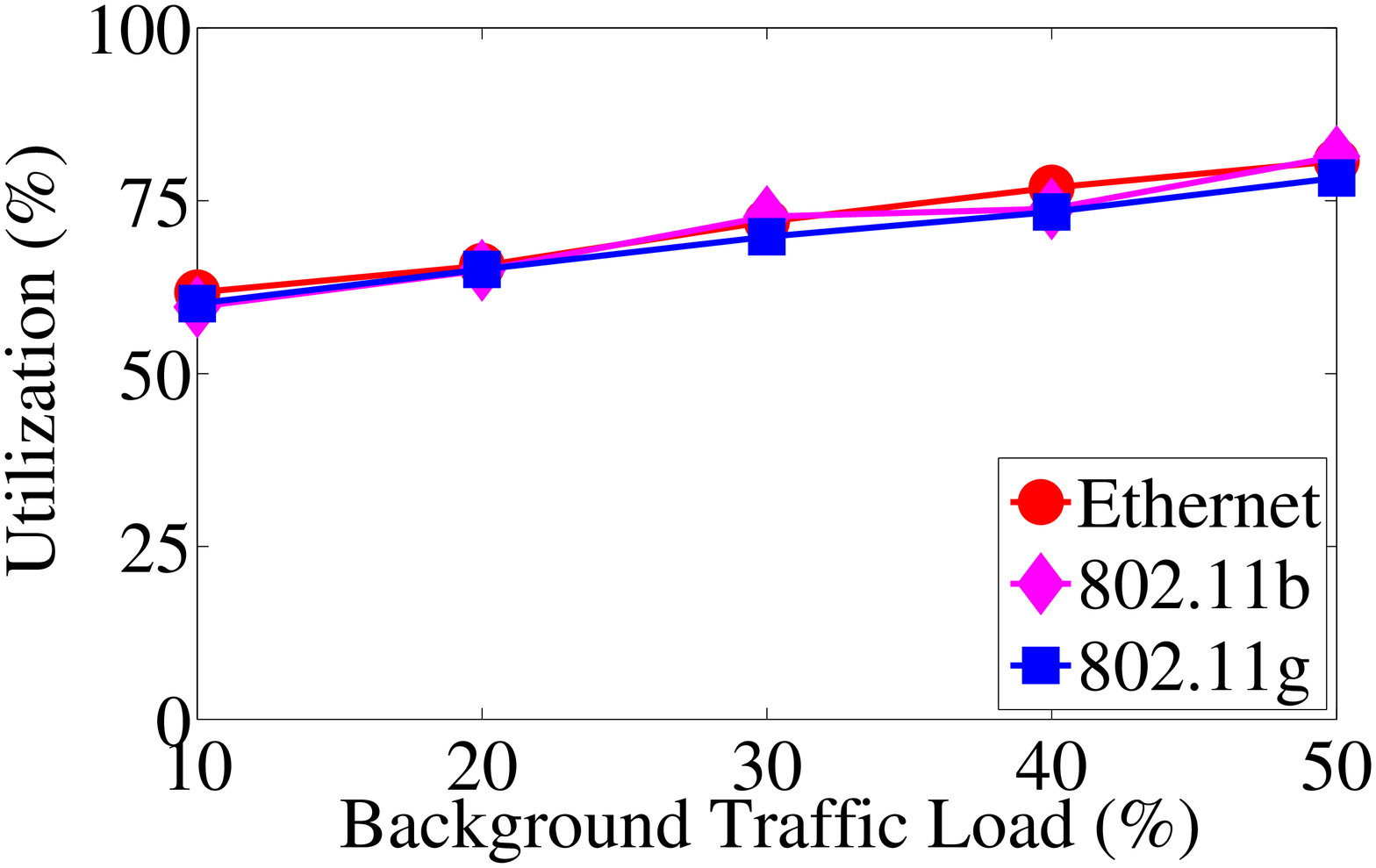}\\
    \includegraphics[width=1.0\columnwidth]{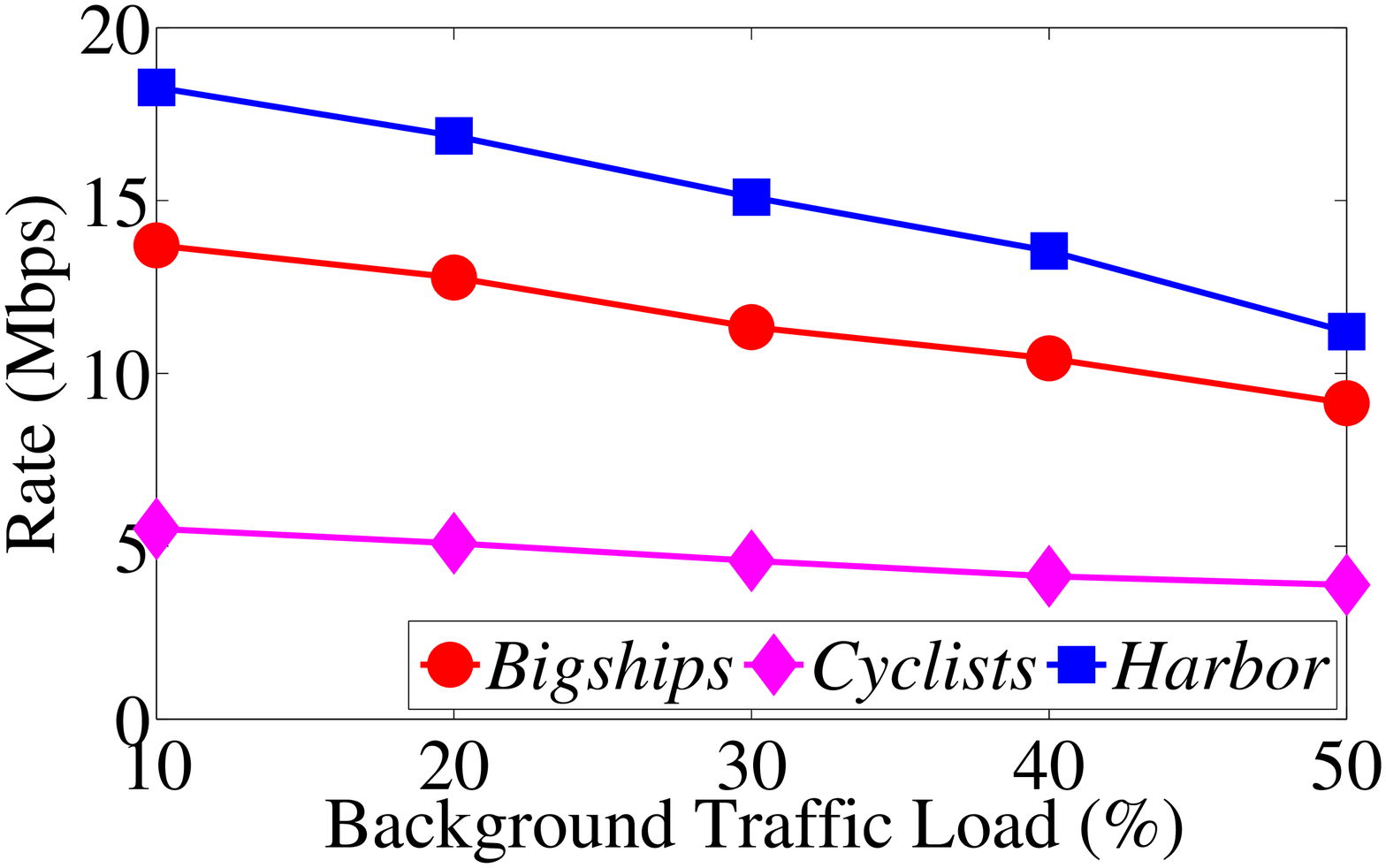}\\
    \includegraphics[width=1.0\columnwidth]{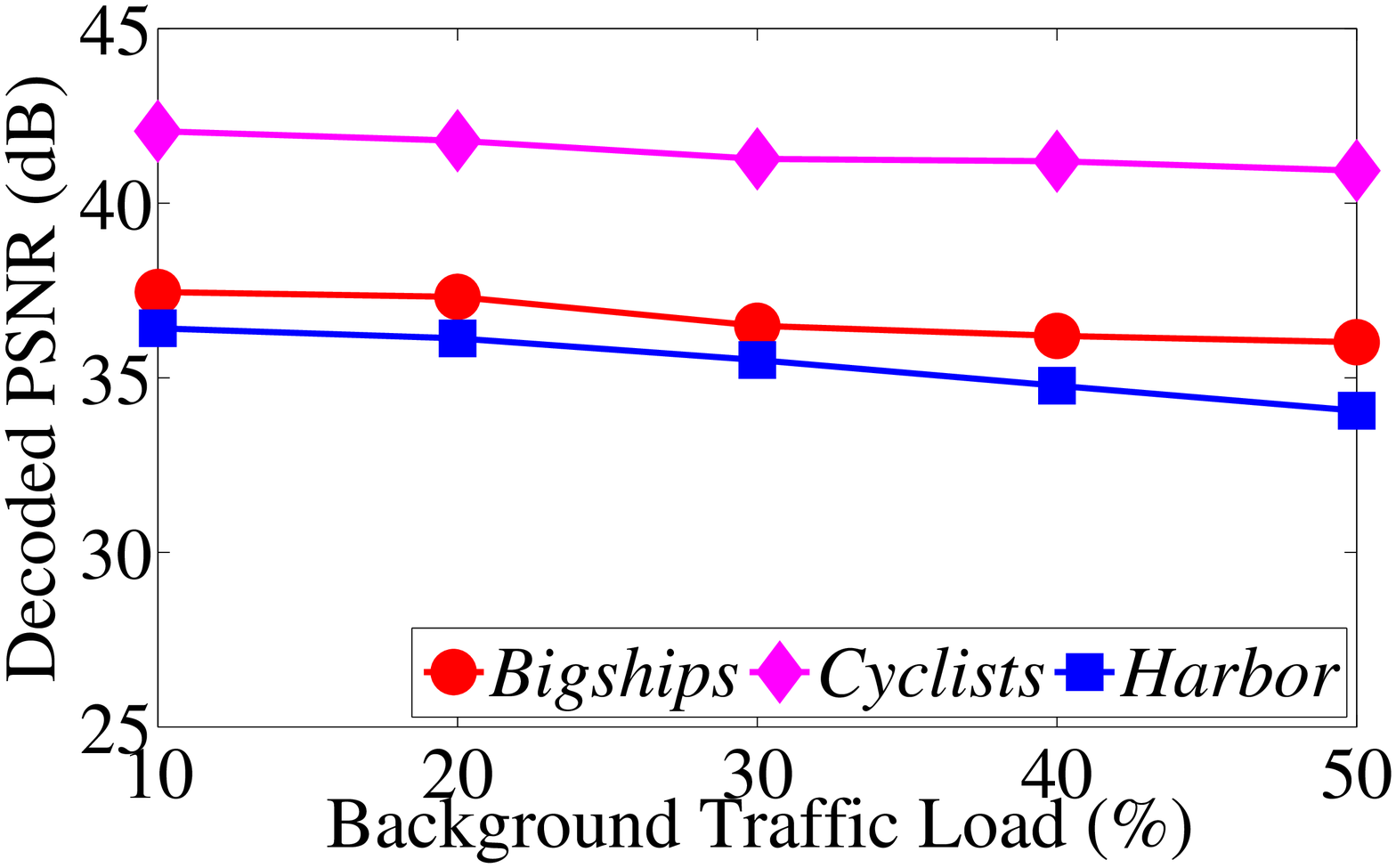}\\
        (a) Media-Aware \\ $\:$
     \end{center}
   \end{minipage}
   \hfill
   \begin{minipage}{0.5\columnwidth}
     \begin{center}
        \leavevmode
    \includegraphics[width=1.0\columnwidth]{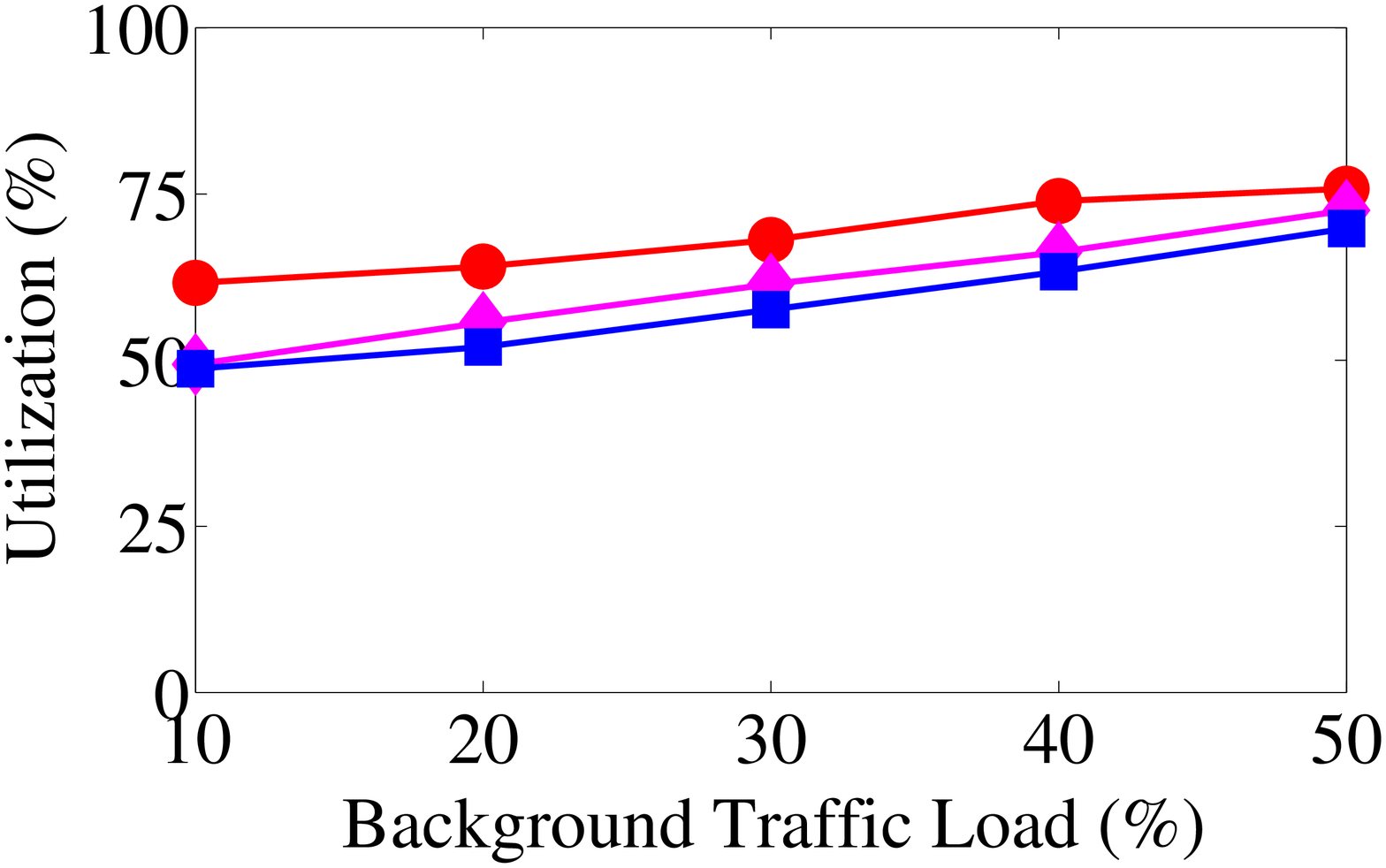}\\
    \includegraphics[width=1.0\columnwidth]{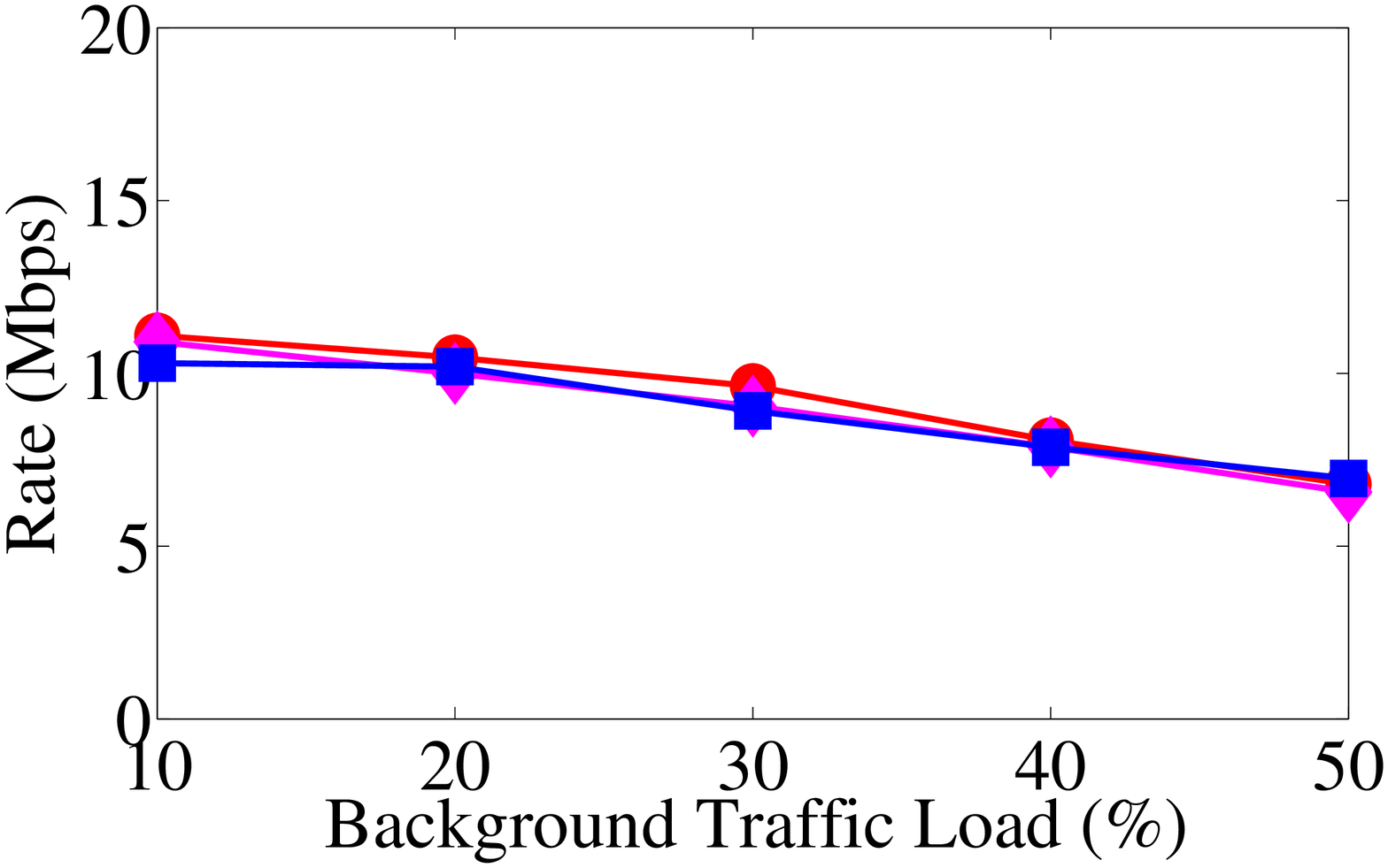}\\
    \includegraphics[width=1.0\columnwidth]{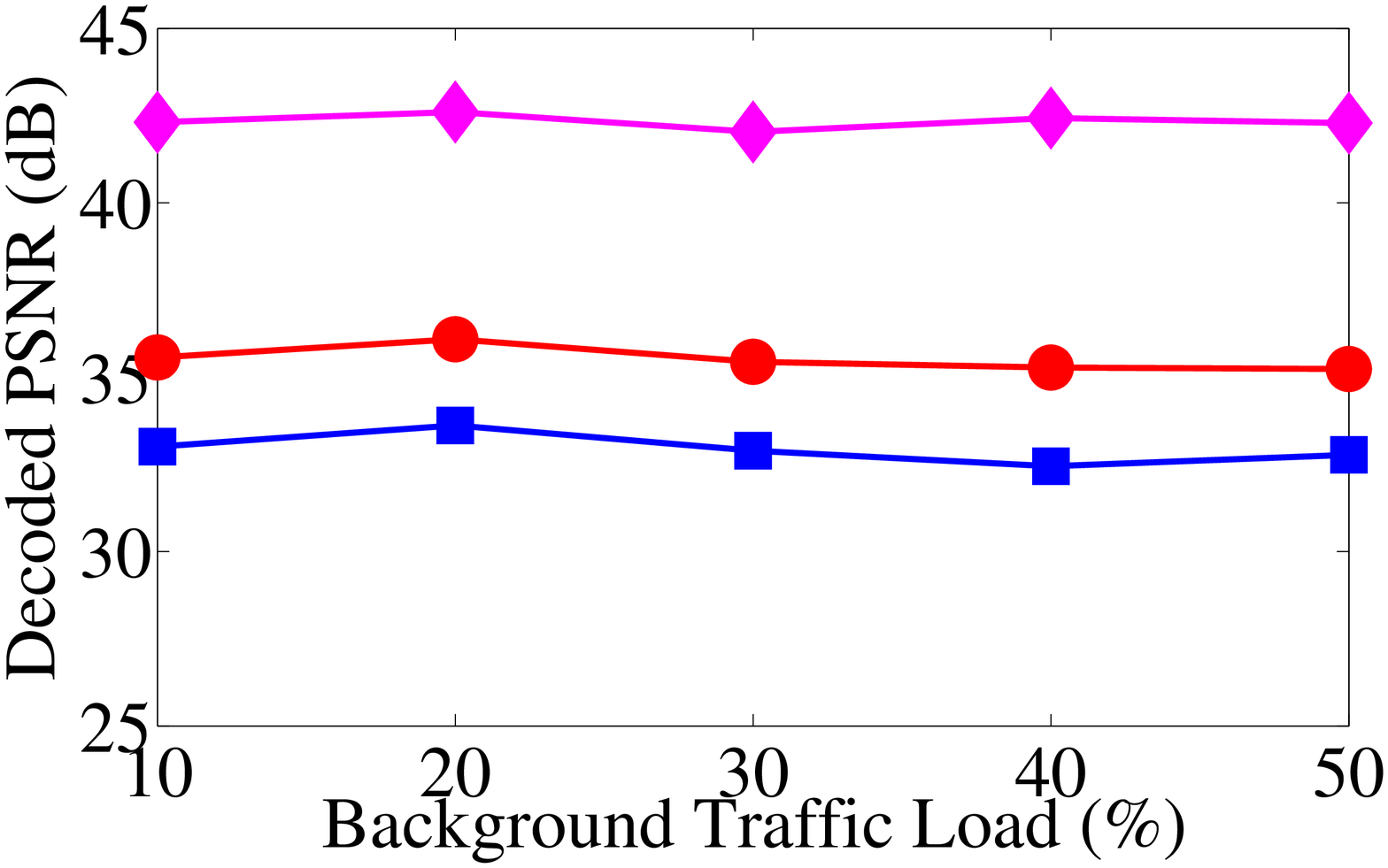}\\
        (b) H$^\infty$-Optimal \\ $\:$
     \end{center}
   \end{minipage}
    \hfill
    \begin{minipage}{0.5\columnwidth}
     \begin{center}
        \leavevmode
    \includegraphics[width=1.0\columnwidth]{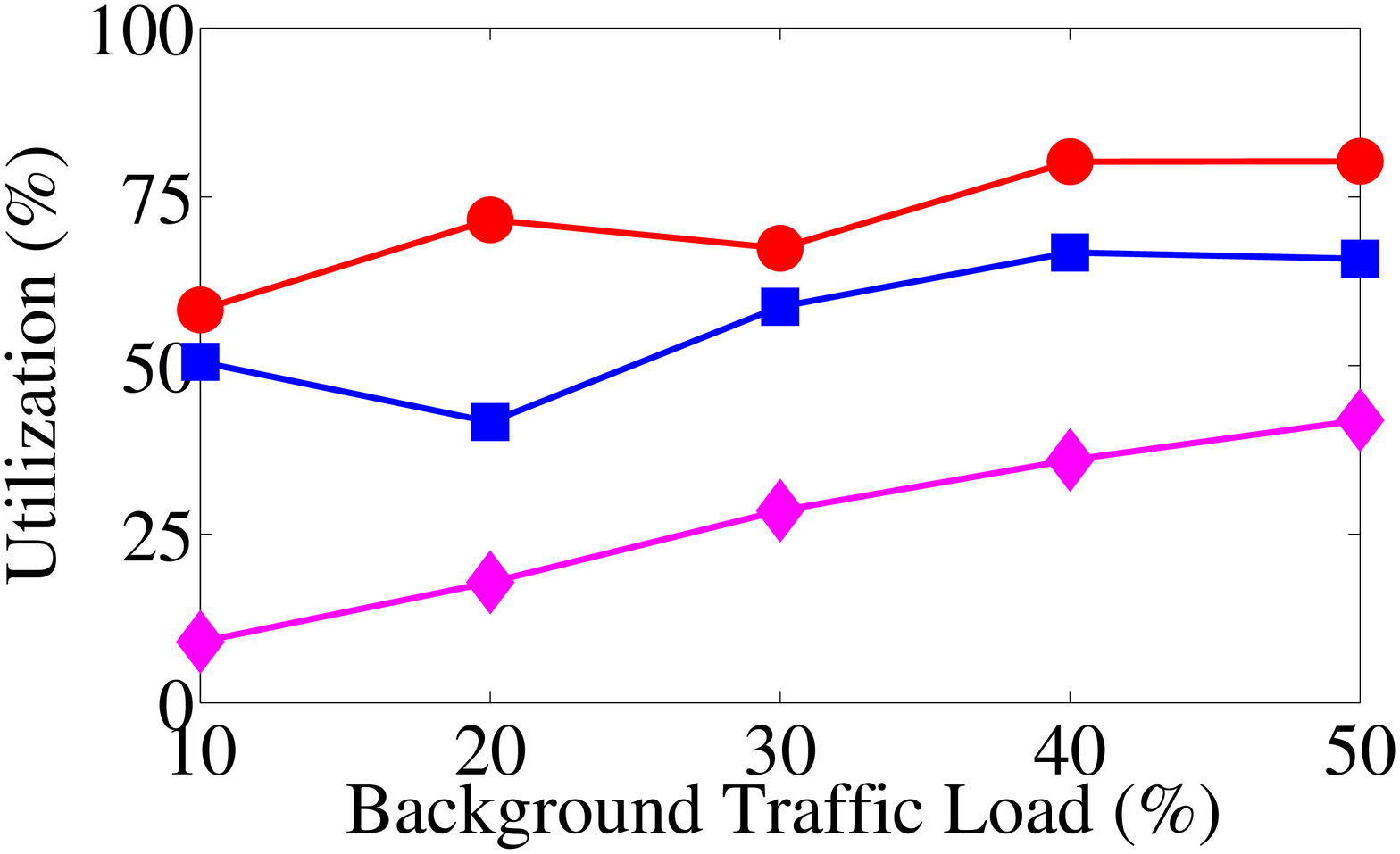}\\
    \includegraphics[width=1.0\columnwidth]{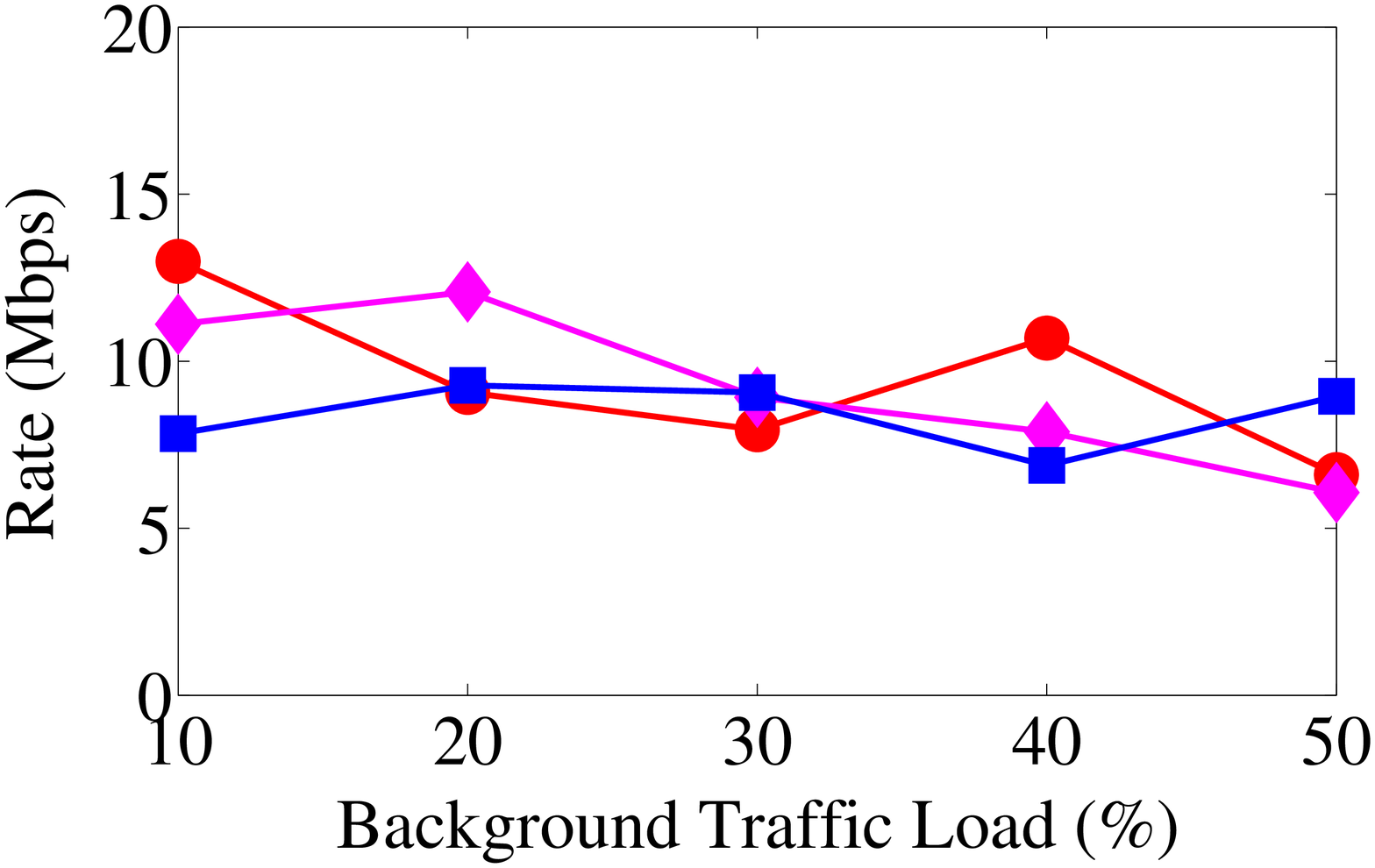}\\
    \includegraphics[width=1.0\columnwidth]{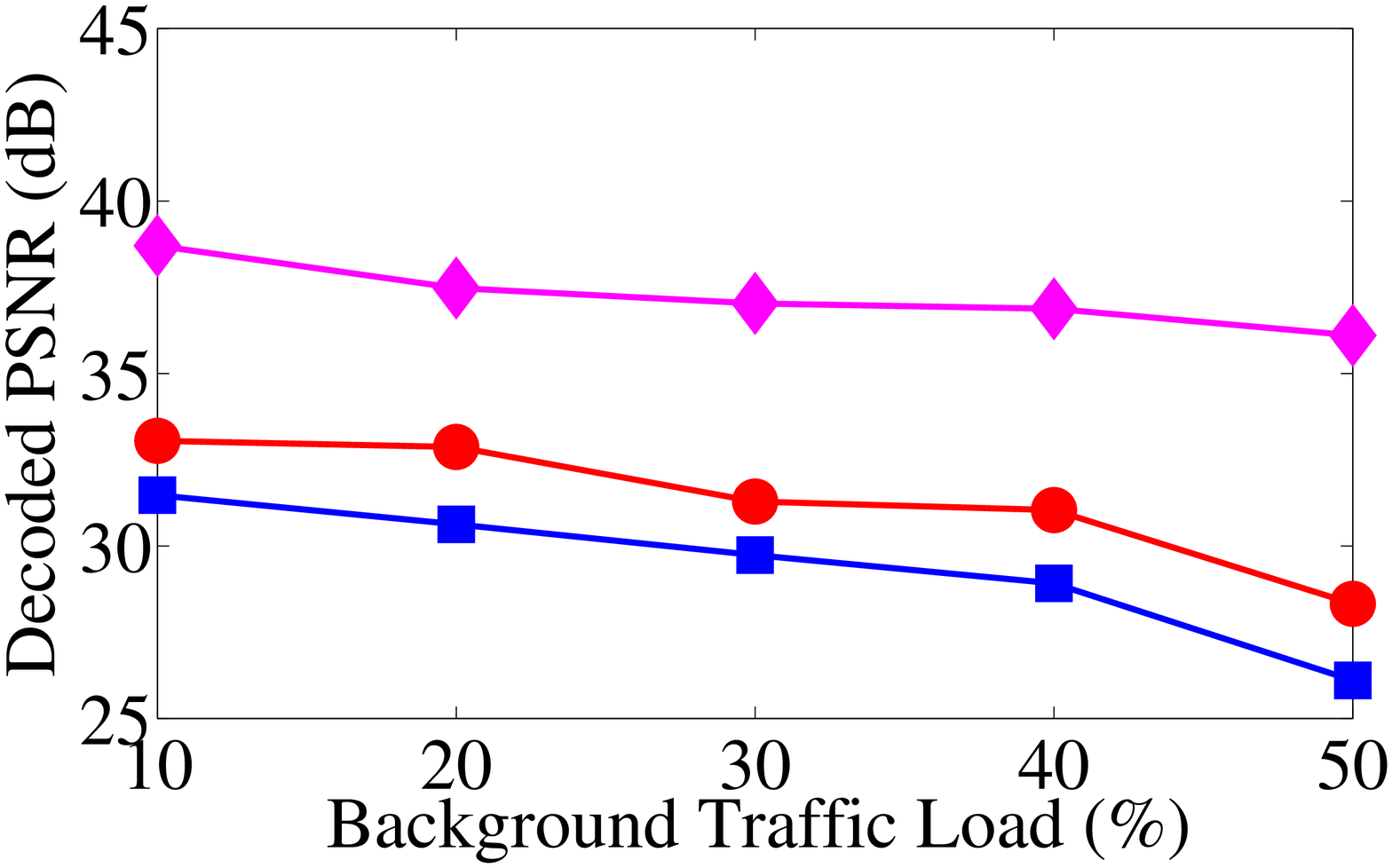}\\
        (c) Greedy AIMD \\ $\:$
     \end{center}
   \end{minipage}
    \hfill
    \begin{minipage}{0.5\columnwidth}
     \begin{center}
        \leavevmode
    \includegraphics[width=1.0\columnwidth]{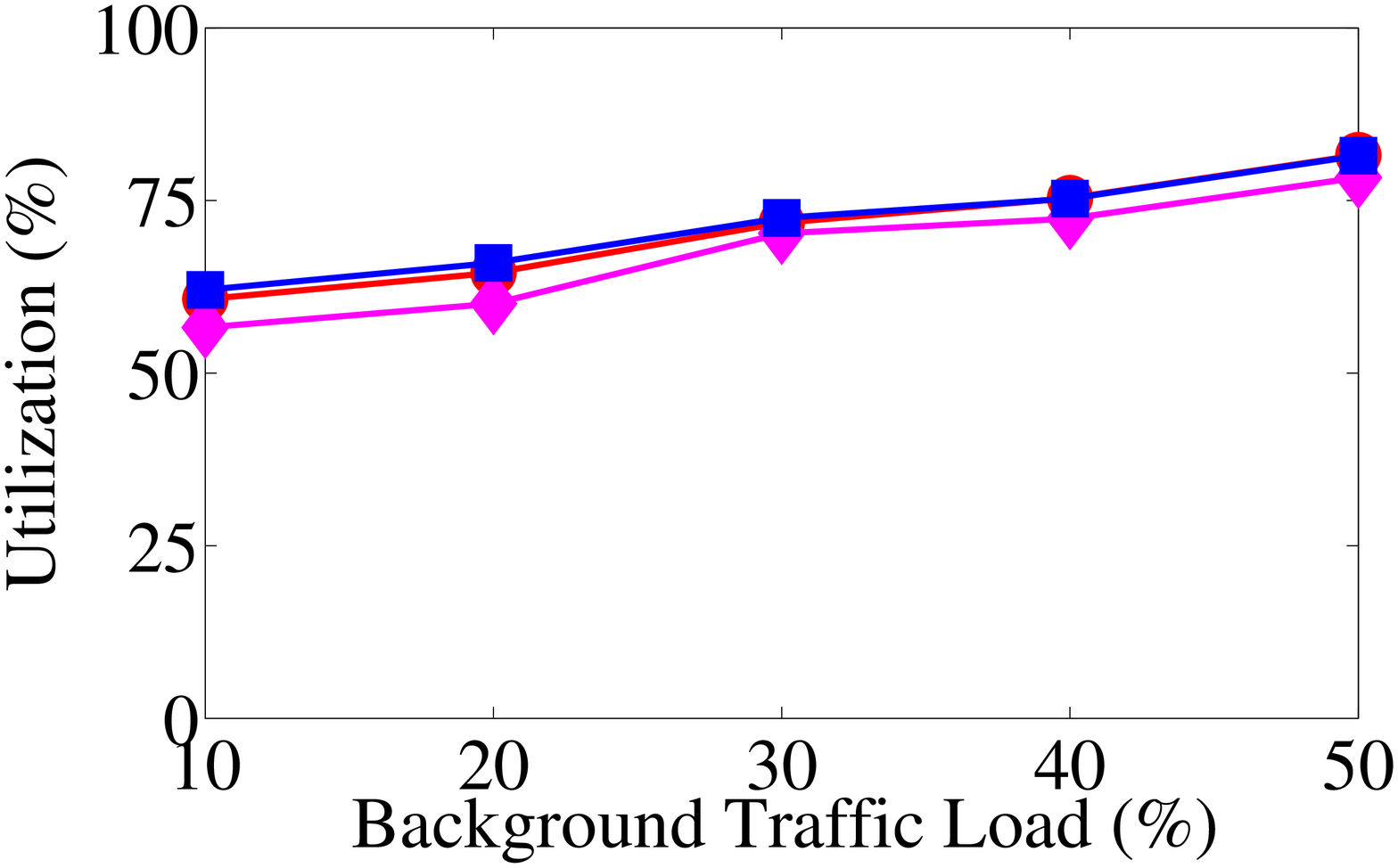}\\
    \includegraphics[width=1.0\columnwidth]{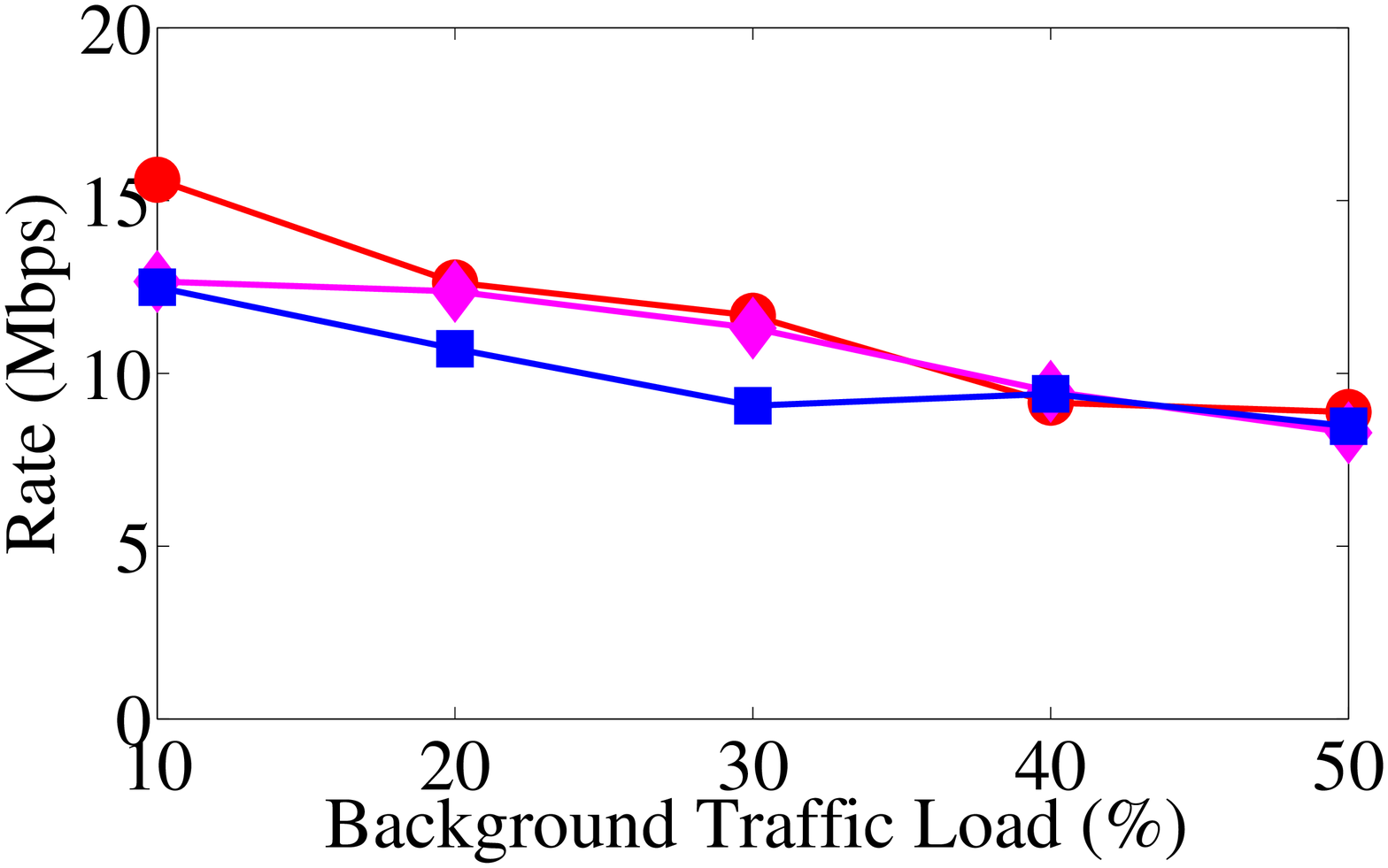}\\
    \includegraphics[width=1.0\columnwidth]{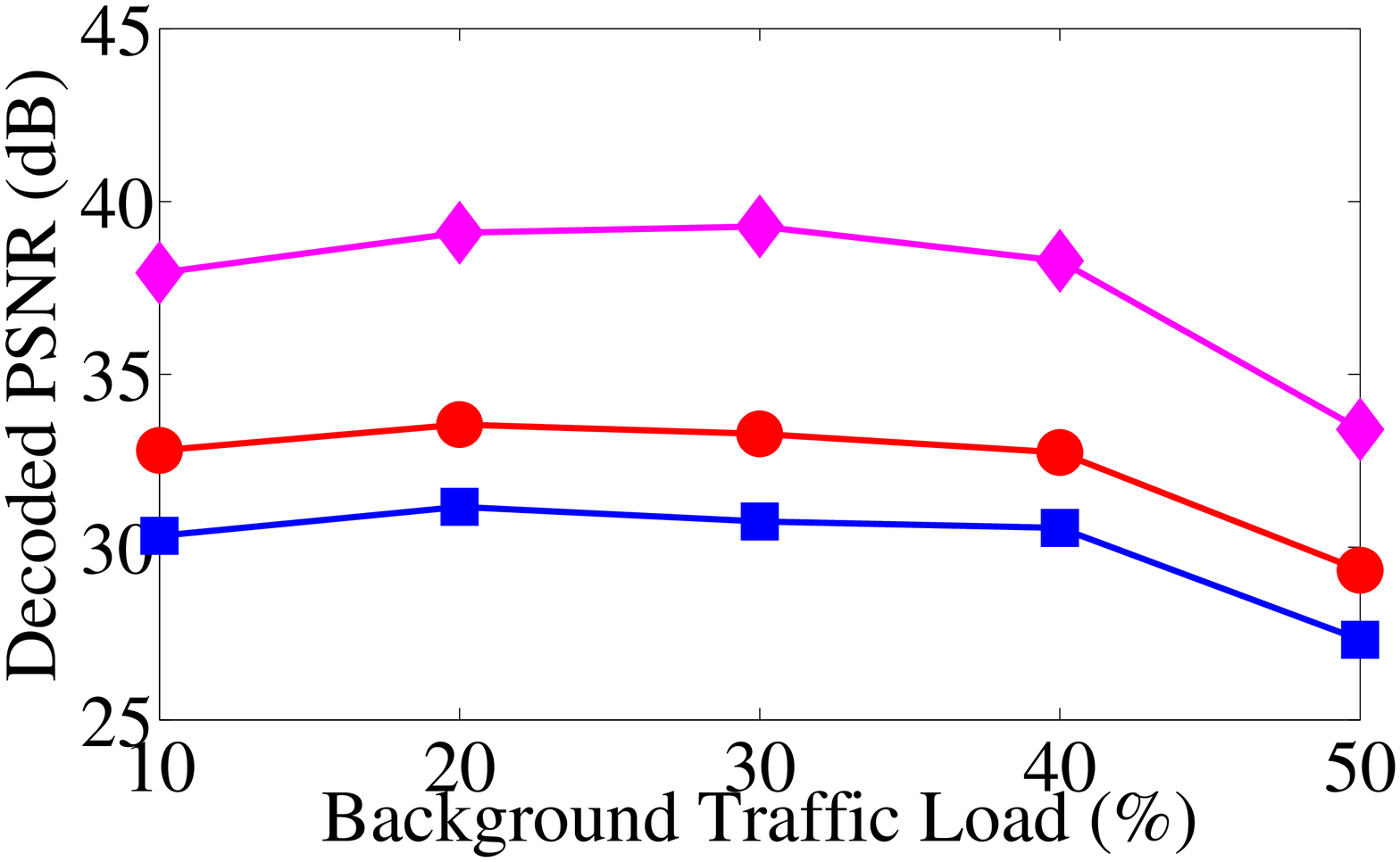}\\
        (d) Rate Proportional \\ AIMD
     \end{center}
   \end{minipage}
 \caption{\small Comparison of allocation results from different
 schemes, as the background traffic load increases.}
 \label{fig:AllocationPerLoad}
\end{figure*}

%
Next, we vary the percentage of background traffic over each network
from 10\% to 50\%, with playout deadline of 300~ms. The impact of
the background traffic load on the allocation results is shown in
Fig.~\ref{fig:AllocationPerLoad}. It can be observed that total
utilization over each interface increases with the background
traffic load. For the media-aware, H$^\infty$-optimal and rate
proportional AIMD schemes, utilization varies between 60\% to 90\%,
whereas for the greedy AIMD scheme, the 802.11b interface is
underutilized.\footnote{Since 802.11b has significantly lower ABR
than the other two interfaces, it is never chosen by the greedy AIMD
scheme.} Note that media-aware allocation ensures balanced
utilization over all three access networks, as dictated by
\eqref{eqn:EqualUtilization}.\\

\begin{figure}
\centering
    \begin{minipage}{0.492\columnwidth}
     \begin{center}
    \includegraphics[width=1.0\columnwidth]{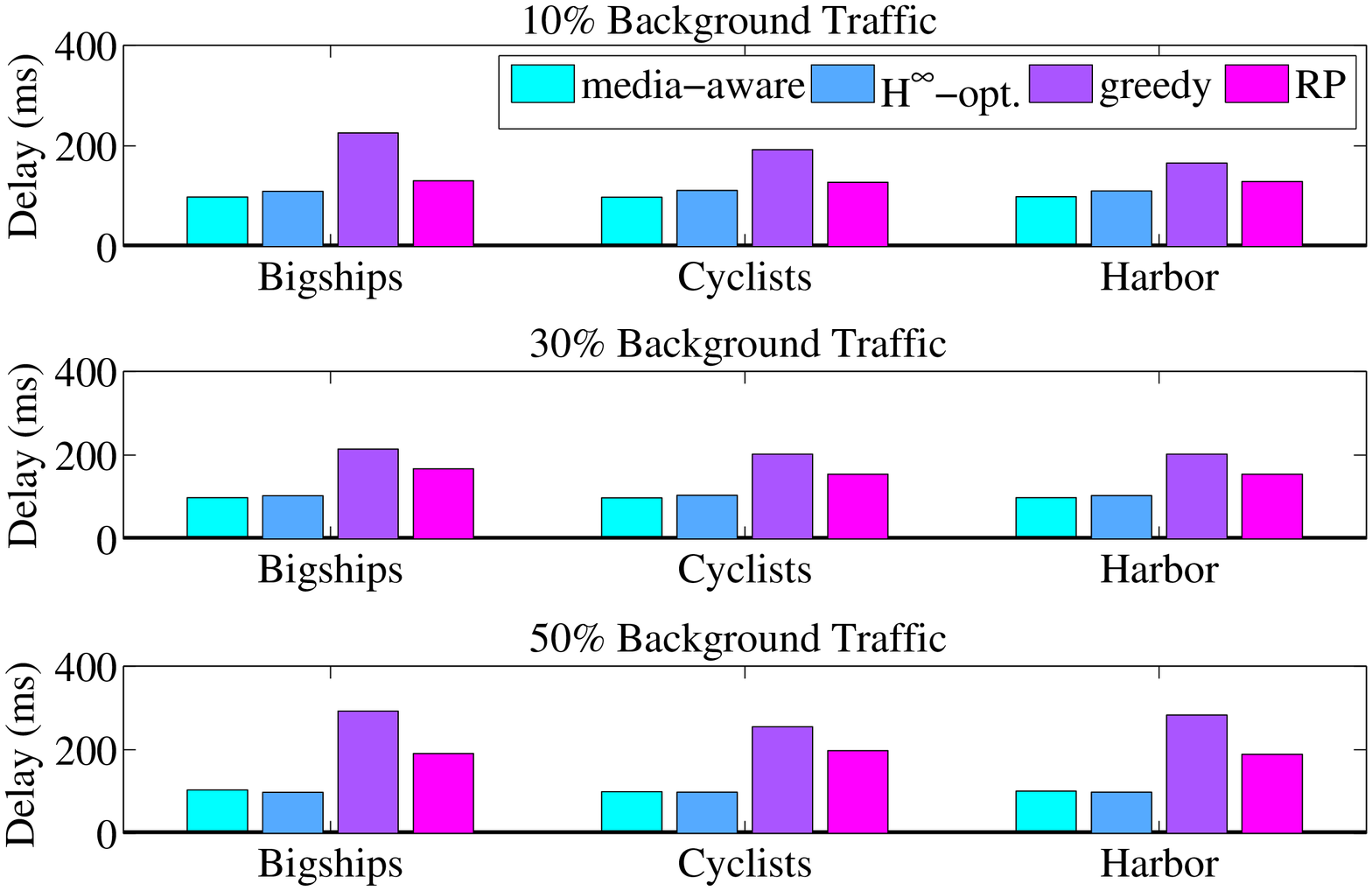}\\ (a)
    \end{center}
    \end{minipage}
    \hfill
    \begin{minipage}{0.492\columnwidth}
     \begin{center}
   \includegraphics[width=1.0\columnwidth]{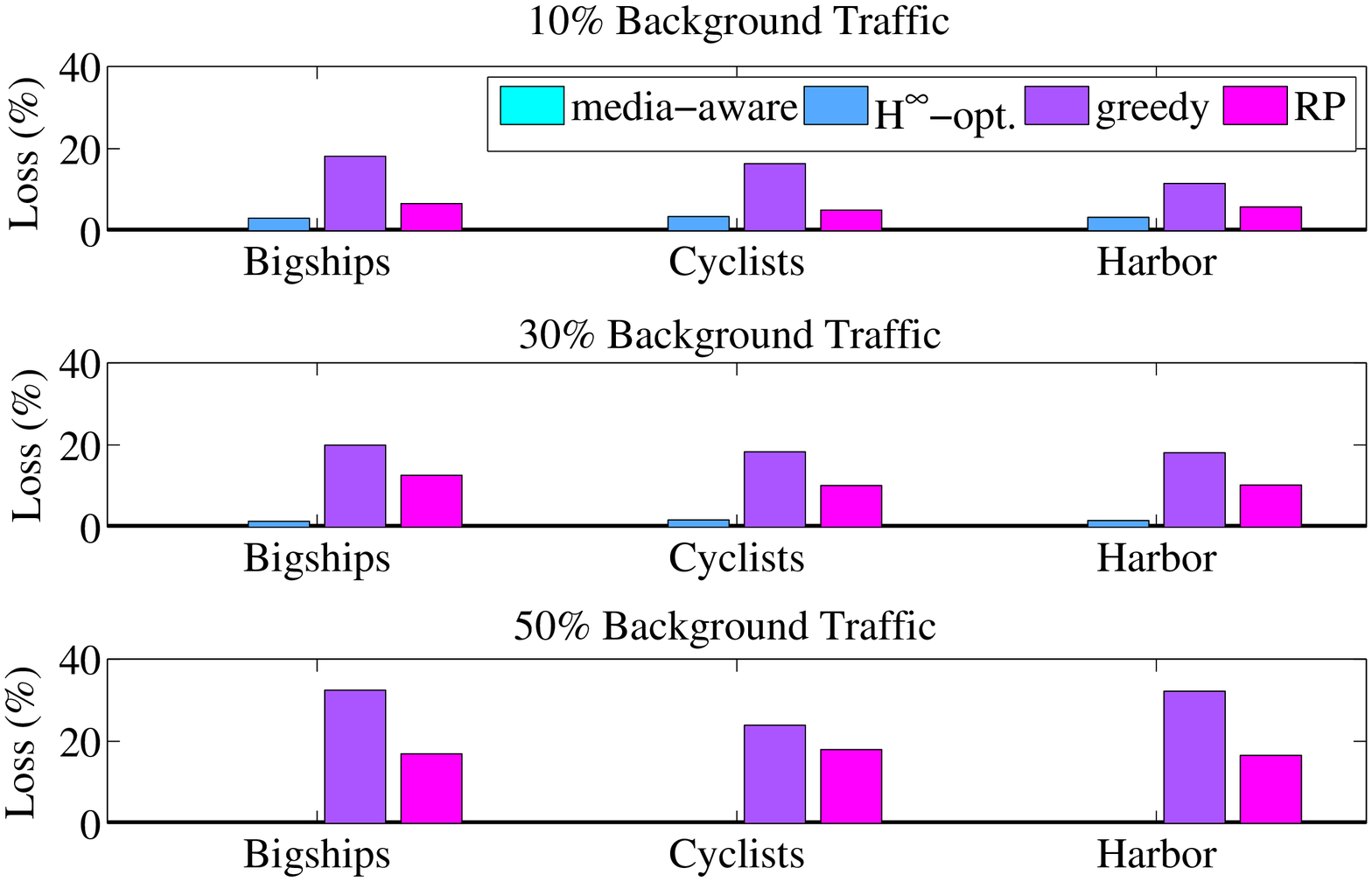}\\ (b)
    \end{center}
    \end{minipage}
\caption{\small Average packet delivery delay (a) and packet loss
ratio (b) at the receiver for each video stream, with background
traffic load at 10\%, 30\% and 50\%, respectively. The playout
deadline is 300~ms.} \label{fig:DelayLossPerLoad}
\end{figure}

\begin{figure}
\centering
   \includegraphics[width=0.7\columnwidth]{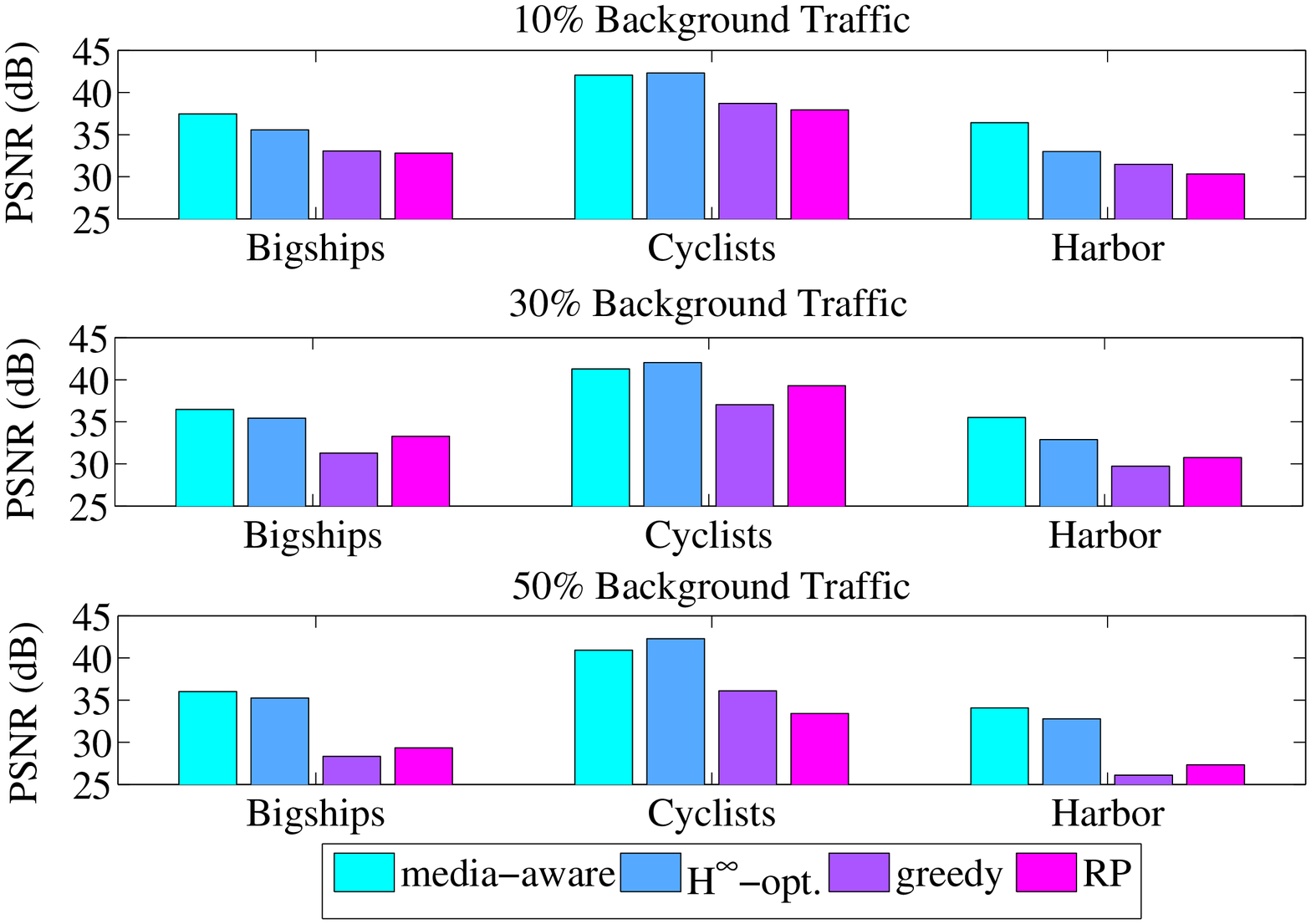}
\caption{\small Received video quality in PSNR for \emph{Bigships},
\emph{Cyclists} and \emph{Harbor}, for background traffic load at
10\%, 30\% and 50\%, respectively. The playout deadline is 300~ms.}
\label{fig:DecodedPSNRPerLoad}
\end{figure}

\indent It can be observed from Fig.~\ref{fig:AllocationPerLoad}
that, increasing background traffic load leads to decreasing
allocated rate in each stream. While the other three schemes treat
the three flows with equal importance, the media-aware allocation
consistently favors the more demanding \emph{Harbor}, thereby
reducing the quality gap between the three sequences. The two
AIMD-based heuristics achieve lower received video quality than
media-aware and H$^\infty$-optimal allocations, especially in the
presence of heavier background traffic load. \\
\indent Figure~\ref{fig:DelayLossPerLoad} compares the average
packet delivery delay and packet loss ratios due to late arrivals.
In the two AIMD-based schemes, allocated rates are reduced only
\emph{after} congestion has been detected. The media-aware
allocation and the H$^\infty$-optimal control schemes, on the other
hand, attempt to avoid network congestion in a proactive manner in
their problem formulations. They therefore yield significantly lower
packet loss ratios and delays. This leads to improved received video
quality, as shown in Fig.~\ref{fig:DecodedPSNRPerLoad}. The
performance gain ranges between 1.5 to 8.8~dB in PSNR of the decoded
video, depending on the sequence content and background traffic
load. Note also, that the packet delivery delays and packet loss
ratios also indicate the impact of each scheme on background traffic
sharing the same access networks. Lower delays and losses achieved
by the media-aware and H$^\infty$-optimal schemes means that they
introduce less disruption to ongoing flows, as a results of
proactive congestion avoidance.\
%
\subsection{Varying Playout Deadline}
\label{subsec:PlayoutDeadline}
\begin{figure*}
   \begin{minipage}{0.5\columnwidth}
     \begin{center}
        \leavevmode
    \includegraphics[width=1.0\columnwidth]{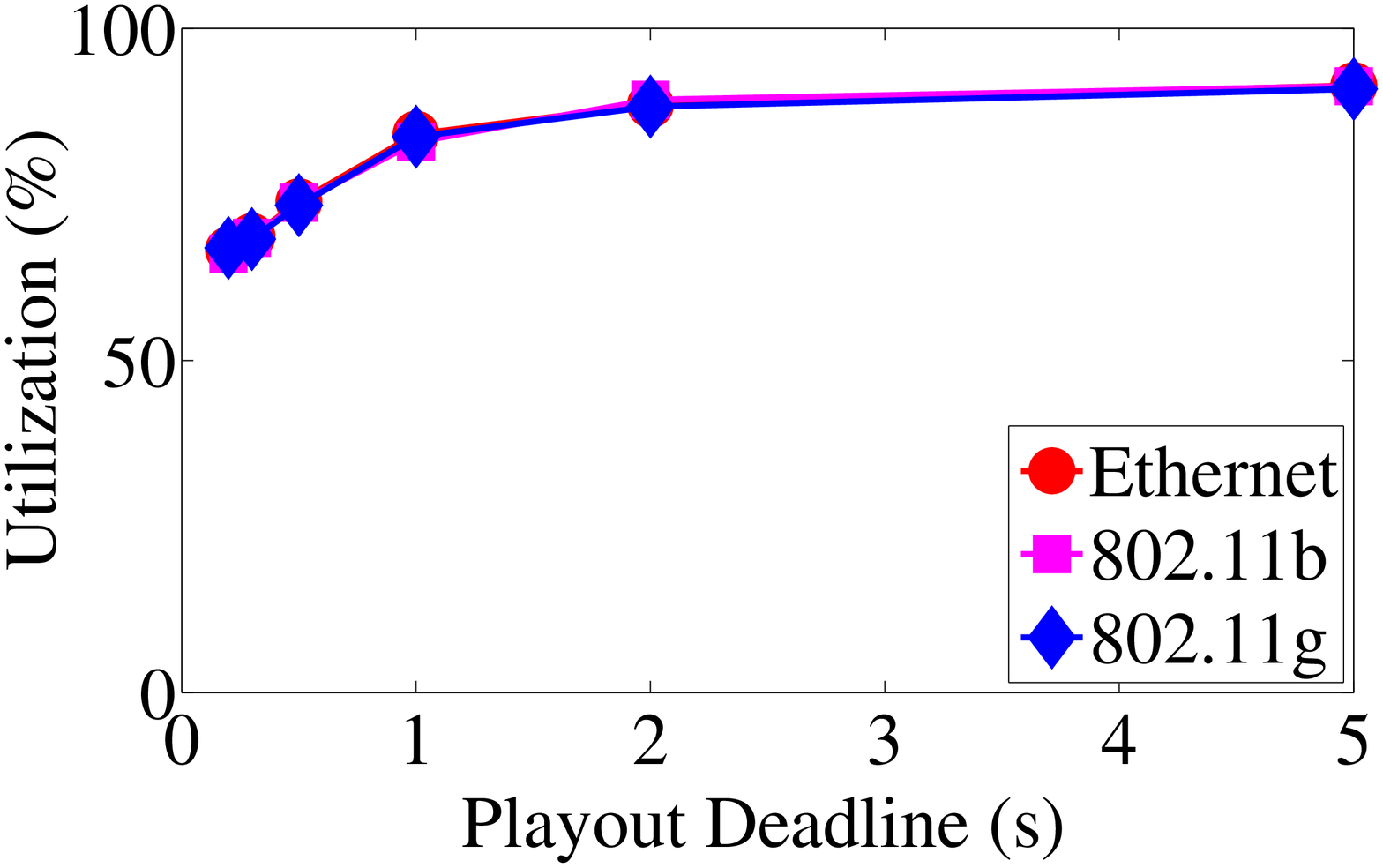}\\
    \includegraphics[width=1.0\columnwidth]{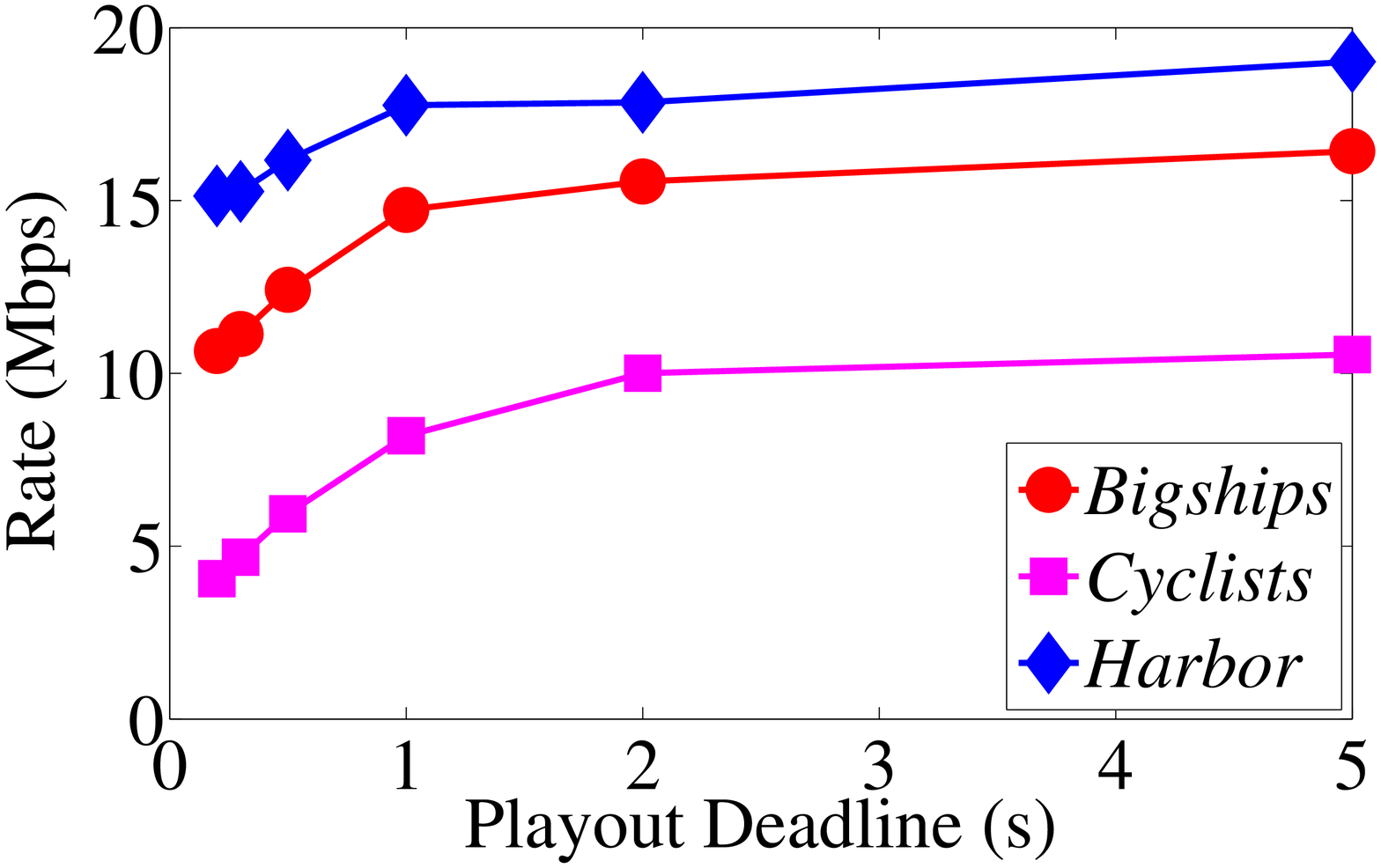}\\
    \includegraphics[width=1.0\columnwidth]{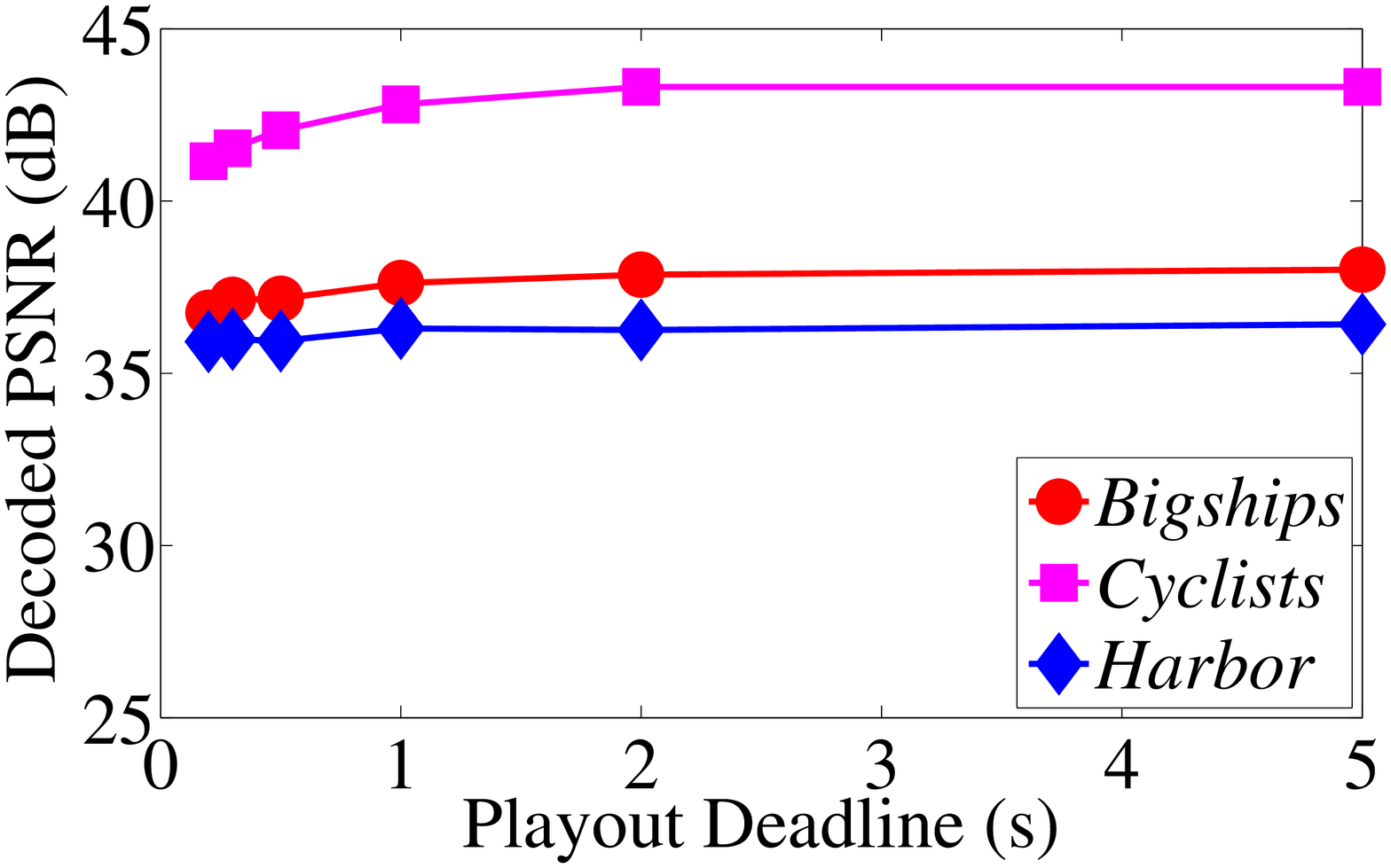}\\
        (a) Media-Aware \\ $\:$
      \end{center}
   \end{minipage}
   \hfill
   \begin{minipage}{0.5\columnwidth}
     \begin{center}
    \includegraphics[width=1.0\columnwidth]{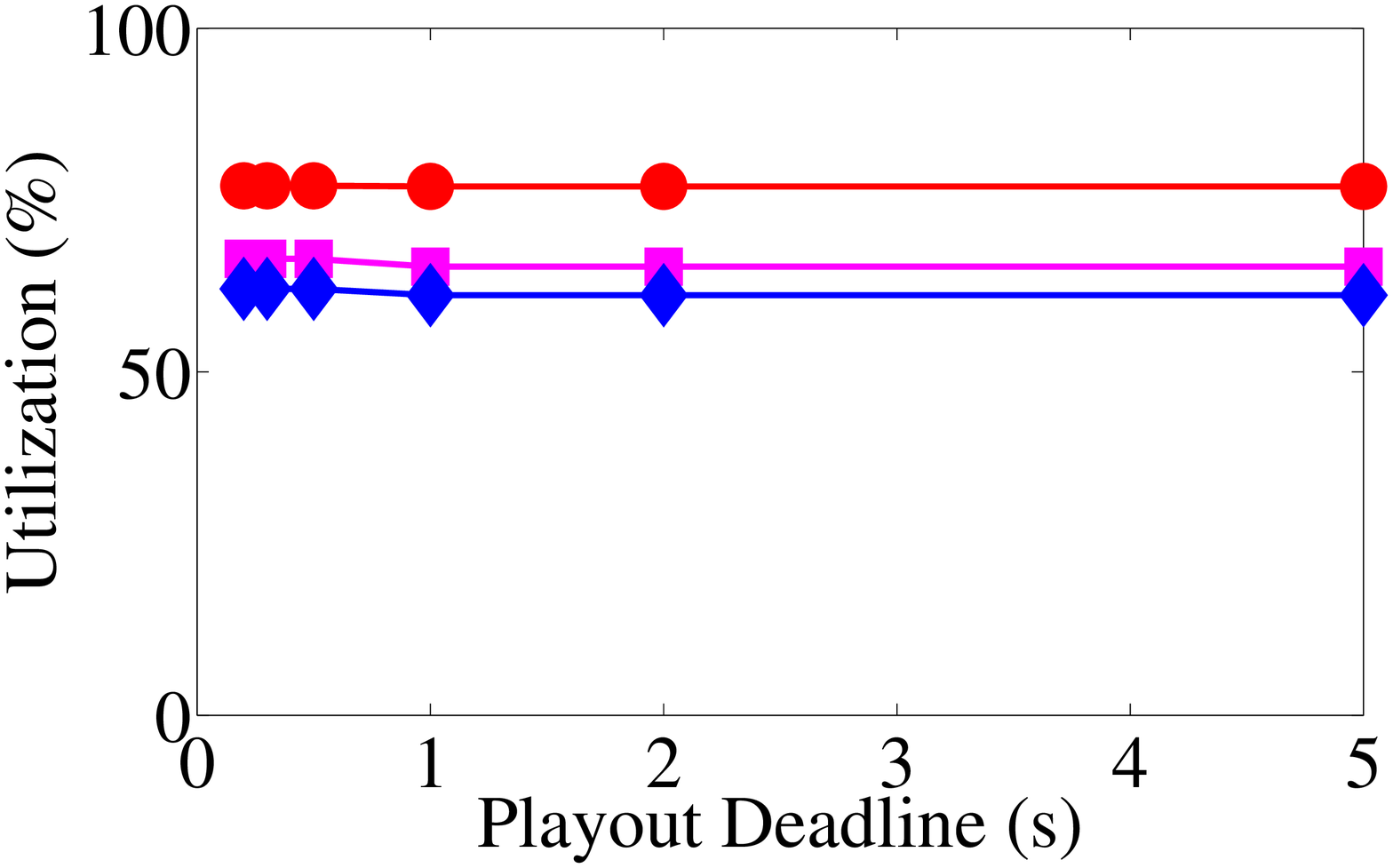}\\
    \includegraphics[width=1.0\columnwidth]{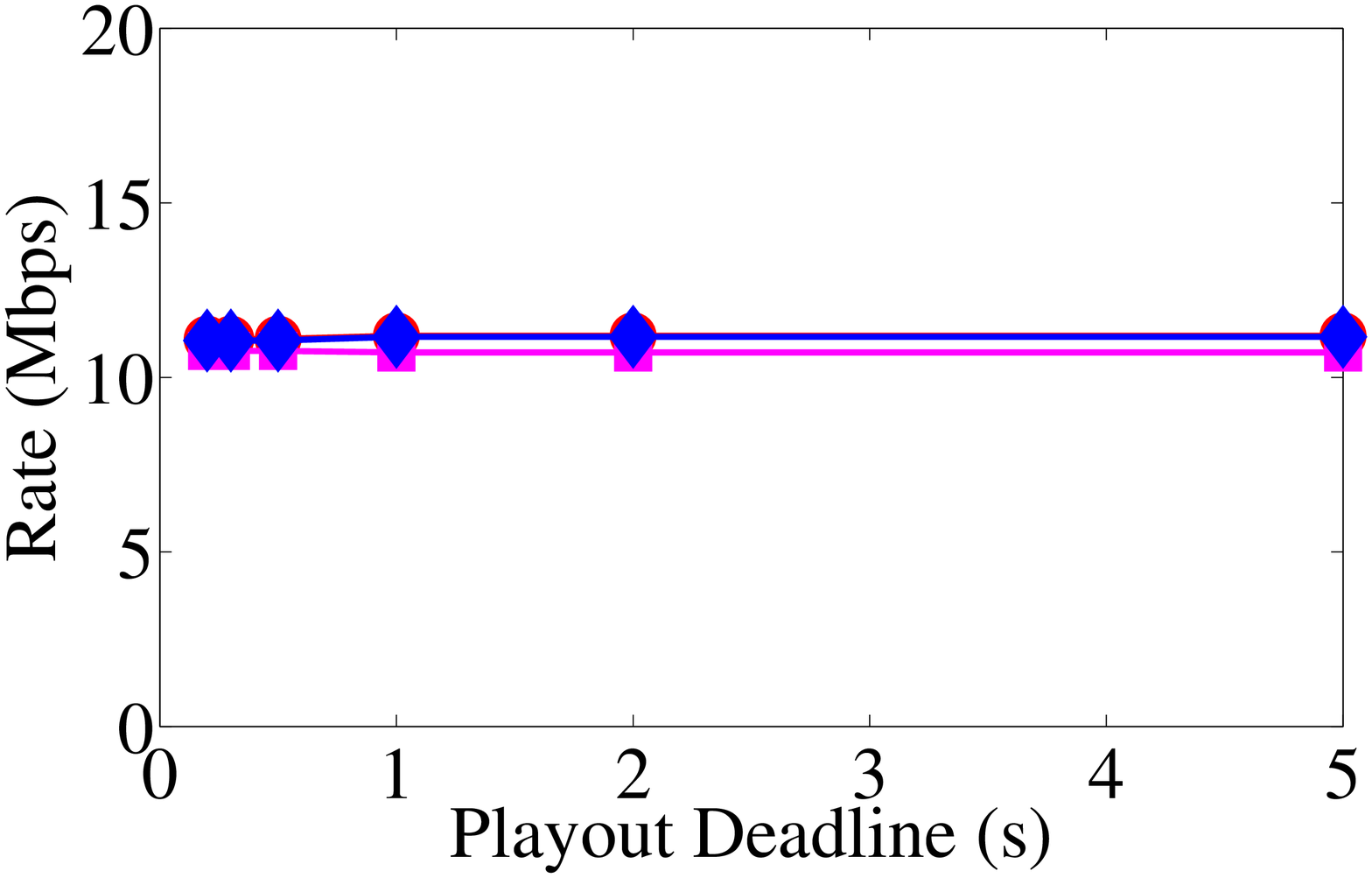}\\
    \includegraphics[width=1.0\columnwidth]{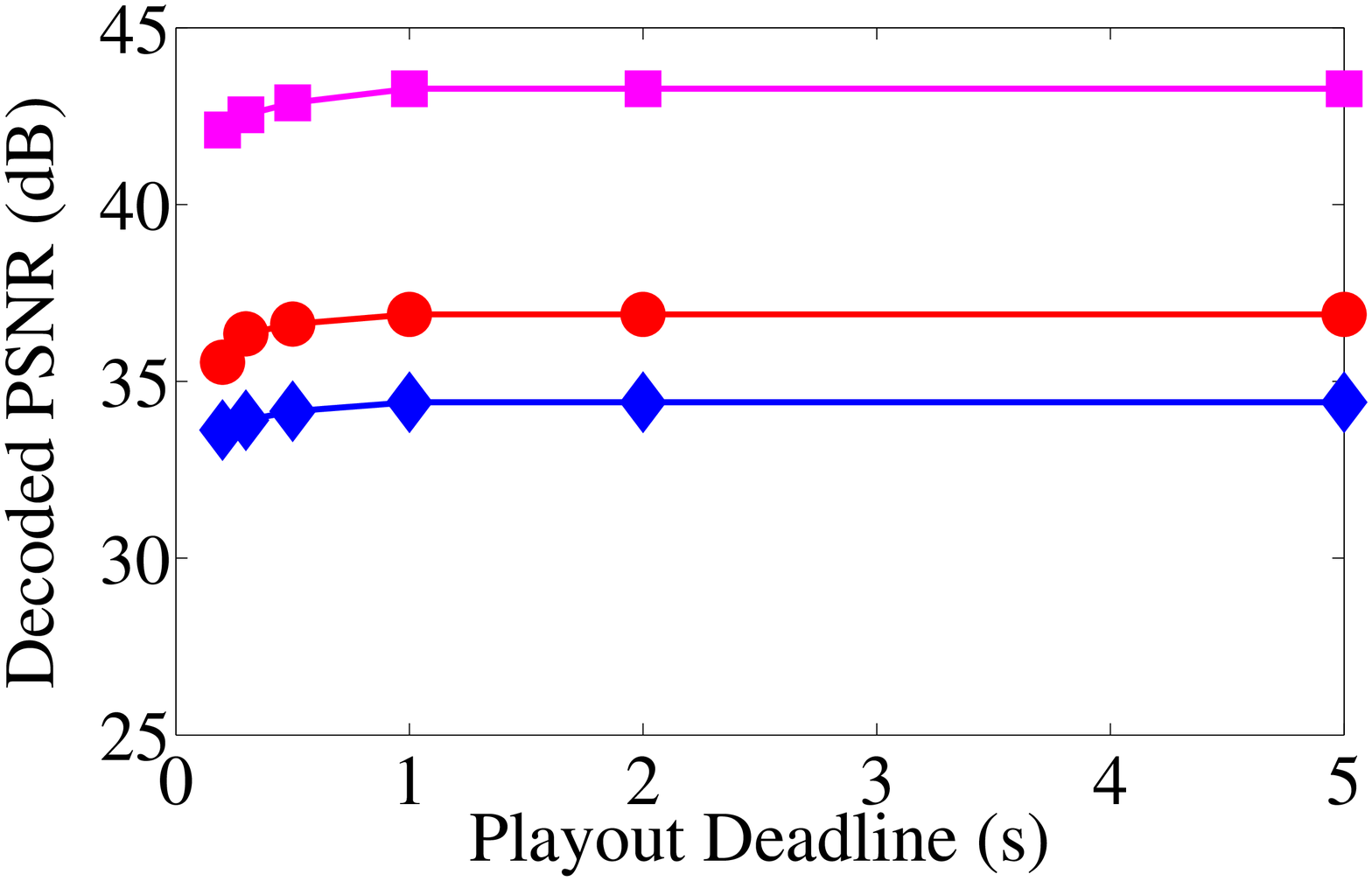}\\
        (b) H$^\infty$-Optimal \\ $\:$
     \end{center}
   \end{minipage}
   \hfill
    \begin{minipage}{0.5\columnwidth}
     \begin{center}
    \includegraphics[width=1.0\columnwidth]{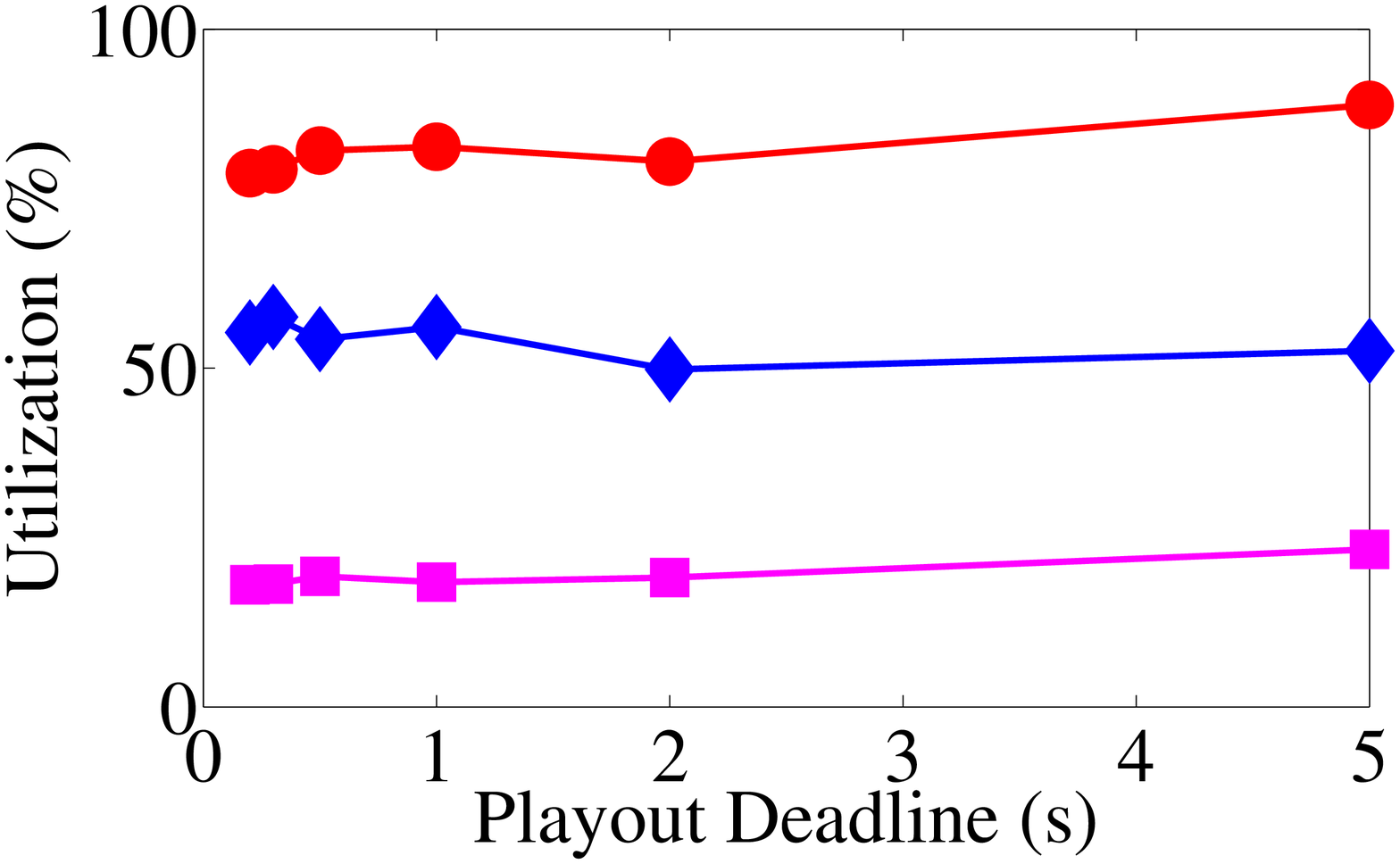}\\
    \includegraphics[width=1.0\columnwidth]{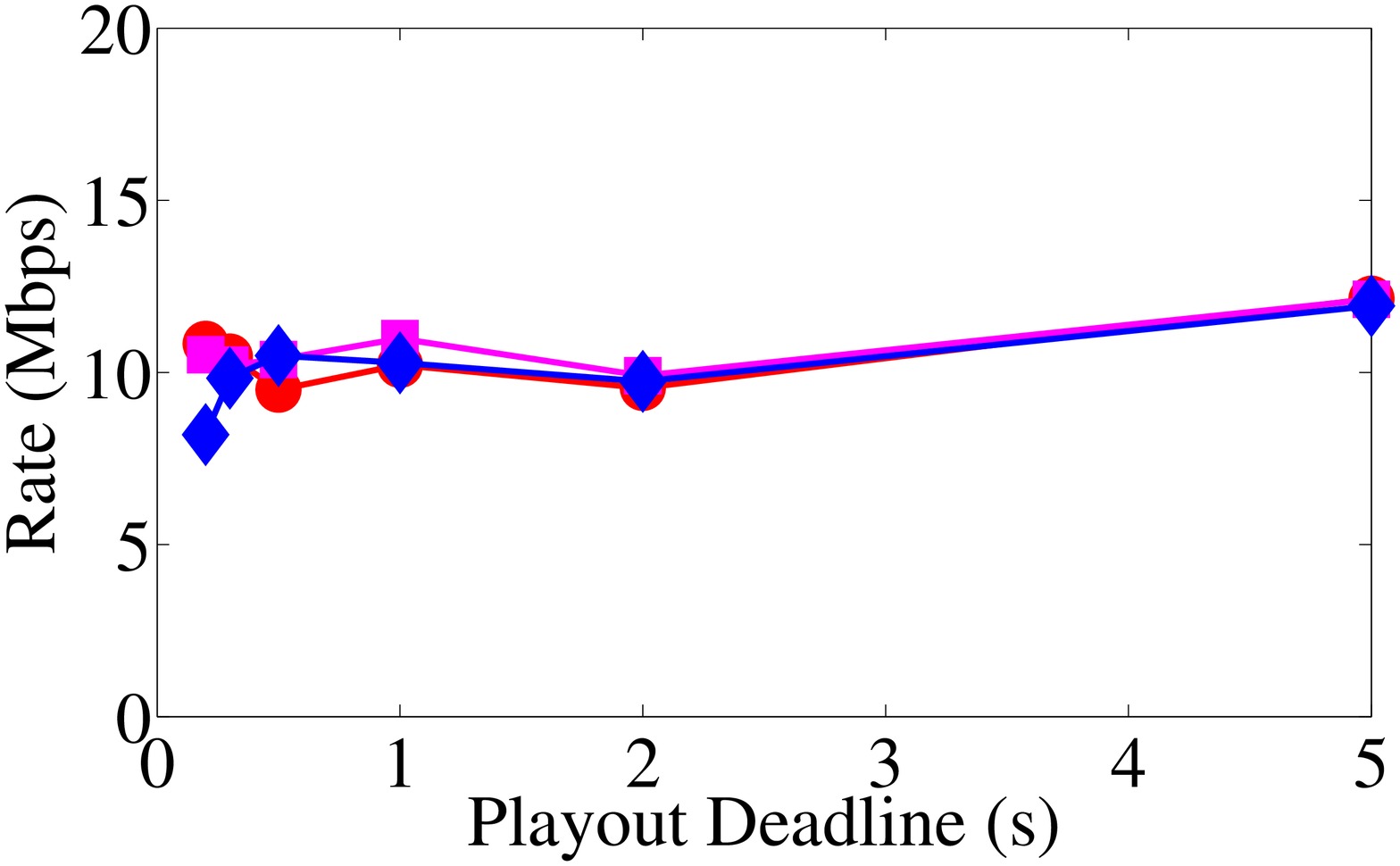}\\
    \includegraphics[width=1.0\columnwidth]{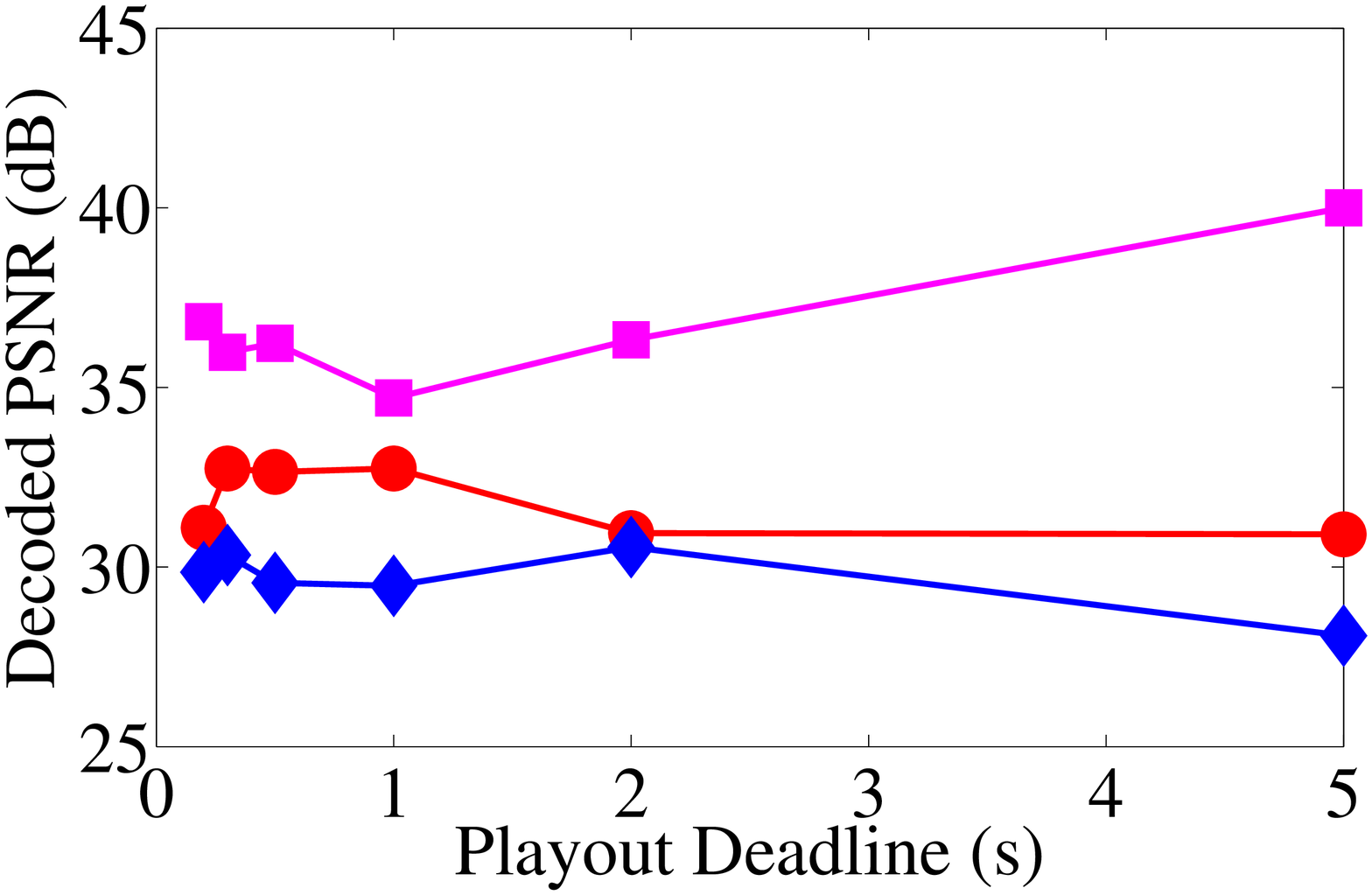}\\
        (c) Greedy AIMD \\ $\:$
     \end{center}
    \end{minipage}
    \hfill
    \begin{minipage}{0.5\columnwidth}
     \begin{center}
    \includegraphics[width=1.0\columnwidth]{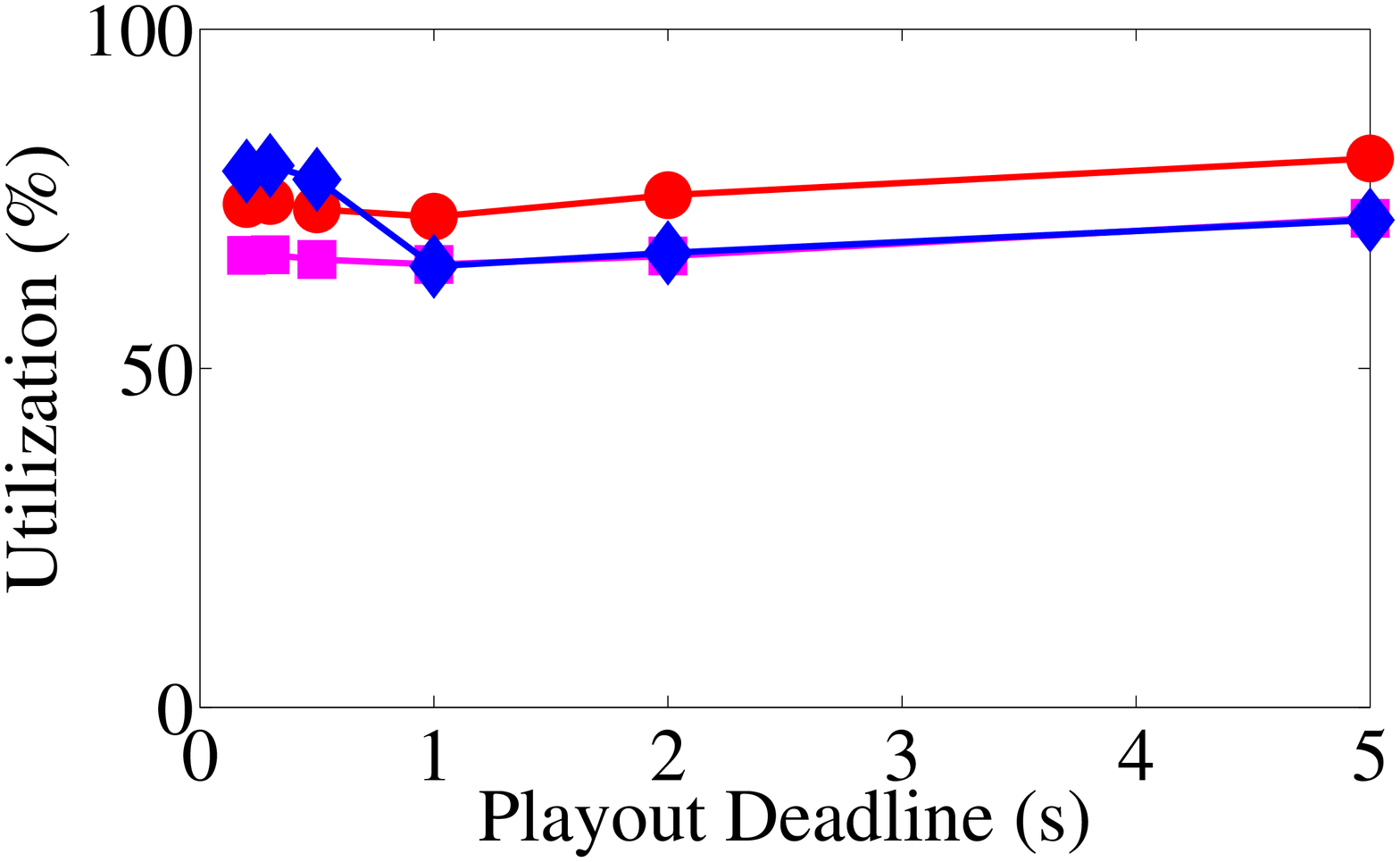}\\
    \includegraphics[width=1.0\columnwidth]{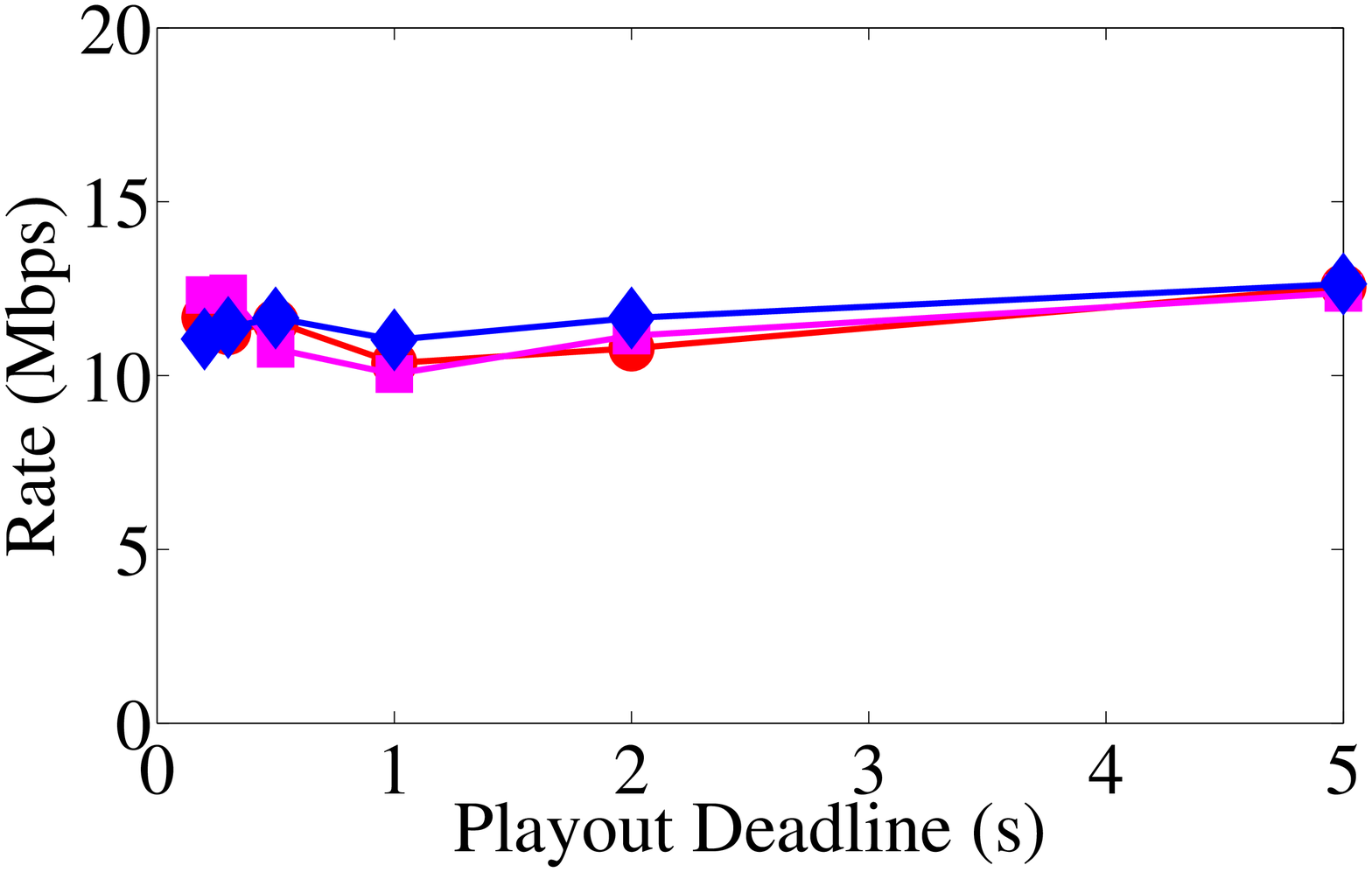}\\
    \includegraphics[width=1.0\columnwidth]{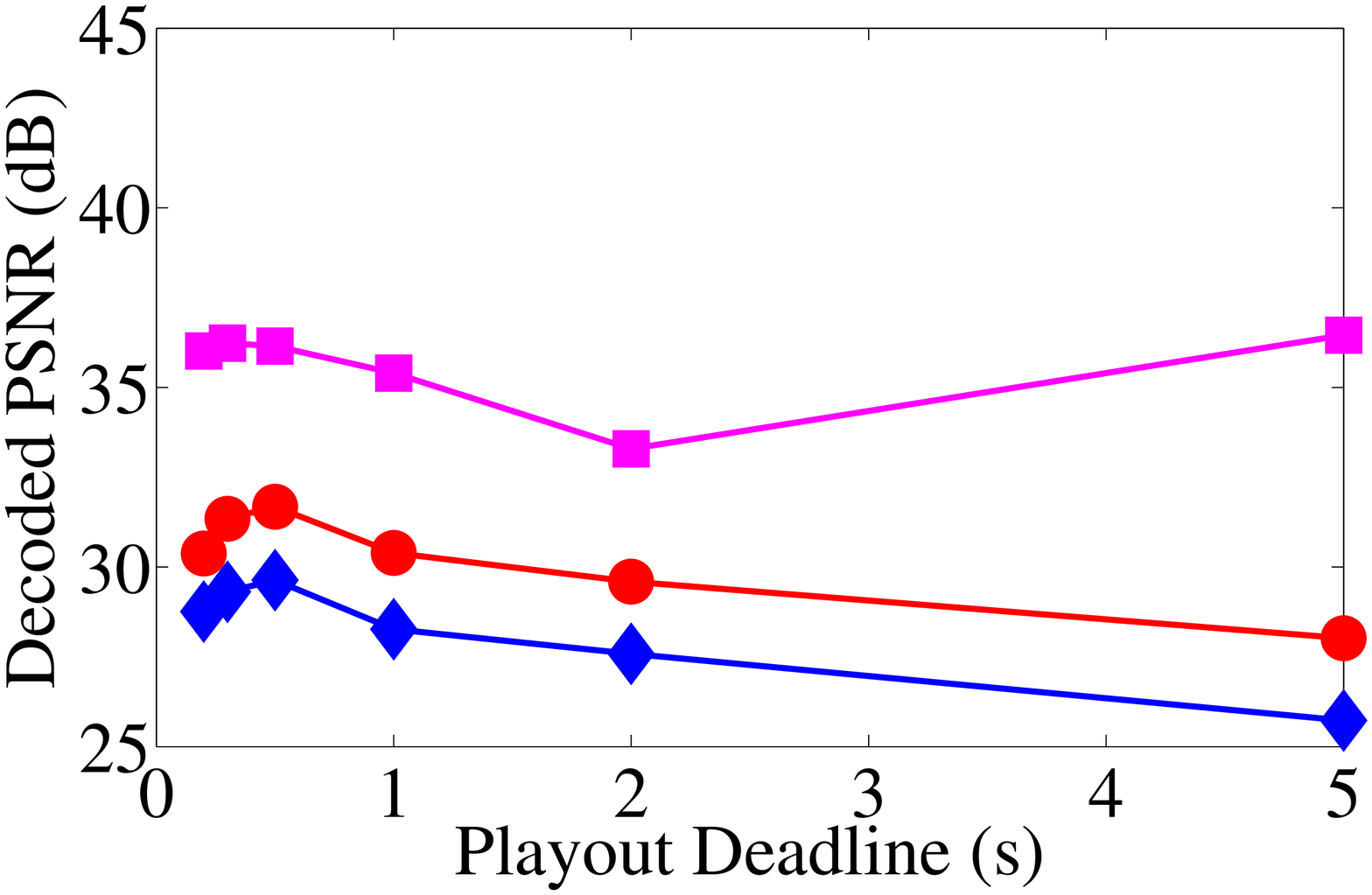}\\
        (d) Rate Proportional \\ AIMD
     \end{center}
   \end{minipage}
\caption{\small  Comparison of aggregated network utilization over
each interface and allocated rate of each stream, as the playout
deadline increases from 200 ms to 5.0 second. Background traffic
load is 20\%.} \label{fig:AllocationPerDeadline}
\end{figure*}

\begin{figure}
\centering
    \includegraphics[width=0.9\columnwidth]{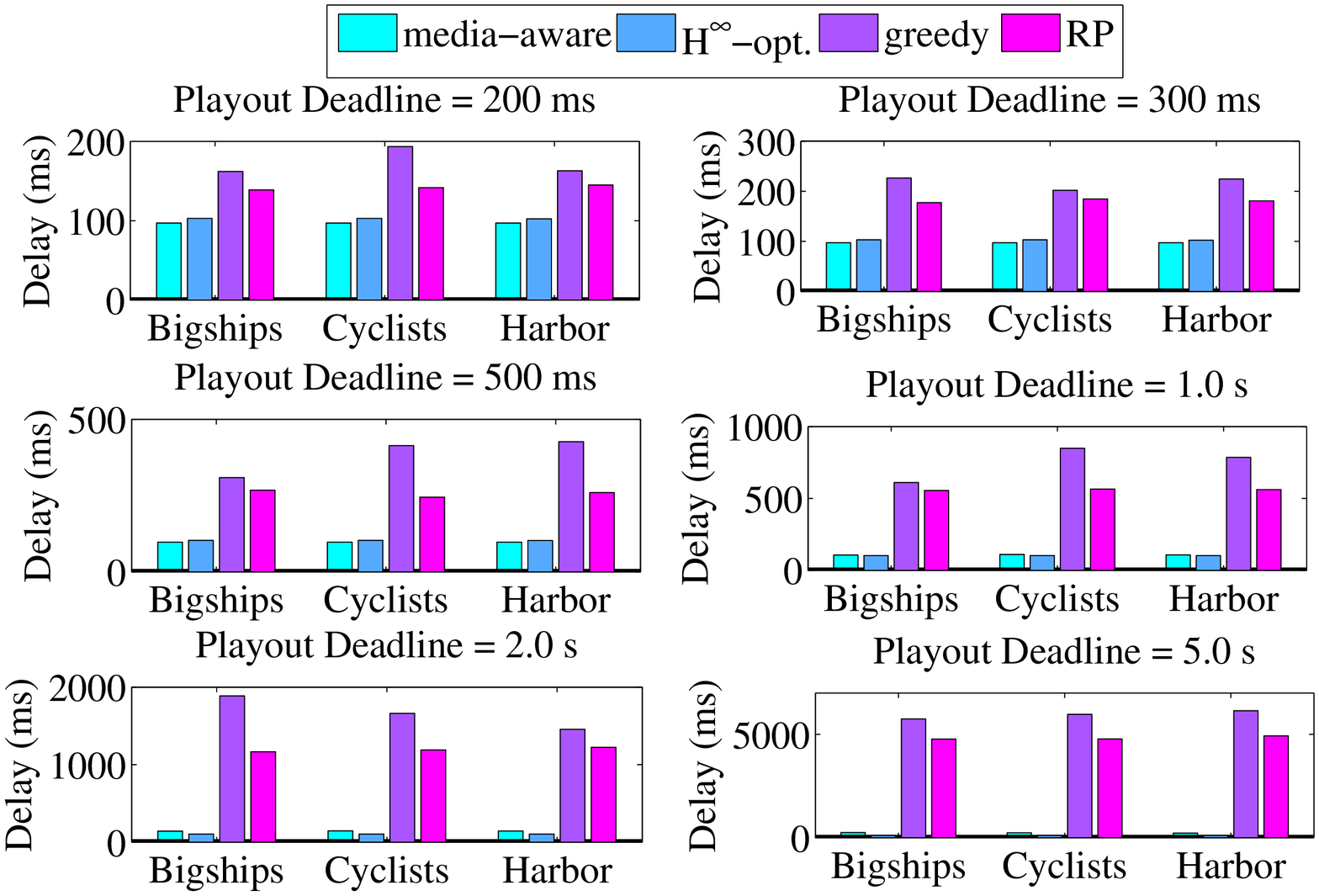}
\caption{\small Average packet delivery delays of each video stream
for playout deadlines ranging from 200~ms to 5.0~s, with 20\%
background traffic load.} \label{fig:DelayPerDeadline}
\end{figure}

\begin{figure}
\centering
    \includegraphics[width=0.9\columnwidth]{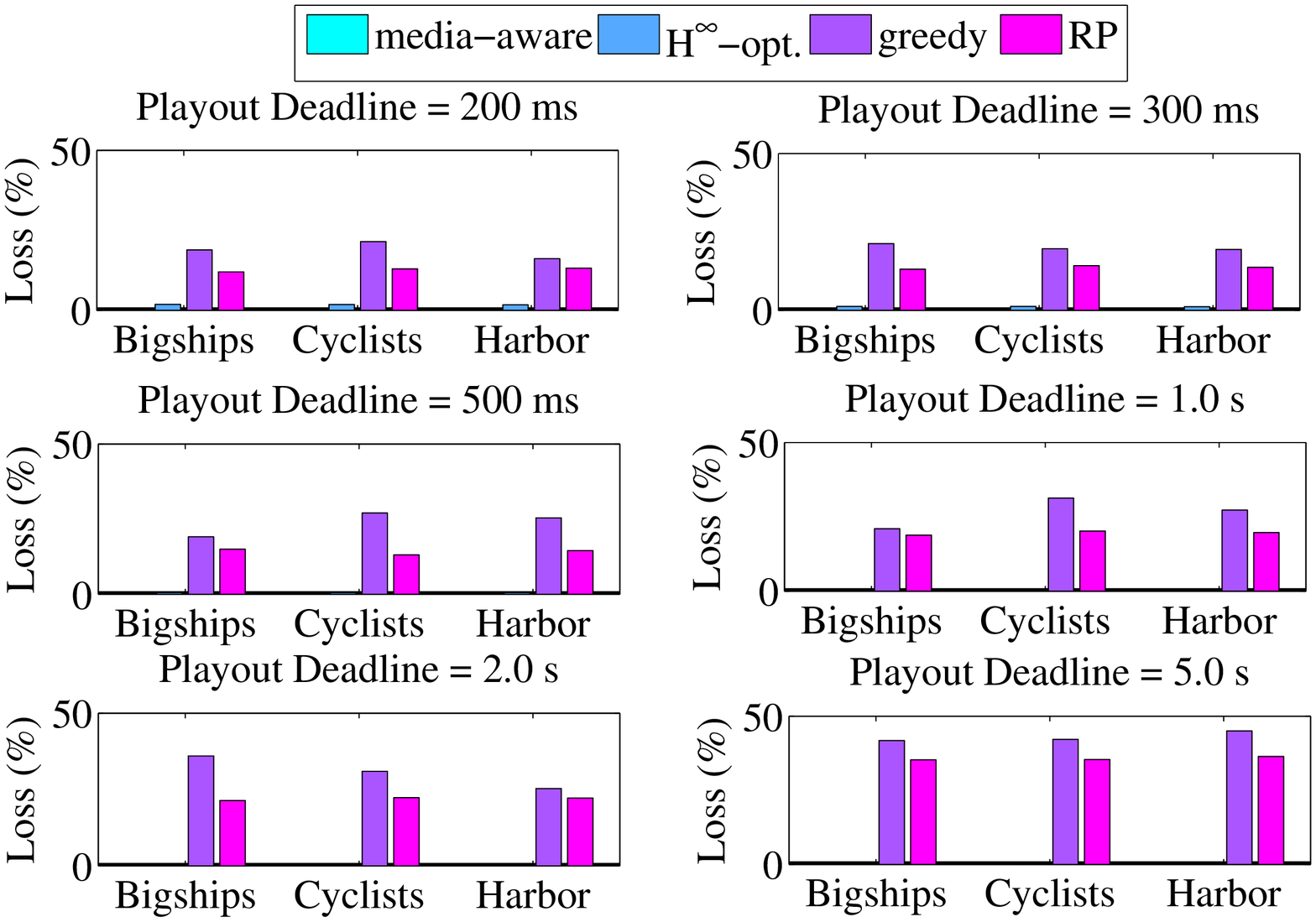}
\caption{\small Average packet loss ratios of each video stream for
playout deadlines ranging from 200~ms to 5.0~s, with 20\% background
traffic load.} \label{fig:LossPerDeadline}
\end{figure}

\begin{figure}
\centering
    \includegraphics[width=0.9\columnwidth]{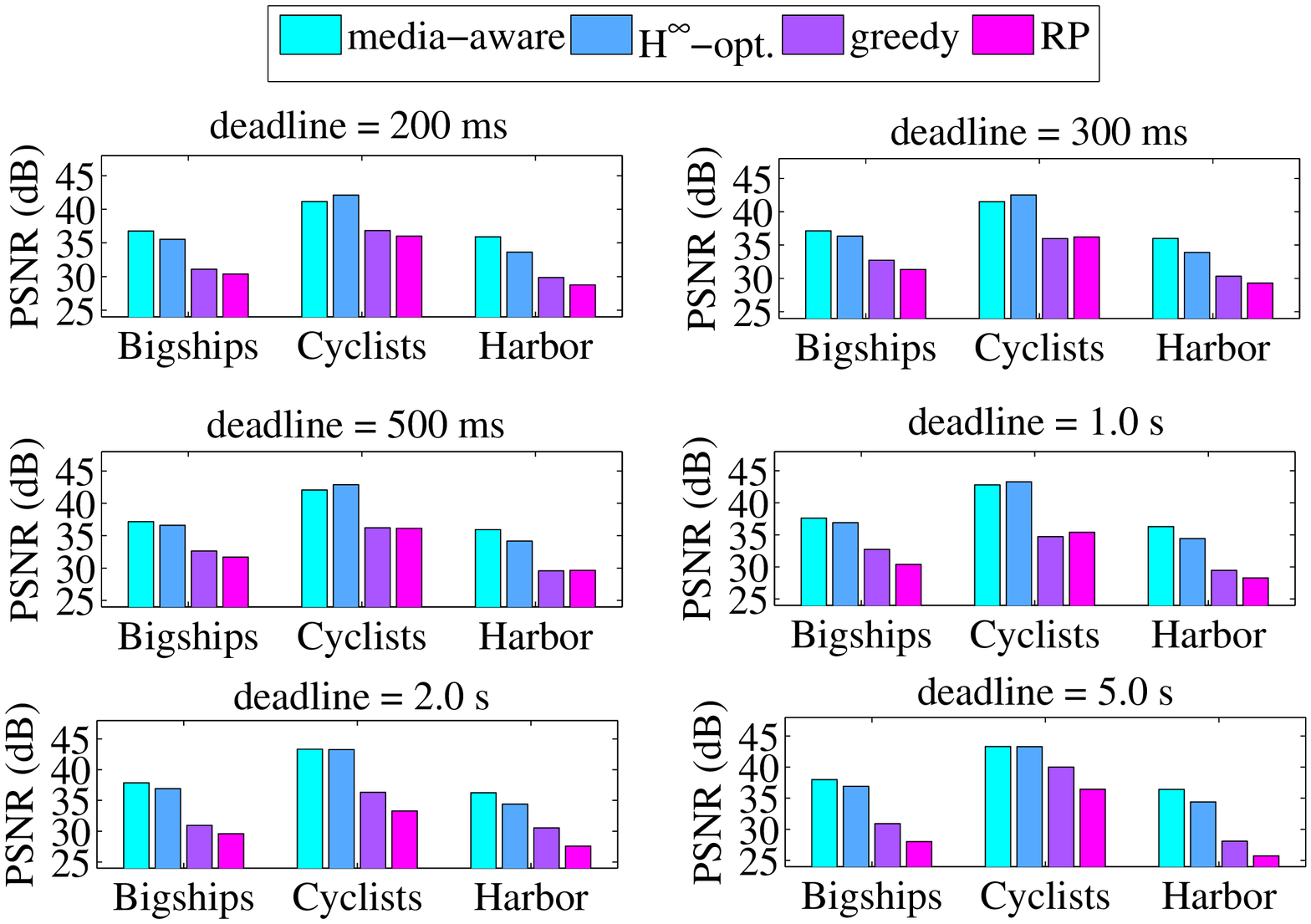}
\caption{\small Received video quality in PSNR of \emph{Bigships},
\emph{Cyclists} and \emph{Harbor} for playout deadlines ranging from
200~ms to 5.0~s, with 20\% background traffic load.}
\label{fig:DecodedPSNRPerDeadline}
\end{figure}

%
In the next set of experiments, we vary the playout deadline for
each video stream from 200~ms to 5.0~seconds, while fixing the
background traffic load at 20\%.
Figure~\ref{fig:AllocationPerDeadline} compares the allocation
results from the four schemes. As the playout deadline increases,
higher network congestion level can be tolerated by each video
stream. The media-aware allocation scheme therefore yields higher
allocated rate and improved video quality, saturating as the playout
deadline exceeds 1.0~second. Allocation from the other three
media-unaware schemes, in comparison, are not so responsive to
changes in the playout deadlines of the video streams.\\
\indent Figures~\ref{fig:DelayPerDeadline} and
\ref{fig:LossPerDeadline} compare the average packet delivery delay
and packet loss ratios due to late arrivals. Similar to results in
the previous section, the media-aware and H$^\infty$-optimal
allocations achieve much lower packet delivery delays and loss
ratios than the two AIMD-based heuristics. The performance gap
increases as the playout deadline becomes more relaxed. The packet
loss ratios are almost negligible (less than 0.1\%) from media-aware
allocation, and very small (less than 2.0\%) from H$^\infty$-optimal
control. In comparison, the packet loss ratios range between
16~-~45\% for greedy AIMD, and between 12~-~37\% for rate
proportional AIMD allocation, far exceeding the tolerance level of
video streaming applications. Consequently, while the average
received video quality of \emph{Bigships} at playout deadline 300~ms
is 34.0~dB and 32.8~dB from the greedy and rate proportional AIMD
schemes, respectively, they are improved to 37.3~dB with media-aware
allocation, and to 36.0~dB with H$^\infty$-optimal control. Similar
results are observed for other sequences with other playout
deadlines, as shown in Fig.~\ref{fig:DecodedPSNRPerDeadline}. The
improvement varies between 3.3~-~10.7 dB in PSNR of the decoded
video. The lower packet delivery delays and packet loss ratios
achieved by the two proposed schemes also indicate that they are
more friendly to ongoing background traffic than the two
AIMD-heuristics, by virtue of more mindful congestion avoidance.\
\section{Conclusions}
%
This paper addresses the problem of rate allocation among multiple
video streams sharing multiple heterogeneous access networks. We
present an analytical framework for optimal rate allocation based on
observed network attributes (available bit rates and round-trip
times) and video distortion-rate~(DR) characteristics, and
investigate a suite of distributed rate allocation policies.
Extensive simulation results demonstrate that both the media-aware
allocation and H$^\infty$-optimal control schemes outperform
AIMD-based heuristics in achieving smaller rate fluctuations, lower
packet delivery delays, significantly reduced packet loss ratios and
improved received video quality. The former benefit from proactive
avoidance of network congestion, whereas the latter adjust the
allocated rates only reactively, \emph{after} detection of packet
drops or excessive delays. The media-aware approach further takes
advantage of explicit knowledge of video DR characteristics, thereby
achieving more balanced video quality and responding to more relaxed
video playout deadline by increasing network utilization. \\
\indent We believe that this work has some interesting implications
for the design of next generation networks in a heterogeneous,
multi-homed environment. Media-aware proactive rate allocation
provides a novel framework for quality-of-service~(QoS) support.
Instead of rigidly reserving the network resources for each
application flow in advance, the allocation can be dynamically
adapted to changes in network conditions and media characteristics.
As the proposed rate allocation schemes are distributed in nature,
they can be easily integrated into wireless devices. Future
extensions to the current work include investigation of measures to
best allocate network resources among different traffic types (e.g.,
web browsing vs. video streaming) and to reconcile their different
performance metrics (e.g., web page refresh time vs. video quality)
as functions of their allocated rates. In addition, our system model
can be further extended to incorporate other types of access
networks employing resource provisioning or admission control.
%
\appendix
%
We now provide the H$^\infty$-optimal control formulation for the
general case of multiple access networks from the perspective of a
single stream $s\in S$. For ease of notation, we drop the
superscript $s$ and define $\mathbf{x}:=[x_n]$, $\mathbf{r}:=[r_n]$,
and $\mathbf{u}:=[u_n]$ for all $n \in {\cal N}$. The counterpart of
the system~\eqref{eqn:SystemStateX} and~\eqref{eqn:RateUpdateR} is
given by:\
%
\begin{equation}
\label{eqn:SystemEquationXR}
\begin{array}{l}
\dot{\mathbf{x}} = A\, \mathbf{x} + B\,\mathbf{u}  + D\,\mathbf{w} \\
\dot{\mathbf{r}} =  -\Phi \,\mathbf{r} + \mathbf{u},
\end{array}
\end{equation}
\noindent where $\mathbf{w}:=[w_n]\;\;\forall n$. Here, the matrices
$A$, $B$, and $\Phi$ are obtained simply by multiplying the identity
matrix by $a$, $b$, and $\phi$, respectively.\\
\indent Correspondingly, system output is:\
%
\begin{equation}
\label{eqn:ControllerOutputMatrix}
\mathbf{z} := H\mathbf{x} + G\mathbf{u},
\end{equation}
%
\noindent The matrix $H$ represents the weight on the cost of
deviation from zero state, i.e. full network utilization. We can
assume that $Q:=H^{T}H$ is positive definite, in that any non-zero
deviation from full utilization leads to a positive cost. Likewise,
the matrix $G$ represents the weight on the cost of deviation from
zero control, i.e., constant allocated rates. We assume that $G^T G$
is positive definite, and that no cost is placed on the product of
control actions and states: $H^{T}G=0$.\\
\indent The cost function is defined as:\
\begin{equation}
\label{eqn:ControllerCostMatrix}
L(\mathbf{x},\mathbf{u},\mathbf{w})=\frac{\|\mathbf{z}\|}{\|
\mathbf{w}\|},
\end{equation}
%
\noindent where $\| \mathbf{z} \|^{2} := \int_{0}^{\infty}|
\mathbf{z}(t)|^{2} dt$ and $\| \mathbf{w} \|^{2} :=
\int_{0}^{\infty}| \mathbf{w}(t)|^{2} dt$. Again, one can
define the worst possible value for cost $L$ as $\gamma^{*}$.\\
\indent Similar to the solutions for the scalar system, we obtain
the H$^\infty$-optimal linear feedback controller for the multiple
network case:\
%
\begin{equation}
\label{eqn:ControllerMatrix}
\mathbf{u} = -(G^TG)^{-1}B^T \Sigma_{\gamma} \mathbf{x}.
\end{equation}
%
\noindent In \eqref{eqn:ControllerMatrix}, the matrix
$\Sigma_{\gamma}$ can be computed by solving the game algebraic
Ricatti equation~(GARE):\
\begin{equation}
\label{eqn:GARE}
A^TZ+ZA-\Sigma(B(G^TG)^{-1}B^T-\gamma^{-2}D D^T)\Sigma+Q = 0.
\end{equation}
%
\noindent It can be verified that a unique minimal nonnegative
definite solution $\Sigma_{\gamma}$ exists for $\gamma >
\gamma^{*}$, if $(A,B)$ is stabilizable and $(A,H)$ is
detectable~\cite{Basar:95}. In our case, since the matrix $B$ is
square and negative definite and the matrix $Q$ is positive
definite, the system is both controllable and observable, hence both
conditions are satisfied.




\end{document}